\documentclass[12pt,reqno]{amsart}

\usepackage{fullpage}

\usepackage{enumitem}
\usepackage{graphicx}
\usepackage{color}
\usepackage{amsmath}
\usepackage{amsfonts}
\usepackage{amssymb}
\usepackage{amscd}
\usepackage{amsthm}
\usepackage{hyperref}

\usepackage{tikz-cd}
\usetikzlibrary{decorations.markings}
\newcommand{\tikzmath}[2][]
{\vcenter{\hbox{\begin{tikzpicture}[#1]#2\end{tikzpicture}}}
}
\tikzstyle{mid}=[decoration={markings, mark=at position 0.55 with {\arrow{>}}}, postaction={decorate}]

\usepackage{mathtools}
\usepackage{mathrsfs}
\usepackage[
backend=biber,
style=alphabetic,
maxnames=9,
maxalphanames=4
]{biblatex}
\renewbibmacro{in:}{}

\DeclareSourcemap{
  \maps[datatype=bibtex]{
    \map{
      \step[fieldsource=note, final]
      \step[fieldset=addendum, origfieldval, final]
      \step[fieldset=note, null]
    }
  }
}

\usepackage{caption}
\usepackage{subcaption}
\DeclareCaptionSubType*{figure}

\numberwithin{equation}{section}

\newtheorem{thm}{Theorem}[section]
\newtheorem*{thm*}{Theorem}
\newtheorem{lem}[thm]{Lemma}
\newtheorem{prop}[thm]{Proposition}
\newtheorem{cor}[thm]{Corollary}
\newtheorem{assum}[thm]{Assumption}

\newtheorem{thmalpha}{Theorem}

\newtheorem{coralpha}[thmalpha]{Corollary}

\theoremstyle{definition}
\newtheorem{defn}[thm]{Definition}
\newtheorem{rem}[thm]{Remark}

\newcommand{\Tr}{\mathop{\mathrm{Tr}}\nolimits}

\newcommand{\caA}{{\mathcal A}}
\newcommand{\caB}{{\mathcal B}}
\newcommand{\caC}{{\mathcal C}}
\newcommand{\caD}{{\mathcal D}}

\newcommand{\caH}{{\mathcal H}}

\newcommand{\caK}{{\mathcal K}}

\newcommand{\caP}{{\mathcal P}}

\newcommand{\caR}{{\mathcal R}}

\newcommand{\caU}{{\mathcal U}}

\newcommand{\caZ}{{\mathcal Z}}

\newcommand{\bbN}{{\mathbb N}}

\newcommand{\bbR}{{\mathbb R}}

\newcommand{\Hom}{\operatorname{Hom}}

\newcommand{\iu}{\mathrm{i}}

\newcommand{\A}{\mathfrak{A}}

\newcommand{\om}{\omega}

\newcommand{\cone}{\Lambda}
\newcommand{\conec}{\Lambda^c}
\newcommand{\ad}{^*}

\newcommand{\qe}{\simeq_{q.e.}}
\newcommand{\restr}{\upharpoonright}
\newcommand{\id}{\mathbb{I}}

\newcommand{\stack}{\otimes_{\mathrm{s}}}

\newcommand{\boxt}{\Delta_{\A_1} \boxtimes \Delta_{\A_2}}
\newcommand{\stacc}{\Delta_{\A_1 \stack \A_2}}

\newcommand{\vN}{\,\overline{\otimes}\,}
\newcommand{\ssc}{{\rm SSC}}
\newcommand{\R}{\mathfrak{R}}

\setlist[enumerate]{label*=(\roman*)}

\title{Stacking and the triviality of invertible phases}

\author{Sven Bachmann}
\address{Department of Mathematics \\ The University of British Columbia \\ Vancouver, BC V6T 1Z2 \\ Canada}
\email{sbach@math.ubc.ca}

\author{Alan Getz}
\address{Department of Mathematics \\ The University of British Columbia \\ Vancouver, BC V6T 1Z2 \\ Canada}
\email{agetz@math.ubc.ca}

\author{Pieter Naaijkens${}^\ast$}
\address{School of Computational and Mathematical Sciences \\ Cardiff University \\ Cardiff, CF24 4AG \\ United Kingdom}
\email{NaaijkensP@cardiff.ac.uk}

\author{Naomi Wray}
\address{School of Computational and Mathematical Sciences \\ Cardiff University \\ Cardiff, CF24 4AG \\ United Kingdom}
\email{WrayNJ@cardiff.ac.uk}

\date{\today}

\begin{document}

\begin{abstract}
We study the superselection sectors of two quantum lattice systems stacked onto each other in the operator algebraic framework. We show in particular that all irreducible sectors of a stacked system are unitarily equivalent to a product of irreducible sectors of the factors. This naturally leads to a faithful functor between the categories for each system and the category of the stacked system. We construct an intermediate `product' category which we then show is equivalent to the stacked system category. As a consequence, the sectors associated with an invertible state are trivial, namely, invertible states support no anyonic quasi-particles.
\end{abstract}

\maketitle
\tableofcontents

\section{Introduction}

In his famous conference~\cite{KitaevSRE}, Kitaev introduced and formalised the dichotomy between short-range and long-range entangled states of quantum lattice systems. Among many other aspects, he defined the notion of invertible states and put them within the class of states that do not have long-range entanglement. He argued in particular that gapped systems with an invertible ground state should not have \emph{anyonic quasi-particle excitations}. They do not exhibit intrinsic topological order and the classification problem is reduced to understanding the chiral central charge on the one hand, and the role of symmetries (namely the symmetry protected topological orders) on the other hand.

Although the first mathematical description of anyons is often attributed to~\cite{LeinaasMyrheim,MGS,WiczekAnyons} in a differential geometric picture, braid statistics appeared simultaneously (in fact, shortly before) in the rigorous quantum field theory literature in~\cite{frohlich1976new}, see also the much more complete~\cite{frohlich2021statistics}. The widespread use of \emph{braided monoidal categories} is more recent although it also was suggested long ago in~\cite{FRS,FroehlichGabbiani}, see also~\cite{Froehlich} and the references therein. It has been widely used in the physics literature since the seminal Appendix~E of~\cite{KitaevAppendices}. The adaptation of the mathematical QFT setting to non-relativistic quantum lattice systems originated in~\cite{naaijkens2011localized,naaijkens2013kosaki} and was fully formalised by~\cite{ogata_derivation_2021} (see also~\cite{MR4927814}). In all of these works, the picture that arises is a classification of anyon types as equivalence classes of certain irreducible representations of the $\rm{C}^*$-algebra of observables, the so-called superselection sectors. 

A key concept in the definition of an invertible state is that of \emph{stacking}.
The stacking operation is the simple fact of considering two physical systems as two layers of one joint system.
If the two quantum systems are described by quasi-local observable algebras $\A_1$ and $\A_2$, we write $\A_1 \stack \A_2$ for the observable algebra of the stacked system.
Mathematically, it is given by the tensor product of $\rm{C}^*$-algebras, which is well-defined since our algebras are nuclear.
For given states $\omega_1$ and $\omega_2$ of $\A_1$ and $\A_2$ respectively, the state $\omega_1 \stack \omega_2$ is defined in the obvious way.
A state is then said to be invertible if, after stacking with a suitable system, the resulting state is in the trivial phase, namely, it can be `adiabatically deformed' to a product state without closing the energy gap.

In this work, we characterise, in full generality, the superselection sectors of stacked systems in terms of those of their components. 
Not surprisingly, we show that they are also obtained as a product of the components: a similar notion to the Deligne product in the language of category theory.
Since product states correspond to a trivial category~\cite{naaijkens2022split}, this immediately reproduces a result by Ogata~\cite{ogata2025mixedstatetopologicalorder}, namely that product states are the neutral elements of the stacking product.
However, our result also applies to stacking with systems in non-trivial phases.
This implies a more interesting corollary of our result, which is nothing else than Kitaev's conclusion, that invertible states have trivial superselection sectors.
That is, an invertible state does not support anyonic excitations, whether abelian or not.

Our setup, which is explained in more details in Section~\ref{Setting}, is that of standard superselection sector theory for 2D quantum spin systems.
The starting point is an irreducible reference representation $\pi_0$.
In the applications that we have in mind, this $\pi_0$ would be obtained as the GNS representation of a pure ground state of a gapped quantum spin system.
We assume that $\pi_0$ satisfies \emph{approximate Haag duality} for cones and that the cone algebras $\pi_0(\cone)''$ are properly infinite (which follows for example if $\pi_0$ comes from a gapped ground state of a finite range interaction~\cite{ogata_derivation_2021}, or if it comes from a pure translation invariant state~\cite{KeylMSW06}).
In this setting, Haag duality was proven first for abelian quantum double models~\cite{naaijkens2015haag}, while the recent~\cite{PerezGarcia} provides a proof of approximate Haag duality for the large class of string-net models, which include all quantum double ground states. Any model that can be adiabatically connected to these models, namely that is in the same phase~\cite{HW2005, BMNS2012}, then satisfies approximate Haag duality by~\cite[Proposition~1.3]{ogata_derivation_2021}.

The assumptions stated above allow us to define a braided $\rm{C}^*$-category $\Delta_\A$ of representations satisfying the superselection criterion~\cite{ogata_derivation_2021}.
Our main results relate $\Delta_{\A_1 \stack \A_2}$ to $\Delta_{\A_1}$ and $\Delta_{\A_2}$. We state them in a slightly informal fashion here and refer to the main text for the details. The first one, Theorem~\ref{thm:SSC}, is rather simple: Stacking superselection sectors of the individual layers yield superselection sectors of the stacked system.

\begin{thmalpha}
    \label{thm:ssstacked}
   If $\rho_1$ and $\rho_2$ are representations of $\A_1$ and $\A_2$ that satisfy the superselection criterion for $\pi_1$ and $\pi_2$, then $\rho_1 \stack \rho_2$ satisfies the superselection criterion for $\pi_1 \stack \pi_2$.
\end{thmalpha}

\noindent In categorical terms, this naturally leads to a faithful functor $F: \Delta_{\A_1} \boxtimes \Delta_{\A_2} \to \Delta_{\A_1 \stack \A_2}$, where $\boxtimes$ is an appropriate notion of the `tensor product' of the categories of superselection sectors, which will be explained in Section~\ref{Deligne}.

The second result, Theorem~\ref{thm:B}, is the technical heart of this work and is concerned with the converse of the above, namely that every irreducible representation in $\Delta_{\A_1 \stack \A_2}$ is of this form. This requires a `Type I condition', namely that every factorial representation $\sigma$ satisfying the superselection criterion for $\pi_1$ is such that $\sigma(\A_1)''$ is a Type I factor.

\begin{thmalpha}
\label{thm:ssstackedirreducible}
If the Type I condition holds for $\pi_1$, then any irreducible representation $\rho$ which satisfies the superselection criterion for~$\pi_1 \stack \pi_2$ is of the form $\rho \cong \rho_1 \stack \rho_2$ with irreducible $\rho_1,\rho_2$ satisfying the superselection criterion for $\pi_1,\pi_2$ respectively.
\end{thmalpha}

Together, these two results show that the sectors of the stacked system are exactly given by products of sectors of the individual layers. Not surprisingly, this can be lifted to the whole braiding structure, yielding the third result, Theorem~\ref{thm:C} in Section~\ref{Deligne}:
\begin{thmalpha}
   \label{thm:deligne}
   The braided tensor $\rm{W}^*$-categories $\Delta_{\A_1} \boxtimes \Delta_{\A_2}$ and $\Delta_{\A_1 \stack \A_2}$ are equivalent.
\end{thmalpha}

Since the superselection theory of a state in the trivial phase is trivial in the sense that the only sectors are direct sums of the reference representation, the above results immediately imply the following corollary, see Corollary~\ref{cor: D}.

\begin{coralpha}
    \label{cor:invstate}
    The superselection structure of invertible phases is trivial. 
\end{coralpha}

One may then wonder whether the operator-algebraic Type I condition follows from any physically reasonable assumption. That this is true, Theorem~\ref{Thm:E}, is an analogue of a result in~\cite{KLM01}, which states that under reasonable conditions, any representation satisfying the superselection criterion is a direct sum of at most countably many irreducible representations satisfying the superselection criterion.

\begin{thmalpha}
   \label{Thm: Type I}
   If $\pi_0$ has only countably many superselection sectors and satisfies the approximate split property, then any representation $\rho$ satisfying $\ssc(\pi_0)$ can be decomposed into a direct sum of irreducible representations, each of which also satisfy $\ssc(\pi_0)$. Moreover, if $\rho$ is factorial, then it is Type I.
\end{thmalpha}

The assumption that there are at most countably many irreducible superselection sectors, or anyon types, is a common physical assumption.  The word `anyon theory' is usually defined to be equivalent to a unitary modular tensor category (UMTC), which implies even finitely many irreducible sectors. The \emph{approximate split property} (defined precisely in Definition~\ref{Approximate Split Property} below) is an assumption on the decay of correlations. The arguments in Section~\ref{sec:teesplit} combine to prove that the approximate split property holds for all quantum phases having one generalized Levin-Wen tensor-network state representative, in particular for all models that satisfy the entanglement bootstrap axioms~\cite{KimRanard}. Indeed, these models satisfy a slightly stronger version of the approximate Haag duality, see again~\cite{PerezGarcia}, and they also satisfy a strict form of the area law for the entanglement entropy. We show, see Corollary~\ref{cor: ASP with xi and R}, that these two assumptions imply the approximate split property. It remains to appeal to~\cite{naaijkens2022split} to deduce the approximate split property for all states in the phase.

Although perhaps not surprising, the claim in Theorem~\ref{Thm: Type I} that any representation satisfying the superselection criterion is a direct sum of irreducible representations is a somewhat subtle result, since a priori one could expect proper direct integrals instead. This emphasizes a technical subtlety about fusion: Without additional assumptions, there may be uncountably many irreducible factors in the decomposition of a product of representations, and they may not individually satisfy the superselection criterion. In the literature one is often interested (sometimes only implicitly) in \emph{dualizable} sectors only, where this issue does not play a role (see Remark~\ref{rem: decomposition} for details).

We shall conclude this work with a few examples in Section~\ref{Examples}. For the quantum double models, our results imply that the superselection sector theory of a model build from a direct product of groups is the stacking product of the sector theories for the models built from each factor. For Levin--Wen models, they imply that the sector theory is an invariant under Morita equivalence of the categories used to define the model. This does not require any \emph{a priori} knowledge of the sector theory, but since Morita-equivalence is defined by the equivalence of the Drinfeld centers, it shows that the Drinfeld center completely characterizes the sector theory. Our last example is that of a symmetry-enriched toric code model.

\textbf{Acknowledgements:} 
We would like to thank Alex Bols, Corey Jones, Kohtaro Kato, Dave Penneys, Frank Verstraete, and Daniel Wallick for insightful and helpful discussions related to this work.
NW acknowledges funding by the Engineering and Physical Sciences Research Council [EP/W524682/1]. 
AG and NW acknowledge funding from the Welsh Government via a Taith Research mobility grant. 
AG and SB acknowledge support of NSERC of Canada as well as the European Commission under the grant \emph{Foundations of Quantum Computational Advantage}.

\section{Setting}\label{Setting}
\subsection{Algebras, states, and representations}

Let $\Gamma \subset \mathbb{R}^2$ be a countable set representing the sites of our quantum spin systems.
In typical applications $\Gamma$ will be for example a lattice, but for our purposes it is sufficient to assume that the number of sites contained in a ball of radius $r$ does not grow too fast with $r$.
This is true if $\Gamma$ is a Delone set.

To each site $x \in \Gamma$ we assign a finite dimensional Hilbert space $\caH_x$.
We assume that $d_x := \dim \caH_x$ is uniformly bounded.
For $P_f(\Gamma)$ the set of finite subsets $\cone_f \subset \Gamma$, we can associate a composite Hilbert space $\caH(\cone_f) := \bigotimes_{x \in \cone_f} \caH_x$ with a corresponding algebra of observables $\A(\cone_f) = B(\caH(\cone_f)) = \bigotimes_{x \in \cone_f} M_{d_x}(\mathbb{C})$.
With the natural inclusion of algebras ($\A(\cone_1) \hookrightarrow \A(\cone_2)$ when $\cone_1 \subset \cone_2 \subset \Gamma$) we can define a net of $\rm{C}^*$-algebras  $\cone_f \mapsto \A(\cone_f)$ which satisfies $[\A(\cone), \A(\widehat{\cone})] = \{0\}$ whenever $\cone \perp \widehat{\cone}$, where the orthogonality relation $\perp$ is given by disjointness.
The local observables are then $\A_{\rm loc} = \bigcup_{\cone_f \subset P(\Gamma)}\A(\cone_f)$, from which we get a $\rm{C}^*$-algebra of observables as the inductive limit of the net: $\A := \overline{\A_{\rm loc}}^{||\cdot||}$.
Note that $\A$ is simple by~\cite[Prop. 11.4.2]{kadison1997fundamentals2}.
The local net $\cone \mapsto \A(\cone)$ of $\rm{C}^*$-subalgebras with the orthogonality relation of disjointness is a (bosonic) quasi-local algebra~\cite[Def.~2.6.3]{Bratteli_Robs}, and we shall refer to any such algebra as a quantum spin system.

For a possibly infinite subset $\cone\subset\Gamma$, the operators localized in this region are given by $\A(\cone) := \overline{\bigcup_{\cone_f \subset P_f(\cone)} \A(\cone_f)}^{||\cdot||} \subset \A$. In this paper, our only consideration is with such infinite regions we call cones, namely sets of the form
\begin{equation*}
    \cone_{\vec z,\theta,\phi} = \Gamma\cap \{\vec z + r \vec e_\beta: r>0, \beta\in(\theta-\phi,\theta+\phi)\}
\end{equation*}
for any $\vec z\in\mathbb{R}^2$, and angles $\theta\in\mathbb{R}$, $\phi\in(0,\pi)$, where $\vec e_\beta = (\cos\beta,\sin\beta)$.
We write $\arg(\Lambda_{\vec{z}, \theta, \phi}) = (\theta-\phi, \theta+\phi)$ and $\operatorname{Arg}(\Lambda_{\vec{z}, \theta, \phi}) = 2 \phi$.
Note that in this definition a cone can have an opening angle greater than $\pi$, as long as it does not cover the whole space. The advantage of this definition is that the interior of the complement of a cone is again a cone. Figure~\ref{fig:cone} visualises this set and its parameters.

\begin{figure}
    \centering
    \begin{tikzpicture}
\clip (-0.25,-0.25) rectangle (4.2,4.2);
\foreach \x in {0,0.5,1,1.5,2,2.5,3,3.5,4}{
        \foreach \y in {0,0.5,1,1.5,2,2.5,3,3.5,4}{
        \filldraw (\x, \y) circle (0.03cm);
        }}
\draw[thick] (4,10.59) -- (1.12,0.63) -- (6.54,5);
\filldraw[blue, opacity=0.25] (4,10.59) -- (1.12,0.63) -- (6.54,5) -- cycle;
\draw[->, very thick] (1.12,0.63) -- (1.9,1.77);
\node[black] at (0.8,0.8) {$\vec{z}$};
\node[black] at (2.25,2.15) {$\vec{e}_\theta$};
\draw[very thick, blue, <->] (1.12,0.63) ++(73.87:3.05) arc (73.87:38.88:3.05);
\node at (3.25,3.25) {$2\phi$};
\end{tikzpicture}
    \caption{A cone at apex $\vec{z}$ with widening angle $\phi$. The points of the lattice inside the cone (blue shaded) make up the set $\cone_{\vec z,\theta,\phi}$. }
    \label{fig:cone}
\end{figure}

The following is a physically relevant family of automorphisms, namely those that preserve locality sufficiently well. One may pick slightly different definitions, but it is important that the spectral flow~\cite{BMNS2012}, or quasi-adiabatic continuation~\cite{HW2005}, falls into the family one chooses. Denote by $d(\cdot, \cdot)$ the metric on $\Gamma$. Then for a finite set $\cone_f \in P_f(\Gamma)$, the diameter of $\cone_f$ is defined as $\mathrm{D}(\cone_f):= \sup_{x,y \in \cone_f} d(x,y)$. 
A (time-dependant) interaction $\Phi$ is a family of maps $\Phi_t:\mathcal{P}_f(\Gamma)\to\A$ such that $\Phi_t(\Lambda_f) = \Phi_t(\Lambda_f)^*\in\A(\Lambda_f)$ and $t\mapsto\Phi_t(\Lambda_f)$ is piecewise continuous.
We further assume some decay of the interactions, namely that
\begin{equation}\label{eq:interaction}
    \sup_{t\in\bbR}\sup_{x\in\Gamma}\sum_{\Lambda_f\ni x}\frac{\Vert\Phi_t(\Lambda_f)\Vert}{\xi(\mathrm{D}(\Lambda_f)+1)}<\infty
\end{equation}
for a positive non-increasing function $\xi$ such that $\sup_{r\geq1}r^n\xi(r)$ is finite for all $n\in\bbN$. 
The interaction $\Phi$ is defined to have finite range if there exists some $d^{\Phi} \geq 1$ such that $\Phi(\cone_f)=0$ whenever $\mathrm{D}(\cone_f)>d^{\Phi}$. In this case, the only non-zero contributions to the energy arise from finite clusters of neighbouring sites. If~\eqref{eq:interaction} holds, then the map $A\mapsto\sum_{\Lambda_f\in\caP_f(\Gamma)}\iu[\Phi_t(\Lambda_f),A]$ is well-defined on $\A_{\rm loc}$ and it can be extended to a densely defined, unbounded *-derivation $\delta^{\Phi_t}$ of $\A$. This derivation generates a strongly continuous one-parameter group of automorphisms $\{\alpha_t^\Phi:t\in\bbR\}$ defined as the solution of
\begin{equation*}
    \frac{d}{dt}\alpha_t^\Phi(A) = \alpha_t^\Phi(\delta^{\Phi_t}(A)),\qquad \alpha_0^\Phi(A) = A,
\end{equation*}
for all $A$ in a dense domain. The existence of $\alpha_t^\Phi$ is a non-trivial fact which follows for example from the Lieb-Robinson bound \cite{Robinson1976Properties,Nachtergaele2006}, see~\cite{Bachmann2022} for the specific case $\xi(r) = \exp(-b x^p)$ with $0<p\leq 1$. A locally generated automorphism (LGA) is an automorphism $\alpha$ such that $\alpha = \alpha_1^\Phi$ for an interaction $\Phi$.

A state is a positive linear functional $\om: \A \rightarrow \mathbb{C}$ such that $\om(A\ad A) \geq 0 $ for all $ A \in \A$ and $\om(\mathbb{I}) = 1$. The set of pure states of $\A$ is denoted $\caP(\A)$. For $\om$ a state of $\rm{C}^*$-algebra $\A$, there exists a representation $\pi_\om: \A \rightarrow B(\caH_\om)$ on a Hilbert space $\caH_\om$, called the GNS representation, such that $\om(A) = \langle \Omega, \pi_\om(A) \Omega \rangle$, where $\Omega$ is a cyclic vector in $\caH_\om$, i.e. $\pi_\om (\A) \Omega$ is dense in $\caH_\om$. We shall refer to $(\caH_\om, \pi_\om, \Omega)$ as the GNS triple of $\om$. A state $\om$ is a gapped ground state of an interaction $\Phi$ if there is $g>0$ such that
\begin{equation*}
    -\iu \om(A^*\delta^\Phi(A))\geq g\omega(A^*A)
\end{equation*}
for all $A\in\A_{\rm loc}$ such that $\omega(A)=0$.

Our goal is to classify certain physically relevant representations of $\A$, which correspond to `charged states' or \emph{anyons}. These are given by the following superselection criterion, see~\cite{BuchholzFredenhagen,ogata_derivation_2021}.

\begin{defn}\label{SSC}
Let $(\caH_0,\pi_0)$ be an irreducible representation of $\A$. A representation $(\caH,\pi)$ of $\A$ satisfies the superselection criterion (SSC) with respect to $\pi_0$ if 
\begin{equation*}
\pi|_{\A(\conec)} \cong \pi_0|_{\A(\conec)}
\end{equation*}
for all cones $\Lambda$, where $\Lambda^c$ is the complement of $\Lambda$ in $\Gamma$.
In this case, we say that $\pi$ satisfies $\ssc(\pi_0)$.
\end{defn} 
\noindent Note that here and below $\cong$ stands for unitary equivalence of representations.

We always make the following assumption on the reference representation.
\begin{assum}
\label{Assum:irrep}
The reference representation $\pi_0$ is irreducible.
\end{assum}
\begin{rem}\label{rem:separability}
Since we consider quantum spin systems, the quasi-local algebras we consider are UHF algebras, and this assumption implies that $\pi_0$ is defined on a separable Hilbert space. Indeed, irreducibility implies that every vector in the representation space is cyclic and we can find a countable dense (in norm) subset of $\A$. Hence every representation $\pi$ satisfying $\ssc(\pi_0)$ is also defined on a separable Hilbert space as the superselection criterion provides a unitary map between the two Hilbert spaces.
\end{rem}

Unitary equivalence splits the set of all representations satisfying $\ssc(\pi_0)$ in disjoint classes called \emph{superselection sectors}, or simply sectors. A representation which is equivalent to $\pi_0$ is in the trivial sector. 
We say that a sector is irreducible if its representatives are irreducible representations.

The aim of superselection sector theory is to show that the set of sectors has a very rich structure, namely that of a braided tensor category.
To ensure that we can take direct sums and subobjects (that is, restrict to an invariant subspace of some representation) as well, we need to make the following technical assumption.
In fact, this assumption is crucial in our analysis, as it will allow us to show certain $*$-isomorphisms are unitarily implemented, allowing us to replace quasi-equivalence of representations with unitary equivalence.
\begin{assum}\label{Assum:GGS}
    The reference representation $\pi_0$ is such that for all cones $\cone$ the algebra $\pi_0(\A(\Lambda))''$ is a properly infinite factor.
\end{assum}
If $\pi_0$ is irreducible, as will always be the case for us, the algebra $\pi_0(\A(\Lambda))''$ is automatically a factor.\footnote{We are relying here on the fact that we are considering quantum spin systems, for which $\A(\Lambda)$ and $\A(\Lambda^c)$ generate the whole of $\A$, so we can argue as in e.g.~\cite[\S 2]{KeylMSW06}.
This need not be true for so-called \emph{abstract} quantum spin systems, in which case the cone algebra may fail to be a factor even if the ground state representation is irreducible.
}
The properly infiniteness assumption is satisfied, for example, if $\pi_0$ is the GNS representation of a gapped ground state $\omega_0$ of a finite range interaction $\Phi$, see~\cite[Lemma~5.3]{ogata_derivation_2021}.
If $\pi_0$ is the GNS representation of a pure translation invariant state, properly infiniteness also follows by adapting the argument of~\cite[Prop. 5.3]{KeylMSW06} to the 2D setting (this needs that cones widen to infinity, so that any local observable can be translated into the cone).
One can always apply a stabilisation procedure to get properly infinite factors~\cite{ogata2025mixedstatetopologicalorder}.

\begin{rem}
One could alternatively use the notion of \emph{absorbing representations} of the cone algebras in stating this assumption~\cite{MR4927814}.
A normal reperesentation $\pi$ of a von Neumann algebra $\mathcal{R}$ is called \emph{absorbing} if for any other normal representation $\rho$, we have $\pi \oplus \rho \cong \pi$.
Assumption~\ref{Assum:GGS} is equivalent to the defining representation of the cone algebras being absorbing by a slight modification of~\cite[Cor. 2.12]{MR4927814} (it is enough to show that $\pi_0(\A(\cone))'$ contains a properly infinite subalgebra, which follows because it contains a cone algebra using approximate Haag duality, cf. Assumption~\ref{Assum:approxhd} below).
\end{rem}

For our construction, we require the reference representations to satisfy a notion of Haag duality -- in this case, approximate Haag duality. For a cone $\Lambda = \cone_{\vec z,\theta,\phi}$, we denote $\textbf{e}_\cone := \textbf{e}_\theta$ and $\cone_\varepsilon := \cone_{\vec z,\theta,\phi+\varepsilon}$. In particular, the cone $\cone_{\varepsilon}-R\textbf{e}_\cone$ is a (backwards) translation and widening of the original cone $\cone$, see Figure~\ref{fig: approximate haag duality visualisation}. Clearly, $\cone \subset \cone_{\varepsilon}-R\textbf{e}_\cone$ whenever $R>0$. 

\begin{defn}[Approximate Haag duality, \cite{ogata_derivation_2021}]\label{def:HaagD}
    Let $(\caH,\pi_0)$ be an irreducible representation of $\A$. Then $\pi_0$ satisfies \textit{approximate Haag duality} if for any $\varphi \in (0,2\pi)$ and $\varepsilon>0$ with $\varphi+4\varepsilon<2\pi$, there is some $R \geq 0$ and decreasing functions $f_\delta(t)$, where $\delta > 0$, with $\lim_{t\to\infty}f_\delta(t)=0$ such that: 
    \begin{enumerate}
        \item \label{it:ahdinclusion} for any cone $\cone \subset \Gamma$ with $|\text{arg}\cone|=\varphi$, there is a unitary $U:\caH\to\caH$ satisfying 
        \begin{equation}\label{eqn: AHDi}
            \pi_0(\A(\conec))'\subset\text{Ad}(U) \left(\pi_0(\A(\cone_\varepsilon-R\textbf{e}_\cone))''    \right),
        \end{equation} and 
        \item \label{it:ahdapprox} for any $\delta>0$ and $t\geq0$ there is a unitary $\tilde{U} \in \pi_0(\A(\cone_{\varepsilon+\delta}-t\textbf{e}_\cone))''$ satisfying \[||U-\tilde{U}|| \leq f_\delta(t).\]
    \end{enumerate}
\end{defn}
\noindent Note that (ii) should be understood as the fact that the unitary $U$ in (i) is approximately localized in the cone $\Lambda$.

\begin{figure}
    \centering
    \begin{tikzpicture}
\clip (-0.25,-0.25) rectangle (4.2,4.2);
\foreach \x in {0,0.5,1,1.5,2,2.5,3,3.5,4}{
        \foreach \y in {0,0.5,1,1.5,2,2.5,3,3.5,4}{
        \filldraw (\x, \y) circle (0.03cm);
        }}
\draw[thick] (4,10.59) -- (1.12,0.63) -- (6.54,5);
\filldraw[blue, opacity=0.25] (4,10.59) -- (1.12,0.63) -- (6.54,5) -- cycle;
\draw[->, very thick] (1.12,0.63) -- (1.9,1.77);
\node[black] at (2.25,2.15) {$\vec{e}_\theta$};
\draw[very thick, blue] (1.12,0.63) ++(73.87:3.05) arc (73.87:38.88:3.05);
\node at (3.25,3.25) {$2\phi$};
\draw[thick] (1,10.59) -- (0.75,0.13) -- (7.54,3);
\filldraw[red, opacity=0.25] (0.75,0.13) -- (1,10.59) -- (4,10.59) -- (1.12,0.63) -- (6.54,5) -- (7.54,3)-- cycle;
\draw[->, very thick, red]  (1.12,0.63) -- (0.75,0.13);
\draw[very thick, red, ->] (0.2241,-0.0923) ++(38.88:3.5) arc (38.88:22.91:3.5);
\draw[very thick, red, ->] (0.7297,-0.7192) ++(73.87:3.65) arc (73.87:88.63:3.65);
\node[red] at (0.5,0.25) {$R$};
\node[red] at (3.5, 1.75) {$\varepsilon$};
\node[red] at (1.25, 3.2) {$\varepsilon$};
\end{tikzpicture}
    \caption{A visualisation of Approximate Haag duality. The algebra on the right hand side of the set inclusion \ref{eqn: AHDi} is of the red, wider cone with opening angle $\phi +\varepsilon$ and shift in the direction of the negative cone vector.}
    \label{fig: approximate haag duality visualisation}
\end{figure}

\begin{assum}\label{Assum:approxhd}
    The reference representation $\pi_0$ satisfies approximate Haag duality.
\end{assum}

Haag duality has been shown for abelian quantum double models~\cite{naaijkens2015haag}.
As pointed out in the introduction, approximate Haag duality has recently been proven~\cite{PerezGarcia} to hold for all string net states, and by stability~\cite[Proposition~1.3]{ogata_derivation_2021} to all states in the topological phases that have a string net representative. This is believed to cover all non-chiral quantum phases in 2D.

We may identify $\pi_0(\A(\Lambda))$ with $\A(\Lambda)$ whenever it is clear from the context which representation is used.
Note that we can do this because $\A$ is simple and hence $\pi$ automatically is injective. 
Going forward, we will always assume that our reference representations satisfy Assumptions~\ref{Assum:irrep}, \ref{Assum:GGS} and~\ref{Assum:approxhd}.
This is enough to guarantee the existence of a braided $\rm{C}^*$-category of superselection sectors~\cite{ogata_derivation_2021}.

\subsection{Stacking}

Given two quantum spin systems $\A_1, \A_2$, we can form a new stacked system denoted $\A_1 \stack \A_2$. By this, we mean the quasi-local algebra constructed from the local net
$$
    P_f(\Gamma)\ni \Lambda_f\mapsto (\A_1 \stack \A_2)(\Lambda_f) := \A_1(\Lambda_f)\otimes \A_2(\Lambda_f),
$$
where the product on the right hand side is the standard tensor product of finite-dimensional matrices. This defines the product $\stack$, which must be well differentiated from the `local' product used in the definition of a quasi-local algebra.
That is, we reserve $\otimes$ for tensor products within a layer, whereas $\stack$ is the tensor product of algebras (or states) between layers.
Here and in the following, we refer to the individual components of a stacked system as its `layers' since this corresponds to the physical picture of putting two systems on top of each other.

\begin{rem}
\label{rem:differentlattices}
Without loss of generality we may assume that both quantum spin systems are defined on the same set of underlying sites.
Indeed, write $\Gamma := \Gamma_{\A_1} \cup \Gamma_{\A_2}$, where $\Gamma_{\A_1}$ (resp. $\Gamma_{\A_2}$) is the underlying set of sites for the quasi-local algebra $\A_1$ (resp. $\A_2$).
Then for the first stacked layer, we can place a trivial Hilbert space $\mathbb{C}$ on each site in $\Gamma_{\A_2} \setminus \Gamma_{\A_1}$, and leave the sites in $\Gamma_{\A_1}$ as before.
The quasi-local algebra $\A_1^\Gamma$ built on this new set is clearly isomorphic to the original $\A_1$.
We can do the same for the second layer.
\end{rem}

We denote by the same symbol the stacking product of states $\omega_1\stack\omega_2$ as well as that of representations $\pi_1\stack\pi_2$ of $\A_1\stack\A_2$. Both are defined on simple products by $(\pi_1\stack\pi_2)(A_1\stack A_2) = \pi_1(A_1) \stack \pi_2(A_2)$, and extended by continuity to all of $\A_1\stack\A_2$.

\subsection{Invertible states}

While stacking is a general operation on any spin systems, it is central in Kitaev's definition of the class of invertible states~\cite{KitaevSRE}.

\begin{defn}
A state $\om$ of a quantum spin system $\A$ is called a product state for the cone $\cone$ if it is of the form \[\om = \om_{\cone} \otimes \om_{\conec}.\] 
where $ \om_{\cone}$ (resp. $\om_{\conec}$) is a state of $\A(\Lambda)$ (resp. $\A(\conec)$).
\end{defn} 
\noindent Note that the factors $\om_{\cone}$ and $\om_{\conec}$ uniquely determine the product state $\omega$. 

An important result regarding product states for a cone is that they have trivial superselection theory~\cite[Theorem 4.5]{naaijkens2022split}.

\begin{thm}
    Let $\om_0$ be a pure state such that its GNS representation $\pi_0$ is equivalent to $\pi_{\cone} \otimes \pi_{\conec}$ for some cone $\Lambda$. Then the corresponding sector theory is trivial, in that every representation $\pi$ satisfying SSC($\pi_0$) is quasi-equivalent to $\pi_0$. 
\end{thm}

Since the superselection theory is stable under application of an LGA \cite{naaijkens2022split,ogata_derivation_2021}, the same is true for any state that can be connected to a product state (for cones) using such an automorphism, and such states do not host anyons.
The application of this to invertible states is direct, as below.

\begin{defn}\label{Invertible state definition}
A pure state $\om$ of a quantum spin system $\A$ is called invertible if there exists a quantum spin system $\bar \A$ and a pure state $\bar\om$ of $\bar\A$, an LGA $\alpha$ of $\A\stack\bar\A$, and a cone $\cone$ such that $(\omega \stack \bar\omega) \circ \alpha$ is a product state for the cone $\cone$.
\end{defn}  

We shall say that $\bar\omega$ is an inverse of $\omega$. Explicitly, this reads
\begin{equation*}
    (\omega \stack \bar\omega) \circ \alpha = \phi_{\cone} \otimes \phi_{\conec}
\end{equation*}
where $\phi_{\cone}$ (resp. $\phi_{\conec}$) is a state of $(\A\stack\bar\A)(\Lambda)$ (resp. $(\A\stack\bar\A)(\conec)$).
We emphasise the difference between the tensor products of the left and right hand side of the equation: the left-hand side refers to the stacking of two layers, where the right-hand side corresponds to a geometric partition into a cone and its complement.

\begin{rem}
Note that we only need to assume that $(\omega \otimes_s \overline{\omega}) \circ \alpha$ is a product state with respect to a \emph{single} cone $\Lambda$.
This is sufficient because the superselection criterion is already a very strong condition, and allows us to localize the representation in the cone $\Lambda$.
\end{rem}
 
\section{Superselection sectors of stacked systems}\label{Main results}
We now are in a position to discuss the main results of our paper, namely the analysis of the superselection sector theory of stacked systems.
Recall that an important class of examples we want to apply the theory to is that of ground states of gapped finite range interactions.
Observe that if $\om_1,\om_2$ are pure gapped ground states of finite range interactions $\Phi_1,\Phi_2$, then $\omega = \om_1\stack\om_2$ is a pure state and is the gapped ground state of the finite range interaction defined by $\Phi(\Lambda) := \Phi_1(\Lambda)\stack \mathbb{I} + \mathbb{I}\stack\Phi_2(\Lambda)$.
In particular, its GNS representation $\pi$ is irreducible.

\subsection{Well-definedness of the sector theory for stacked systems}

In order to be able to discuss the superselection sectors of stacked systems, we first show that the stacked system again satisfies our assumptions.
In particular, approximate Haag duality is inherited from the individual layers.

Used frequently below is a result regarding the commutants of tensor products of $\rm{C}^*$-algebras, which we state here for convenience of the reader.
The proof can be found in~\cite[Prop IV.4.13]{TakesakiI}.

\begin{lem}\label{lem:double commutants}
    Let $\caA_i \subset B(\mathcal{H}_i)$ be a unital $\rm{C}^*$-algebra for $i=1,2$ and $\vN$ the usual von Neumann-algebraic tensor product.  Then
    \begin{equation*}
        (\caA_1 \otimes \caA_2)'' = \caA_1'' \vN \caA_2''.
    \end{equation*}
    Equivalently,
    \begin{equation*}
        (\caA_1 \otimes \caA_2)' = (\caA_1'' \vN \caA_2'' )'.
    \end{equation*}
\end{lem}

As expected, approximate Haag duality is preserved under stacking.
\begin{thm}\label{thm: Haag duality}
    Let $\pi_1, \pi_2$ be irreducible representations of $\A_1, \A_2$ on Hilbert spaces $\caH_1, \caH_2$ respectively, and assume they both satisfy approximate Haag duality. Then $\pi_1\stack\pi_2$ also satisfies approximate Haag duality.
\end{thm}

\begin{proof}
    Each $\pi_i$ is assumed to satisfy approximate Haag duality, and so let $R_i, f^{(i)}_{\delta}(t), \text{ and } U_i:\caH_i\to\caH_i$ be the corresponding values, decreasing functions, and unitaries, see Definition~\ref{def:HaagD}.   
    Set $R = \max R_i$.
    Then we have the following: for any cone $\cone$,
    \begin{align*}
        (\pi_1 \stack \pi_2\left( (\A_1 \stack \A_2)(\conec))\right))' 
        &= (\pi_1(\A_1(\conec))\stack \pi_2(\A_2(\conec)))'  \\
        &= (\pi_1(\A_1(\conec))'' \vN \pi_2(\A_2(\conec))'')' \\
        &= (\pi_1(\A_1(\conec))' \vN \pi_2(\A_2(\conec))')''
        \\ & \subset (\text{Ad}(U_1)\pi_1(\A_1(\cone_\varepsilon-R{\textbf{e}_\Lambda}))'' \vN \text{ Ad}(U_2) \pi_2(\A_2(\cone_\varepsilon-R{\textbf{e}_\Lambda})'')''
        \\&= \text{  Ad}(U_1 \stack U_2) (\pi_1 \stack \pi_2((\A_1 \stack \A_2)(\cone_\varepsilon-R{\textbf{e}_\Lambda})))''. 
    \end{align*}
    We first used Lemma~\ref{lem:double commutants}, then the commutation theorem for tensor products of von Neumann algebras (see ~\cite[Theorem 11.2.16]{kadison1997fundamentals2}), while the inclusion is the definition of approximate Haag duality. The last equality follows again from the lemma. This shows condition~\ref{it:ahdinclusion} of the duality.

    We turn to condition~\ref{it:ahdapprox}. Let $\widetilde{U}_1, \widetilde{U}_2$ be the unitaries given since $\pi_1, \pi_2$ satisfy approximate Haag duality. Then $\widetilde{U}_1 \stack \widetilde{U}_2 \in (\pi_1 \stack \pi_2((\A_1 \stack \A_2)(\cone_{\varepsilon+\delta}-t\textbf{e}_\cone)))''$, again by Lemma~\ref{lem:double commutants}. Moreover,
    \begin{align*}
        &||(U_1 \stack U_2) - (\widetilde{U}_1 \stack \widetilde{U}_2)|| 
        \\& \leq ||(U_1 \stack U_2)-(\widetilde{U}_1 \stack U_2)|| + ||(\widetilde{U}_1 \stack U_2)-(\widetilde{U}_1 \stack \widetilde{U}_2)||
        \\&\leq ||(U_1 - \widetilde{U}_1)|| \cdot||U_2|| + ||\widetilde{U}_1|| \cdot ||(U_2 - \widetilde{U}_2)|| < f_{\delta}^{(1)}(t) + f_{\delta}^{(2)}(t)
    \end{align*}
    and the function $f_\delta(t) := f^{(1)}_{\delta}(t) + f^{(2)}_{\delta}(t)$ is a decreasing function with $\lim_{t\to\infty}f_\delta(t)=0$.
\end{proof}

With this, we conclude that the sector theory for the stacked system is well-defined if it is well-defined for each layer:
\begin{prop}\label{prop:stacked assumptions}
    Assume that $\pi_j, j=1,2$, satisfy Assumptions~\ref{Assum:irrep}, \ref{Assum:GGS}, and \ref{Assum:approxhd}. Then $\pi_1\stack\pi_2$ satisfies the same assumptions.
\end{prop}
\begin{proof}
    The tensor product of two irreducible representations is irreducible again (see Remark~\ref{rem:irreducibility} below).
    We have just proved that the stacked representation satisfies Assumption~\ref{Assum:approxhd}. The properly infinite factor assumption~\ref{Assum:GGS} follows immediately from~Lemma~\ref{lem:double commutants}, noting that the von Neumann-algebraic tensor product with a properly infinite von Neumann algebra is again properly infinite.
\end{proof}

\subsection{Tensor decomposition of the stacked sectors}
Our first main result says that for every sector of the stacked system, we get sectors for each of the layers by restricting the representation.

We first recall a standard result.
\begin{prop}\label{prop: GNS of product}
For $j = 1,2$, let the GNS representation of the state $\om_j$ of $\A_j$ be given as $(\caH_j, \pi_j, \Omega_j)$. Then $(\caH_1 \otimes \caH_2, \pi_1 \stack \pi_2, \Omega_1 \otimes \Omega_2)$ is the GNS representation of the product state $\om_1 \stack \om_2$. 
\end{prop}
\begin{proof}
It suffices to check that $\om_1 \stack \om_2 (\cdot) = \langle \Omega_1 \otimes \Omega_2 , (\pi_1 \stack \pi_2 ) (\cdot) (\Omega_1 \otimes \Omega_2 ) \rangle$ and that $\Omega_1 \otimes \Omega_2$ is cyclic under the action of $\pi_1 \stack \pi_2 (\A_1 \stack \A_2)$.  Indeed,
\begin{align*}
\langle \Omega_1 \otimes \Omega_2, &(\pi_1\stack \pi_2)(A_1 \otimes A_2)(\Omega_1 \otimes \Omega_2) \rangle
= \langle \Omega_1 \otimes \Omega_2, \pi_1(A_1)\Omega_1 \otimes \pi_2(A_2)\Omega_2 \rangle \\
 = &\langle \Omega_1, \pi_1(A_1)\Omega_1 \rangle \langle \Omega_1, \pi_2(A_2)\Omega_2 \rangle = \om_1(A_1)\om_2(A_2) = (\om_1 \stack \om_2)(A_1 \otimes A_2)
\end{align*}
for all $A_i\in\A_j, j=1,2$.  Moreover, since $\mathcal{H}_j = \overline{\pi_j(\mathfrak{A}_j) \Omega_j}$, it is clear that the Hilbert space $\overline{ \pi_1 \stack \pi_2 (\mathfrak{A}_1 \stack \mathfrak{A}_2 ) (\Omega_1 \otimes \Omega_2) }$ is the closed subspace of $\mathcal{H}_1 \otimes \mathcal{H}_2$ that is generated by the pure tensors; namely, all of $\mathcal{H}_1 \otimes \mathcal{H}_2$.  Hence, $\Omega_1 \otimes \Omega_2$ is cyclic.
\end{proof}

\begin{rem}\label{rem:irreducibility}
    Note that unlike in the case of representations of groups, the tensor product of irreducible representations of $\rm{C}^*$-algebras is again irreducible.  The reason for this is that a representation is irreducible if and only if the von Neumann algebra generated by its image is the entire collection of bounded operators.  Thus, for irreducible representations $\rho_i : \mathfrak{B}_i \rightarrow B(\mathcal{K}_i)$, $i=1,2$, we have
    $$(\rho_1 (\mathfrak{B}_1) \otimes \rho_2 (\mathfrak{B}_2))'' = \rho_1 (\mathfrak{B}_1)'' \overline{\otimes} \rho_2 (\mathfrak{B}_2)'' = B(\mathcal{K}_1) \overline{\otimes} B(\mathcal{K}_2) = B(\mathcal{K}_1 \otimes \mathcal{K}_2),$$
    where the first equality follows from Lemma~\ref{lem:double commutants} and the last follows from the commutation theorem for tensor products of von Neumann algebras.
\end{rem}

An operator $A_1 \in \A_1$ can be extended to an operator of the stacked system by stacking with the identity: $A_1 \stack \id \in \A_1 \stack \A_2$. In a similar fashion, we can restrict a representation of the stacked system to a representation of one layer. The natural question is then whether this new restricted representation satisfies the SSC with respect to the corresponding tensor product factor. 
The following result depends crucially on Assumption~\ref{Assum:GGS}.

\begin{thm}\label{thm: SSC for restriction}
    Let $\sigma$ be a representation satisfying SSC($\pi$), where $\pi = \pi_1 \stack \pi_2$.
    Assume moreover that the $\pi_i$ satisfy Assumptions~\ref{Assum:irrep} and~\ref{Assum:GGS}.
    Let $\sigma_1 := \sigma \upharpoonright \mathfrak{A}_1$, namely $\sigma_1 (A) = \sigma (A \stack \mathbb{I})$ for all $A \in \mathfrak{A}_1$. Then $\sigma_1$ satisfies SSC($\pi_1$).
    Similarly, we get a representation $\sigma_2 := \sigma \restr \A_2$ satisfying SSC($\pi_2$).
\end{thm}

Before we give the proof of this theorem we will note some useful properties of this restricted representation.
Note that we will be in the setting of Theorem~\ref{thm: SSC for restriction} for the remainder of this subsection, in Lemma~\ref{quasi equivalence} to Corollary~\ref{cor:quasiequiv}.
Clearly, analogous statements hold for the representation $\sigma_2$.

\begin{lem}\label{quasi equivalence}
    For any cone $\Lambda$, we have that $\sigma_1 \upharpoonright \mathfrak{A}_1 (\Lambda^c) \qe \pi_1 \upharpoonright \mathfrak{A}_1 (\Lambda^c)$, where $\qe$ denotes quasi-equivalence of representations.
\end{lem}

\begin{proof}
    Note that $\pi_1 \upharpoonright \mathfrak{A}_1(\Lambda^c)$ is quasi-equivalent to $\mathfrak{A}_1(\Lambda^c) \ni A \mapsto \pi_1(A) \stack \mathbb{I}$, since the latter is equivalent to a direct sum of countably many copies of $\pi_1 \upharpoonright \mathfrak{A}_1(\Lambda^c)$.  By the superselection criterion, there is some unitary $U_\Lambda$ such that $U_\Lambda \sigma (A) U_\Lambda^* = \pi(A)$ for each $A \in \mathfrak{A} (\Lambda^c) := \mathfrak{A}_1 (\Lambda^c) \stack \mathfrak{A}_2 (\Lambda^c)$.  In particular, for $A \in \mathfrak{A}_1 (\Lambda^c)$, we have
    $$U_\Lambda \sigma_1(A) U_\Lambda^* = U_\Lambda \sigma(A \stack \mathbb{I}) U_\Lambda^* = \pi(A \stack \mathbb{I}) = \pi_1(A) \stack \mathbb{I}.$$
    Hence, $\sigma_1$ is unitarily equivalent to $\mathfrak{A}_1(\Lambda^c) \ni A \mapsto \pi_1(A) \stack \mathbb{I}$, and so $\sigma_1 \upharpoonright \mathfrak{A}_1 (\Lambda^c) \qe \pi_1 \upharpoonright \mathfrak{A}_1 (\Lambda^c)$.
\end{proof}

\begin{lem}\label{lem:properly infinite}
The von Neumann algebras
    $\sigma_1 (\mathfrak{A}_1(\Lambda^c))'$ and $\pi_1(\A_1(\conec))'$ are properly infinite.
\end{lem}

\begin{proof}
    The von Neumann algebra $\pi_2 (\mathfrak{A}_2(\Lambda^c))''$ is properly infinite by assumption, and hence so is $\mathbb{I} \stack \pi_2 (\mathfrak{A}_2(\Lambda^c))''$.  
    Because $\sigma$ satisfies the superselection criterion, there is a unitary  $U_\Lambda$ such that $U_\Lambda \sigma (A) U_\Lambda^* = \pi (A)$ for all $A \in \mathfrak{A}(\Lambda^c)$. It follows that $U_\Lambda^* (\mathbb{I} \stack \pi_2 (\mathfrak{A}_2 (\conec) ) )'' U_\Lambda$ is also properly infinite. 
    Then for $A_1 \in \A_1(\conec), A_2 \in \A_2(\conec)$,
    \begin{align*}
    \sigma_1(A_1) U_\Lambda^* (\mathbb{I} \stack \pi_2(A_2)) U_\Lambda &= \sigma (A_1 \stack \mathbb{I}) \sigma (\mathbb{I} \stack A_2) \\
        &= \sigma ( \mathbb{I} \stack A_2) \sigma ( A_1 \stack \mathbb{I}) = U_\Lambda^* (\mathbb{I} \stack \pi_2 (A_2)) U_\Lambda \sigma_1 (A_1),
    \end{align*}
    and so $U_\Lambda^* (\mathbb{I} \stack \pi_2 (\mathfrak{A}_2 (\conec) ) ) U_\Lambda \subset \sigma_1 (\mathfrak{A}_1 (\Lambda^c))'$.  
    Taking double commutants of both sides yields that
    $\sigma_1 (\mathfrak{A}_1 (\Lambda^c))'$ contains a properly infinite subalgebra, and therefore is properly infinite.
    
    Now choose a cone $\widehat{\Lambda}$ for which $\Lambda^c \cap \widehat{\Lambda} = \emptyset$. By locality, $\pi_1(\mathfrak{A}_1(\Lambda^c)) \subset \pi_1(\mathfrak{A}_1(\widehat{\Lambda}))'$, and so $\pi_1(\mathfrak{A}_1(\widehat{\Lambda}))'' \subset \pi_1(\mathfrak{A}_1(\Lambda^c))'$.  Since $\pi_1(\mathfrak{A}_1(\widehat{\Lambda}))''$ is properly infinite by assumption, $\pi_1(\mathfrak{A}_1(\Lambda^c))'$ must be as well.
\end{proof}

These preliminary results are enough to see that the restriction of a representation to one layer of the stacked system does indeed satisfy the SSC($\pi_1$).

\begin{proof}[Proof of Theorem \ref{thm: SSC for restriction}]
    Let $\Lambda$ be any cone.  By Lemma \ref{quasi equivalence}, it follows that $\sigma_1 \upharpoonright \mathfrak{A}_1 (\Lambda^c) \qe \pi_1 \upharpoonright \mathfrak{A}_1 (\Lambda^c)$.  There is therefore a $*$-isomorphism
    $$\tau_\Lambda: \sigma_1 (\mathfrak{A}_1 (\Lambda^c))'' \rightarrow \pi_1 (\mathfrak{A}_1 (\Lambda^c))''$$
    such that $\tau_\Lambda ( \sigma_1 (A)) = \pi_1 (A)$ for all $A \in \mathfrak{A}_1 (\Lambda^c)$.  But since $\sigma_1 (\mathfrak{A}_1 (\Lambda^c))'$ and $\pi_1 (\mathfrak{A}_1 (\Lambda^c))'$ are both properly infinite and $\caH_{\sigma_1},\caH_{\pi_1}$ are both separable, see Remark~\ref{rem:separability}, $\tau_\Lambda$ is unitarily implemented by~\cite[Corollary 8.12]{StratilaZsido} (see also Exercise 9.6.32 of~\cite{kadison1997fundamentals2}), namely $\pi_1 \upharpoonright \A_1(\conec) \cong \sigma_1 \upharpoonright \A_1(\conec)$. This concludes the proof of the theorem since this holds for all cones.
\end{proof}

To conclude, we note the following properties of $\sigma_i$ in case $\sigma$ is irreducible, which will be of use later.
\begin{lem}\label{lem:factors}
Suppose that $\sigma \in \ssc(\pi)$ is irreducible.
Then the representation $\sigma_1$ satisfies the following properties:
    \begin{enumerate}
        \item \label{it:factorrep} $\sigma_1$ is a factor representation.
        \item \label{it:factorcone} For any cone $\Lambda$, $\sigma_1 \upharpoonright \mathfrak{A}_1 (\Lambda)$ is a factor representation.
    \end{enumerate}
\end{lem}

\begin{proof}
    To prove~\ref{it:factorrep}, note that $\sigma (\mathbb{I} \stack \mathfrak{A}_2) \subset \sigma_1(\mathfrak{A}_1)'$ since observables in different layers commute.
    Hence we have
    $$\sigma_1 (\mathfrak{A}_1) \vee \sigma_1 (\mathfrak{A}_1)' \supset \sigma (\mathfrak{A}_1 \stack \mathbb{I} ) \vee \sigma (\mathbb{I} \stack \mathfrak{A}_2) = \sigma (\mathfrak{A}_1 \stack \mathfrak{A}_2) '' = B(\mathcal{H}),$$
    where $\sigma_1 (\mathfrak{A}_1) \vee \sigma_1 (\mathfrak{A}_1)'$ denotes the smallest von Neumann algebra generated by $\sigma_1 (\mathfrak{A}_1)$ and $\sigma_1(\mathfrak{A}_1)'$. Thus, $\sigma_1 (\mathfrak{A}_1)'' \vee \sigma_1 (\mathfrak{A}_1)' = B(\mathcal{H})$, or equivalently, $\sigma_1 (\mathfrak{A}_1)'' \cap \sigma_1 (\mathfrak{A}_1)' = \mathbb{C}\mathbb{I}$.

    The proof of~\ref{it:factorcone} is similar, using $\mathfrak{A}_1 = \mathfrak{A}_1(\Lambda) \otimes \mathfrak{A}_1 (\Lambda^c)$, and $\sigma_1 (\mathfrak{A}_1 (\Lambda^c)) \subset \sigma_1 (\mathfrak{A}_1 (\Lambda))'$.
\end{proof}

Using Lemma~\ref{quasi equivalence} and Lemma~\ref{lem:factors} then gives the following.
\begin{cor}
    \label{cor:quasiequiv}
    Suppose that $\sigma$ is irreducible.
    Then every subrepresentation of $\pi_1 \upharpoonright \mathfrak{A}_1 (\Lambda^c)$ is quasi-equivalent to every subrepresentation of $\sigma_1 \upharpoonright \mathfrak{A}_1 (\Lambda^c)$.
\end{cor}
\begin{proof}
    Immediate from Proposition 10.3.12 of \cite{kadison1997fundamentals2}.
\end{proof}

\subsection{The superselection sectors of a stacked system}

We are now in a position to prove, by utilising the results above, our first two main theorems. 
The first theorem says that we can construct superselection sectors of the stacked system from sectors of the individual layers as expected.

\begin{thm}[Theorem~\ref{thm:ssstacked}]\label{thm:SSC}
    Let $\A_1$ and $\A_2$ be quantum spin systems on $\Gamma$ and suppose that we have irreducible reference representations $\pi_1$ and $\pi_2$ satisfying Assumptions~\ref{Assum:GGS} and \ref{Assum:approxhd}.
    Then $\pi_1 \stack \pi_2$ also satisfies the assumptions. If $\rho_1$ and $\rho_2$ satisfy the superselection criterion for $\pi_1$ and $\pi_2$ respectively, then $\rho_1 \stack \rho_2$ satisfies the superselection criterion for $\pi_1 \stack \pi_2$.
\end{thm}

\begin{proof}
    The first claim is Proposition~\ref{prop:stacked assumptions}. Let $\Lambda$ be a cone and $U_i$ be a unitary implementing the equivalence from the superselection criterion, i.e.
    $$U_i \rho_i ( A_i ) = \pi_i ( A_i ) U_i$$
    for all $A_i \in \A_i (\Lambda^c)$ and $i =1,2$.
    Then clearly
    $$U_1 \otimes U_2 (\rho_1 \stack \rho_2 (A_1\stack A_2)) = (\pi_1 \stack \pi_2 (A_1 \stack A_2)) U_1 \otimes U_2$$
    for all $A_1\stack A_2 \in (\A_1 \stack \A_2 )(\Lambda^c)$, so $\rho_1 \stack \rho_2$ satisfies SSC($\pi_1 \stack \pi_2$).
    \end{proof}

Our second main result concerns the inverse of this construction.
In particular, a natural question is which sectors of the stacked system come from stacking sectors in the individual layers.
Note that Theorem~\ref{thm: SSC for restriction} gives (representatives of) sectors $\sigma_1$ and $\sigma_2$ for each of the layers, but in general $\sigma \not\cong \sigma_1 \stack \sigma_2$.\footnote{This is analogous to the observation that a quantum state on a bipartite system in general cannot be recovered from its reduced density matrices.} However, if $\sigma$ is irreducible, this does turn out to be the case, provided the Type I condition holds.

The following proposition is seen as Example 11.2.5 in \cite{kadison1997fundamentals2}, and hence we point the reader towards the source and omit the proof.
\begin{prop}\label{prop: description of Type I factors}
    A factor $\mathcal{R} \subset B(\mathcal{K})$ is Type I if and only if there exist Hilbert spaces $\mathcal{K}_1$ and $\mathcal{K}_2$ and a unitary $U : \mathcal{K} \rightarrow \mathcal{K}_1 \otimes \mathcal{K}_2$ such that $U \mathcal{R} U^* = B(\mathcal{K}_1) \otimes \mathbb{I}_{\mathcal{K}_2}$.
\end{prop} 

\begin{thm}[Theorem~\ref{thm:ssstackedirreducible}]\label{thm:B}
    Let $\pi_1$ and $\pi_2$ be as above and assume that every factor representation $\sigma_1$ satisfying the superselection criterion for $\pi_1$ is such that $\sigma_1(\A_1)''$ is a Type~I factor.
    Let $\pi = \pi_1\stack\pi_2$. If $\rho$ is an irreducible representation that satisfies SSC($\pi$), then $\rho \cong \rho_1 \stack \rho_2$ where $\rho_1,\rho_2$ are irreducible representations satisfying SSC($\pi_1$), SSC($\pi_2$) respectively.
\end{thm}

    \begin{proof}
    Let $\rho : \A_1 \stack \A_2 \rightarrow B(\mathcal{H})$ be an irreducible representation satisfying SSC($\pi$).
    For $i=1,2$, let $\sigma_i = \rho \upharpoonright \mathfrak{A}_i$, namely $\sigma_1 (A_1) = \rho(A_1 \stack \mathbb{I})$ and $\sigma_2 (A_2) = \rho(\mathbb{I} \stack A_2)$ for all $A_i \in \mathfrak{A}_i$.
    By Theorem~\ref{thm: SSC for restriction} and Lemma~ \ref{lem:factors}, $\sigma_1$ is a factor representation which satisfies SSC($\pi_1$), and thus by our assumption we have that $\sigma_1(\mathfrak{A}_1)''$ is a Type I factor.  Therefore there are Hilbert spaces $\mathcal{K}_1$ and $\mathcal{K}_2$ and a unitary $U : \mathcal{H} \rightarrow \mathcal{K}_1 \otimes \mathcal{K}_2$ such that
    $$U \sigma_1 (\mathfrak{A}_1)'' = (B(\mathcal{K}_1) \otimes \mathbb{I}_{\mathcal{K}_2} )U.$$
    Since $\sigma_2(\mathfrak{A}_2) \subset \sigma_1(\mathfrak{A}_1)'$, we have
    $$U \sigma_2 (\mathfrak{A}_2) U^* \subset U \sigma_1(\mathfrak{A}_1)' U^* = (U \sigma_1(\mathfrak{A}_1)'' U^*)' = (B(\mathcal{K}_1) \otimes \mathbb{I}_{\mathcal{K}_1} )' = \mathbb{I}_{\mathcal{K}_1} \otimes B(\mathcal{K}_2),$$
    where the last equality follows from the commutation theorem for tensor products of von Neumann algebras.
    
    For $A_1 \in \mathfrak{A}_1$ and $A_2 \in \mathfrak{A}_2$, this implies that there exist $B_1 \in B(\mathcal{K}_1)$ and $B_2 \in B(\mathcal{K}_2)$ such that $U \sigma_1 (A_1) U^* = B_1 \otimes \mathbb{I}_{\mathcal{K}_2}$ and $U \sigma_2(A_2) U^* = \mathbb{I}_{\mathcal{K}_1} \otimes B_2$.  Thus, we define $\rho_i : \mathfrak{A}_i \rightarrow B(\mathcal{K}_i)$ as 
    $$\rho_1 (A_1) = B_1 \ \text{and} \ \rho_2 (A_2) = B_2.$$
    It is easy to see that $\rho_1$ and $\rho_2$ are indeed representations, and
    $$U \rho (A_1 \stack A_2) U^*  = U \sigma_1(A_1) U^* U\sigma_2(A_2) U^* = B_1 \stack B_2 = \rho_1(A_1) \stack \rho_2 (A_2),$$
    so $\rho \cong \rho_1 \stack \rho_2$.  Moreover, since $\rho$ is irreducible and for a $T \in \rho_1(\mathfrak{A}_1)'$ we have $U^*(T \stack \mathbb{I}_{\mathcal{K}_2}) U \in \rho(\mathfrak{A}_1 \stack \mathfrak{A}_2)'$,  then $\rho_1$ must also be irreducible and similarly for $\rho_2$. 

    Finally, for any cone $\Lambda$, Theorem~\ref{thm: SSC for restriction} yields a unitary $V_\Lambda$ such that $V_\Lambda^*\sigma_1(A)V_\Lambda = \pi_1(A)$ for all $A\in\mathfrak{A}_1(\Lambda^c)$, and by the above,
    \begin{equation*}
        (UV_\Lambda)\pi_1(A)(UV_\Lambda)^* = \rho_1(A)\stack\mathbb{I}_{\mathcal{K}_2},
    \end{equation*}
    namely, $\pi_1\upharpoonright \mathfrak{A}_1 (\Lambda^c)$ and $\rho_1\upharpoonright \mathfrak{A}_1 (\Lambda^c)$ are quasi-equivalent.
    We want to show that this is in fact a unitary equivalence.
    By quasi-equivalence, there exists a $*$-isomorphism $\tau : \rho_1(\mathfrak{A}_1(\Lambda^c))'' \to \pi_1(\mathfrak{A}_1(\Lambda^c))''$.
    By replacing $\Lambda^c$ with $\Lambda$, it also follows that $\rho_1(\mathfrak{A}_1(\Lambda))''$ and $\pi_1(\mathfrak{A}_1(\Lambda))''$ are $*$-isomorphic.
    Since the latter is properly infinite by assumption, so is $\rho_1(\mathfrak{A}_1(\Lambda))''$.
    From locality it follows that $\rho_1(\mathfrak{A}_1(\Lambda))'' \subset \rho_1(\mathfrak{A}_1(\Lambda^c))'$, hence the right-hand side of the inclusion is also a properly infinite von Neumann algebra.
    By Lemma~\ref{lem:properly infinite}, $\pi_1(\mathfrak{A}_1(\Lambda^c))'$ is properly infinite, and hence $\tau$ is unitarily implemented, again by~\cite[Corollary 8.12]{StratilaZsido}, Remark~\ref{rem:separability} and the fact that $\caH\cong\caK_1\otimes\caK_2$. Therefore, $\pi_1 \restr \A_1(\conec) \cong \rho_1 \restr \A_1(\conec)$ and $\rho_1$ satisfies $\text{SSC}(\pi_1)$. 
\end{proof}

\subsection{Type I property}
In this section we show that the Type I assumption made in Theorem~\ref{thm:ssstackedirreducible} is true under natural conditions.
We will prove that if
\begin{enumerate}
    \item \label{it:assapprox} $\pi_1$ satisfies the \textit{approximate split property}, defined below in Definition~\ref{Approximate Split Property}, and
    \item $\label{it:asscount} \pi_1$ has at most countably many super-selection sectors,
\end{enumerate}
then every factor representation $\rho_1$ satisfying SSC($\pi_1$) is Type I. By this, we mean the von Neumann algebra $\rho_1(\A_1)''$ is of Type I.  These two conditions are in fact satisfied by a wide variety of physical models of interest.
For example, \ref{it:assapprox} is satisfied for every quantum double model and is stable under the application of any locally generated automorphism (see Section~\ref{sec:teesplit} for the validity of the split property and its relation with the area law for the entanglement entropy), and \ref{it:asscount} must be satisfied for any physically realistic model, since we would not expect a physical quantum spin system to give rise to uncountably many particle types (and indeed, in axiomatizations of anyon theories, the number of anyon types is usually assumed to be finite, see e.g.~\cite{KitaevAppendices}).

The key technical step in showing this result is, following an approach by~\cite{KLM01}, a decomposition result of representations satisfying the superselection criterion.
For any irreducible representation $\rho$ of $\mathfrak{A}$ that satisfies SSC($\pi$), the restriction $\rho_1 = \rho\upharpoonright\mathfrak{A}_1$ again satisfies SSC($\pi_1$), see Theorem~\ref{thm: SSC for restriction}.
The representation $\rho_1$ will certainly not be irreducible, since its image has a non-trivial commutant, however it can be decomposed into a direct integral of irreducible representations,
$$\rho_1 \cong \int_X^\oplus \rho_x d \mu (x)$$
for some measure space $(X, \Sigma, \mu)$, see e.g.~\cite[Thm. IV.8.32]{TakesakiI}.
The question remains whether each $\rho_x$ (or at least all $\rho_x$ for $x$ outside of some zero-measure set) satisfies SSC($\pi_1$).  Theorem~\ref{thm:direct integral} below shows that when $\pi_1$ satisfies the approximate split property, this is indeed the case. 

\begin{defn}[Approximate split property]\label{Approximate Split Property}
    Let $\mathfrak{B}$ be a quasi-local algebra on a planar lattice $\Gamma$, and let $\sigma$ be a representation. Let $d > 0$.  For two cones $\Lambda_1 \subset \Lambda_2$, we shall write $\Lambda_1 \Subset \Lambda_2$ if $\overline{\mathrm{arg}(\Lambda_1)}\subset \mathrm{arg}(\Lambda_2)$ (i.e., the boundaries of $\Lambda_2$ are not parallel to the boundaries of $\Lambda_1$)
and $\text{dist}(\Lambda_1, \Lambda_2^c) > d$.  We say $\sigma$ satisfies the \emph{approximate split property} if for any pair of cones $\Lambda_1 \Subset \Lambda_2$, there exists a Type I factor $\mathcal{N}$ satisfying
    $$\sigma (\mathfrak{B} (\Lambda_1))'' \subset \mathcal{N} \subset \sigma (\mathfrak{B} (\Lambda_2))''.$$
\end{defn}
Note that the definition implicitly depends on $d$, which will be model-dependent.
It holds for example for the ground state representation of abelian quantum double models with $d=1$~\cite{naaijkens2015haag}.
In the next section we show that it holds for a wide class of models of interest.

\begin{thm}\label{thm:direct integral}
    Suppose $\varphi$ is an irreducible representation of $\mathfrak{B}$ satisfying the approximate split property, $\varphi(\mathfrak{B}(\Lambda))''$ is properly infinite for every cone $\Lambda$, and $\rho : \mathfrak{B} \rightarrow B (\mathcal{H})$ is a representation satisfying SSC($\varphi$).  Let $(X, \Sigma, \mu)$ be a measure space and $\mathcal{H}_x$ a Hilbert space for each $x \in X$ so that
    $$\mathcal{H} \cong \int_X^\oplus \mathcal{H}_x d \mu (x)$$
    is a direct integral decomposition of $\mathcal{H}$ for which
    $$\rho \cong \int_X^\oplus \rho_x d\mu (x)$$
    is a decomposition of $\rho$ into irreducible representations $\rho_x :\mathfrak{B} \rightarrow B(\mathcal{H}_x)$.  Then $\rho_x$ satisfies SSC($\varphi$) for $\mu$-almost every $x\in X$.
\end{thm}

Before proving this result, we make some comments and show some helpful lemmas.
To simplify notation, we assume without loss of generality that $\varphi$ and $\rho$ are represented on the same Hilbert space $\varphi, \rho : \mathfrak{B} \rightarrow B(\mathcal{H})$.\footnote{This can be done because our representations satisfy the superselection criterion.}
We also identify $\mathfrak{B}$ with its image under $\varphi$, so that $\mathfrak{B} \equiv \varphi(\mathfrak{B})$ and $\mathfrak{B}(X) \equiv \varphi(\mathfrak{B}(X))$ for every $X \subset \Gamma$.  This means that studying representations of $\mathfrak{B}$ satisfying SSC($\varphi$) is equivalent to studying $*$-homomorphisms $\mathfrak{B} \rightarrow B(\mathcal{H})$ that can be unitarily implemented when restricted to any cone.

We begin with some preliminaries.  Choose a cone $\Lambda_a$ and define the family of admissible cones with respect to this choice by
\begin{equation}
    \label{eq:admissable}
    \mathcal{C}^{\Lambda_a} := \{ \Lambda \in \mathcal{C} \mid \operatorname{Arg}(\Lambda) < \operatorname{Arg} (\Lambda_a^c) \ \text{and} \ \Lambda + x \cap \Lambda_a = \emptyset \ \text{for some} \ x \in \mathbb{R}^2 \}.
\end{equation}
The \textit{auxiliary algebra} corresponding to a choice of $\Lambda_a$ is the $\rm{C}^*$-algebra $\mathfrak{B}^{\Lambda_a}$ generated by the von Neumann algebras $\mathfrak{B}(\Lambda)''$ as $\Lambda$ ranges over $\mathcal{C}^{\Lambda_a}$.  Explicitly,
\begin{equation}
    \label{eq:auxalgebra}
\mathfrak{B}^{\Lambda_a} = \overline{\bigcup_{\Lambda \in \mathcal{C}^{\Lambda_a}} \mathfrak{B}(\Lambda)''}^{\| \cdot \|}.
\end{equation}
This is well-defined since the set $\mathcal{C}^{\Lambda_a}$ is directed.

\begin{rem}
    We should note that our definition of the auxiliary algebra differs slightly from the conventional one, since one typically only requires that the generating admissible cones only satisfy $\text{Arg}(\Lambda) \leq \text{Arg}(\Lambda_a^c)$, rather than $\text{Arg}(\Lambda) < \text{Arg}(\Lambda_a^c)$.  The reason for this change is that we will require that for any $\Lambda \in \mathcal{C}^{\Lambda_a}$, there must be some $\widehat{\Lambda} \in \mathcal{C}^{\Lambda_a}$ for which $\Lambda \Subset \widehat{\Lambda}$.  This would not be the case for the cone $\Lambda_a^c$ itself or any of its translations, so it is necessary to exclude these possibilities.  We also emphasize that our purpose for introducing the auxiliary algebra in this section is to have access to a C*-algebra that is generated by a directed set of cone von Neumann algebras, not to define the tensor product of the superselection category.  Consequently, whenever the auxiliary algebra is referenced in the sections that follow, one is free to follow either convention, since the theory does not depend on the precise choice.
\end{rem}

If $\rho$ is a representation of $\mathfrak{B}$ satisfying SSC($\varphi$), then by definition for any cone $\Lambda \in \mathcal{C}$, $\rho_\Lambda : = \rho|_{\mathfrak{B}(\Lambda)}$ is unitarily implemented, and therefore extends to a normal (namely, $\sigma$-weakly continuous) representation of $\mathfrak{B}(\Lambda)''$.  By letting $\Lambda$ range over every cone in $\mathcal{C}^{\Lambda_a}$ and completing with respect to the norm we obtain a representation of $\mathfrak{B}^{\Lambda_a}$ that restricts to a normal representation on $\mathfrak{B}(\Lambda)''$ for every $\Lambda \in \mathcal{C}^{\Lambda_a}$.  The representation obtained from $\rho$ in this way is referred to as the representation of $\mathfrak{B}^{\Lambda_a}$ generated by $\rho$.  If some representation of $\mathfrak{B}^{\Lambda_a}$ restricts to a representation of $\mathfrak{B}$ that satisfies SSC($\varphi$), we call it localizable.  It is easy to see that the localizable representations of $\mathfrak{B}^{\Lambda_a}$ are precisely those that are generated by representations of $\mathfrak{B}$ satisfying SSC($\varphi$).

\begin{defn}
(i) Let $S \subset \mathcal{C}$ be a set of cones in $\Gamma$.  A family of representations $\rho_\Lambda : \mathfrak{B}(\Lambda)'' \rightarrow B(\mathcal{H})$, for $\Lambda \in S$, is a \emph{locally normal family} of representations with respect to $S$ if
\begin{itemize}
\item $\rho_\Lambda$ is normal for each $\Lambda \in S$
\item $\rho_\Lambda = \rho_{\tilde{\Lambda}}|_{\mathfrak{B}(\Lambda)''}$ whenever $\Lambda \subset \tilde{\Lambda}$
\end{itemize}
(ii) A representation of the auxiliary algebra $\mathfrak{B}^{\Lambda_a}$ is a \emph{locally normal representation} if it restricts on cones in $\mathcal{C}^{\Lambda_a}$ to a locally normal family of representations with respect to $\mathcal{C}^{\Lambda_a}$.
\end{defn}

We immediately note the relation between the two notions just introduced. Every locally normal family of representations with respect to $\mathcal{C}^{\Lambda_a}$ generates a locally normal representation of $\mathfrak{B}^{\Lambda_a}$ by ranging over the cones in $\mathcal{C}^{\Lambda_a}$ and completing with respect to the norm.  Reciprocally, in fact by definition, every locally normal representation of $\mathfrak{B}^{\Lambda_a}$ restricts on cones in $\mathcal{C}^{\Lambda_a}$ to a locally normal family of representations with respect to $\mathcal{C}^{\Lambda_a}$.  Thus, the notions of a locally normal representation and a locally normal family of representations with respect to $\mathcal{C}^{\Lambda_a}$ are interchangeable. In fact, from the assumption that $\mathfrak{B}(\Lambda)''$ is a properly infinite factor, any  representation of $\mathfrak{B}(\Lambda)''$ is normal (and hence any  representation of $\mathfrak{B}^{\Lambda_a}$ is locally normal) provided it is on a separable Hilbert space, see~\cite[Thm. V.5.1]{TakesakiI}.

For the purpose of this section, the interest of locally normal families is that they are unitarily implemented.

\begin{lem}
    For any cone $\Lambda$ and separable Hilbert space $\mathcal{H}$, a representation $\mathfrak{B}(\Lambda)'' \rightarrow B(\mathcal{H})$ is normal if and only if it is unitarily implemented.  In particular, a locally normal family $\{\rho_\Lambda:\Lambda \in\caC\}$ with respect to $\caC$ gives rise to a representation of $\mathfrak{B}$ satisfying the SSC.
\end{lem}
\begin{proof}
    We claim that $\mathfrak{B}(\Lambda)'$ is properly infinite for any $\Lambda\in\caC$. Indeed, $\mathfrak{B}(\Lambda_1)\subset\mathfrak{B}(\Lambda)'$ for any disjoint cones $\Lambda_1,\Lambda$. Taking double commutants yields $\mathfrak{B}(\Lambda_1)''\subset\mathfrak{B}(\Lambda)'$ and since $\mathfrak{B}(\Lambda_1)''$ is a properly infinite factor by assumption, so is $\mathfrak{B}(\Lambda)'$. With this, the claim is an immediate consequence of~\cite[Proposition~V.3.1]{TakesakiI}, where $\sigma$-finiteness follows from separability of $\mathcal{H}$ and faithfulness follows from the fact that the kernel must be generated by a central projection, since it is a $\sigma$-weakly closed two sided ideal.
\end{proof}

\begin{rem}\label{rem: locally normal SSC}
While every localizable representation of $\mathfrak{B}^{\Lambda_a}$, meaning it is generated by a representation of $\mathfrak{B}$ satisfying the SSC, is locally normal, the converse may in principle not be true as it may fail to be unitarily implementable on cones not in $\mathcal{C}^{\Lambda_a}$.
\end{rem}

In order to proceed, we will need a suitable countable subset of the cones in $\mathcal{C}$.  Note that the possible cones in the plane can be labelled by the points of the manifold
$$\mathcal{M} := \mathbb{R}^2 \times \mathbb{S}^1 \times (0, 2 \pi),$$
where the locations of the vertex of each cone are given by $\mathbb{R}^2$, the orientations of the cones are labelled by $\mathbb{S}^1$, and the opening angle of the cones are in $(0, 2 \pi)$.  For any $p \in \mathcal{M}$, let $\Lambda_p$ be its corresponding cone.  Let $\mathcal{M}_\mathbb{Q} = \mathcal{M}\cap\mathbb{Q}^4$.  Now define
$$\mathcal{C}_\mathbb{Q} := \{ \Lambda_p \in \mathcal{C} \mid p \in 
\mathcal{M}_\mathbb{Q}  \} \ \text{and} \ \mathcal{C}^{\Lambda_a}_\mathbb{Q} := \{ \Lambda_p \in \mathcal{C}^{\Lambda_a} \mid p \in 
\mathcal{M}_\mathbb{Q}  \}.$$
We remark that this labelling is not injective, since the vertex of any cone that opens at an angle of $\pi$ can be taken to be any point along its boundary.

Since a normal representation of the von Neumann algebra of a cone restricts to a normal representation of the von Neumann algebra of any sub-cone, it is easy to see that any locally normal family of representations with respect to $\mathcal{C}^{\Lambda_a}_\mathbb{Q}$ can be extended uniquely to a locally normal family of representations with respect to $\mathcal{C}^{\Lambda_a}$.

\begin{defn}
    Let $\mathcal{N}$ be a Type I factor.  The compact operators relative to $\mathcal{N}$ is the $\rm{C}^*$-subalgebra of $\mathcal{N}$ generated by the finite projections in $\mathcal{N}$.  
\end{defn}

\noindent Note that these are precisely the operators corresponding to the compact operators tensored with the identity under the unitary conjugation given in Proposition~\ref{prop: description of Type I factors}.

By the approximate split property, for each pair of cones $\Lambda_1 \Subset \Lambda_2$, let $\mathcal{N}( \Lambda_1, \Lambda_2)$ be a choice of Type I von Neumann factor satisfying
$$\mathfrak{B}(\Lambda_1)'' \subset \mathcal{N}(\Lambda_1, \Lambda_2) \subset \mathfrak{B}(\Lambda_2)''$$
and let $K(\Lambda_1, \Lambda_2)$ be the compact operators relative to $\mathcal{N}(\Lambda_1, \Lambda_2)$.  For readers uncomfortable with the Axiom of Choice, we remark that the choice of $\mathcal{N}(\Lambda_1, \Lambda_2)$ can be made canonical, as shown in \cite{doplicher1984standard}, however the exact choice is unimportant.  We define a new unital $\rm{C}^*$-algebra $\mathfrak{K}^{\Lambda_a}$ to be generated by the identity and the algebras $K(\Lambda_1, \Lambda_2)$ for $\Lambda_1 \Subset \Lambda_2$ both belonging to $\mathcal{C}^{\Lambda_a}_{\mathbb{Q}}$.  Note that since $\mathcal{C}^{\Lambda_a}_\mathbb{Q}$ is countable and each $K(\Lambda_1, \Lambda_2)$ is separable, $\mathfrak{K}^{\Lambda_a}$ is separable as well.

Clearly, if $\Lambda_1 \Subset \Lambda_2 \subset \Lambda_3 \Subset \Lambda_4$, then $\mathcal{N}(\Lambda_1, \Lambda_2) \subset \mathcal{N}(\Lambda_3, \Lambda_4)$.  However, we may not have that $K(\Lambda_1, \Lambda_2) \subset K(\Lambda_3, \Lambda_4)$.  To remedy this issue, for $\Lambda_1 \Subset \Lambda_2$ we define
$$\mathfrak{L}(\Lambda_1, \Lambda_2) = \mathcal{N}(\Lambda_1, \Lambda_2) \cap \mathfrak{K}^{\Lambda_a}.$$
We therefore have the inclusions $K(\Lambda_1, \Lambda_2) \subset \mathfrak{L} (\Lambda_1, \Lambda_2) \subset \mathcal{N}(\Lambda_1, \Lambda_2)$, and also that $\mathfrak{K}^{\Lambda_a}$ is the $\rm{C}^*$-subalgebra of $\mathfrak{B}^{\Lambda_a}$ generated by the identity and the algebras $\mathfrak{L} (\Lambda_1, \Lambda_2)$ as $\Lambda_1 \Subset \Lambda_2$ run in $\mathcal{C}^{\Lambda_a}_\mathbb{Q}$.  Importantly, $\mathfrak{K}^{\Lambda_a}$ is not a $\rm{C}^*$-subalgebra of $\mathfrak{B}$, which is why it is necessary to carry out this analysis in the context of normal representations rather than restricting our attention to representations on $\mathfrak{B}$, even though it is the latter that we are really interested in.

\begin{lem}
Suppose $\Lambda_1 \Subset \Lambda_2 \subset \Lambda_3 \Subset \Lambda_4$ for $\Lambda_1,..., \Lambda_4 \in \mathcal{C}^{\Lambda_a}$.  Then
$$\mathfrak{L}(\Lambda_1, \Lambda_2) \subset \mathfrak{L} ( \Lambda_3, \Lambda_4).$$
\end{lem}

\begin{proof}
Since $\mathfrak{L}(\Lambda_1, \Lambda_2) \subset \mathcal{N}(\Lambda_1, \Lambda_2) \subset \mathcal{N}(\Lambda_3, \Lambda_4)$, we have
\begin{equation*}
\mathfrak{L}(\Lambda_1, \Lambda_2) = \mathcal{N}(\Lambda_1, \Lambda_2) \cap \mathfrak{K}^{\Lambda_a} \subset \mathcal{N}(\Lambda_3, \Lambda_4) \cap \mathfrak{K}^{\Lambda_a} = \mathfrak{L} (\Lambda_3, \Lambda_4).\qedhere
\end{equation*}
\end{proof}

The idea is to look at representations which are non-degenerate when restricted to the compact operators.
The following lemma is Corollary 53 in \cite{KLM01}, where the proof can be found as well. 

\begin{lem}
\label{lem:normalext}
Let $\mathcal{N}$ be a Type I factor with separable predual, $K \subset \mathcal{N}$ the ideal of compact operators relative to $\mathcal{N}$, and $\mathfrak{L}$ a $\rm{C}^*$-algebra satisfying $K \subset \mathfrak{L} \subset \mathcal{N}$.  If $\rho$ is a representation of $\mathfrak{L}$ for which $\rho|_K$ is non-degenerate, then $\rho$ is $\sigma$-weakly continuous and therefore extends uniquely to a normal representation of $\mathcal{N}$.
\end{lem}

We can now show that locally normal representations of $\mathfrak{B}^{\Lambda_a}$ are uniquely determined by their restrictions to $\mathfrak{K}^{\Lambda_a}$.

\begin{lem}\label{lem: non-degenerate reps}
Let $\rho$ be a locally normal representation of $\mathfrak{B}^{\Lambda_a}$.  Then $\rho$ restricts to a representation of $\mathfrak{K}^{\Lambda_a}$ and $\rho|_{K(\Lambda_1 , \Lambda_2)}$ is non-degenerate for every pair $\Lambda_1, \Lambda_2 \in \mathcal{C}^{\Lambda_a}$ with $\Lambda_1 \Subset \Lambda_2$. 

Conversely, if $\sigma$ is a representation of $\mathfrak{K}^{\Lambda_a}$ such that $\sigma|_{K(\Lambda_1, \Lambda_2)}$ is non-degenerate for all $\Lambda_1, \Lambda_2 \in \mathcal{C}^{\Lambda_a}_\mathbb{Q}$ with $\Lambda_1 \Subset \Lambda_2$, then there exists a unique locally normal representation $\widetilde{\sigma}$ of $\mathfrak{B}^{\Lambda_a}$ that extends $\sigma$.
\end{lem}

\begin{proof}
We begin by showing that $\rho|_{K(\Lambda_1, \Lambda_2)}$ is non-degenerate.  It is easy to see that the compact operators act non-degenerately on the Hilbert space on which they are defined, and so must the compact operators relative to any Type I factor.  Thus, $K(\Lambda_1, \Lambda_2)$ must act non-degenerately on $\mathcal{H}$.  Moreover, $\rho$ is locally normal, and so as before, it follows from the assumption that our cone von Neumann algebras are properly infinite that $\rho$ is implemented by conjugation with a unitary on $K(\Lambda_1,\Lambda_2)$.  Since $K(\Lambda_1, \Lambda_2)$ acts non-degenerately on $\mathcal{H}$, it must also do so after unitary conjugation.  Hence, $\rho|_{K(\Lambda_1,\Lambda_2)}$ is non-degenerate.

We now show that a representation $\sigma$ of $\mathfrak{K}^{\Lambda_a}$ extends to a locally normal representation $\tilde{\sigma}$ when $\sigma|_{K(\Lambda_1, \Lambda_2)}$ is non-degenerate for $\Lambda_1,\Lambda_2 \in \mathcal{C}^{\Lambda_a}$, $\Lambda_1 \Subset \Lambda_2$.  For any such pair $\Lambda_1, \Lambda_2$, we denote by $\widetilde{\sigma}_{\Lambda_1, \Lambda_2}$ the unique normal extension of $\sigma|_{\mathfrak{L}(\Lambda_1, \Lambda_2)}$ to $\mathcal{N}(\Lambda_1, \Lambda_2)$ given by Lemma~\ref{lem:normalext}.

We will only show $\sigma$ can be extended to a locally normal family of representations with respect to $\mathcal{C}^{\Lambda_a}_\mathbb{Q}$, since this can always be extended to a family with respect to $\mathcal{C}^{\Lambda_a}$, and thus determines a locally normal representation of $\mathfrak{B}^{\Lambda_a}$.  Given $\Lambda \in \mathcal{C}^{\Lambda_a}_\mathbb{Q}$, choose $\Lambda_1 \in \mathcal{C}^{\Lambda_a}_\mathbb{Q}$ for which $\Lambda \Subset \Lambda_1$, and set
$$\widetilde{\sigma}_\Lambda = \widetilde{\sigma}_{\Lambda, \Lambda_1}|_{\mathfrak{B}(\Lambda)''}.$$
We claim that the mapping $\Lambda \in \mathcal{C}^{\Lambda_a} \mapsto \widetilde{\sigma}_\Lambda$ is a well defined locally normal representation.  By construction each $\widetilde{\sigma}_\Lambda$ is normal, so we need only show that $\widetilde{\sigma}$ is well defined.

To see why this is, let $\Lambda_2 \in \mathcal{C}^{\Lambda_a}_\mathbb{Q}$ be such that $\Lambda \Subset \Lambda_2$.  We need to show that
$$\widetilde{\sigma}_{\Lambda, \Lambda_1} |_{\mathfrak{B}(\Lambda)''} = \widetilde{\sigma}_{\Lambda, \Lambda_2} |_{\mathfrak{B}(\Lambda)''}.$$
Choose $\Lambda_3, \Lambda_4 \in \mathcal{C}^{\Lambda_a}_\mathbb{Q}$ for which
$$\Lambda_j \subset \Lambda_3 \Subset \Lambda_4$$
for $j = 1,2$.  Then
$$\mathcal{N}(\Lambda, \Lambda_j) \subset \mathfrak{B}(\Lambda_j)'' \subset \mathfrak{B}(\Lambda_3)'' \subset \mathcal{N}(\Lambda_3, \Lambda_4),$$
and intersecting with $\mathfrak{K}^{\Lambda_a}$ gives us $\mathfrak{L}(\Lambda, \Lambda_j) \subset \mathfrak{L}(\Lambda_3, \Lambda_4)$.  Thus, $\widetilde{\sigma}_{\Lambda_3, \Lambda_4}|_{\mathcal{N}(\Lambda, \Lambda_j)}$ is a normal extension of $\sigma|_{\mathfrak{L}(\Lambda, \Lambda_j)}$ to $\mathcal{N}(\Lambda, \Lambda_j)$.  But by uniqueness, this is just $\widetilde{\sigma}_{\Lambda, \Lambda_j}$.  Thus we have that
$$\widetilde{\sigma}_{\Lambda, \Lambda_1} |_{\mathfrak{B}(\Lambda)''} = \widetilde{\sigma}_{\Lambda_3, \Lambda_4}|_{\mathfrak{B}(\Lambda)''} = \widetilde{\sigma}_{\Lambda, \Lambda_2} |_{\mathfrak{B}(\Lambda)''}$$
as desired.
\end{proof}

\begin{lem}
Let $\rho :\mathfrak{B}^{\Lambda_a} \rightarrow B(\mathcal{H})$ be a locally normal representation.  Then
$$\rho \left(\mathfrak{B}^{\Lambda_a}\right)' = \rho \left(\mathfrak{K}^{\Lambda_a}\right)'.$$
\end{lem}

\begin{proof}
Since $\mathfrak{K}^{\Lambda_a} \subset \mathfrak{B}^{\Lambda_a}$, we have $\rho \left(\mathfrak{B}^{\Lambda_a}\right)' \subset \rho \left(\mathfrak{K}^{\Lambda_a}\right)'$.
To see the other direction, let $\Lambda \in \mathcal{C}^{\Lambda_a}$, and choose $\Lambda_1, \Lambda_2 \in \mathcal{C}^{\Lambda_a}_\mathbb{Q}$ with
$$\Lambda \subset \Lambda_1 \Subset \Lambda_2.$$
Let $\langle \mathbb{I}, K(\Lambda_1, \Lambda_2) \rangle$ denote the $\rm{C}^*$-subalgebra of $\mathfrak{K}^{\Lambda_a}$ generated by the unit and $K(\Lambda_1, \Lambda_2)$.  Then $\langle \mathbb{I}, K(\Lambda_1, \Lambda_2) \rangle$ is unital, unlike $K(\Lambda_1, \Lambda_2)$, which implies that $\langle \mathbb{I}, K(\Lambda_1, \Lambda_2) \rangle '' = \mathcal{N}(\Lambda_1, \Lambda_2)$, and therefore contains $\mathfrak{B}(\Lambda)''$.  To see why, note that $\mathcal{N}(\Lambda_1, \Lambda_2)$ is Type I, and so by Proposition \ref{prop: description of Type I factors} there exists a unitary $U : \mathcal{H} \rightarrow \mathcal{K}_1 \otimes \mathcal{K}_2$ for Hilbert spaces $\mathcal{K}_1$ and $\mathcal{K}_2$ for which
$$\mathcal{N}(\Lambda_1, \Lambda_2) = U^* ( B(\mathcal{K}_1) \otimes \mathbb{C} ) U \ \text{and} \ K(\Lambda_1, \Lambda_2) = U^* (K_1 \otimes \mathbb{C} ) U,$$
where $K_1$ denotes the compact operators on $\mathcal{K}_1$.  Since $K_1 '' = B(\mathcal{K}_1)$, we also have $\langle \mathbb{I}_{\mathcal{K}_1}, K_1 \rangle '' = B (\mathcal{K}_1)$.  Moreover, since $\langle \mathbb{I}_{\mathcal{K}_1}, K_1 \rangle$ is unital, 
$$(\langle \mathbb{I}_{\mathcal{K}_1} , K_1 \rangle \otimes \mathbb{C} \mathbb{I}_{\mathcal{K}_2})'' = B(\mathcal{K}_1) \otimes \mathbb{C} \mathbb{I}_{\mathcal{K}_2}.$$
Conjugating with $U$ gives $\langle \mathbb{I}, K(\Lambda_1, \Lambda_2) \rangle '' = \mathcal{N}(\Lambda_1, \Lambda_2)$.

Since $\mathcal{N}(\Lambda_1, \Lambda_2) \subset \mathfrak{B}(\Lambda_2)''$ and $\rho|_{\mathfrak{B}(\Lambda_2)''}$ is unitarily implemented, it follows directly that $\rho ( \langle \mathbb{I}, K(\Lambda_1, \Lambda_2) \rangle '' ) = \rho(\langle \mathbb{I}, K(\Lambda_1, \Lambda_2) \rangle )''$.  Thus,
$$\rho \left(\mathfrak{B}(\Lambda) ''\right) \subset \rho (\langle \mathbb{I}, K(\Lambda_1, \Lambda_2) \rangle '' ) = \rho(\langle \mathbb{I}, K(\Lambda_1, \Lambda_2) \rangle )''  \subset \rho (\mathfrak{K}^{\Lambda_a})''.$$
Since $\rho \left(\mathfrak{B}(\Lambda)''\right) \subset \rho \left(\mathfrak{K}^{\Lambda_a}\right)''$ for each $\Lambda \in \mathcal{C}^{\Lambda_a}$, we have that $\rho \left(\mathfrak{B}^{\Lambda_a}\right) \subset \rho\left(\mathfrak{K}^{\Lambda_a}\right)''$, and so $\rho\left(\mathfrak{K}^{\Lambda_a}\right)' \subset \rho\left(\mathfrak{B}^{\Lambda_a}\right)'$.
\end{proof}

\begin{rem}\label{rem:equivalent commutants}
    Note that in the case, $\rho: \mathfrak{B}^{\Lambda_a} \rightarrow B(\mathcal{H})$ is \textit{localizable}, meaning that $\rho|_{\mathfrak{B}}$ satisfies the SSC, we also have that $\rho(\mathfrak{B})' = \rho (\mathfrak{B}^{\Lambda_a})'$.  Indeed, $\rho(\mathfrak{B}) \subset \rho(\mathfrak{B}^{\Lambda_a})$, so $\rho(\mathfrak{B}^{\Lambda_a})' \subset \rho (\mathfrak{B})'$.  Conversely, for every $\Lambda \in \mathcal{C}^{\Lambda_a}$,
    $$\rho (\mathfrak{B}(\Lambda)'') \subset \rho (\mathfrak{B}(\Lambda))'' \subset \rho (\mathfrak{B})'',$$
    so $\rho(\mathfrak{B}^{\Lambda_a})'' \subset \rho (\mathfrak{B})''$, and thus $\rho (\mathfrak{B})' = \rho (\mathfrak{B}^\Lambda_a)'$.  Hence, in this case we have
    $$\rho(\mathfrak{B})' = \rho (\mathfrak{B}^{\Lambda_a}) ' = \rho (\mathfrak{K}^{\Lambda_a})' .$$
\end{rem}

Recall that Theorem~\ref{thm:direct integral} states that every representation $\rho$ satisfying the SSC can be written as a direct integral of irreducible representations $\rho_x$ which themselves satisfy the SSC for almost every $x$.
We now give a proof of this result.

\begin{proof}[Proof of Theorem~\ref{thm:direct integral}]
Let $\Lambda_a$ be any cone in $\mathbb{R}^2$.  Denote the extension of $\rho$ to $\mathfrak{B}^{\Lambda_a}$ by $\rho^{\cone_a}$, and the restriction of $\rho^{\cone_a}$ to $\mathfrak{K}^{\Lambda_a}$ by $\rho^{\mathfrak{K}}$.  As pointed out in Remark~\ref{rem: locally normal SSC}, that any $\rho_x$ is unitarily equivalent to $\varphi$ when both are restricted to any cone in $\mathcal{C}^{\Lambda_a}$ is equivalent to the claim that $\rho_x$ can be extended to locally normal representation $\rho_x^{\Lambda_a}$ of $\mathfrak{B}^{\Lambda_a}$ on $\mathcal{H}_x$.  We will show that for any choice of $\Lambda_a$ this can be done for a full measure set of $x \in X$.

Recall that the direct integral decomposition of $\rho$ into irreducibles given by
$$\mathcal{H} \cong \int_X^\oplus \mathcal{H}_x d \mu (x) \ \text{and} \ \rho \cong \int_X^\oplus \rho_x d \mu (x)$$
corresponds to a choice of maximally abelian sub-von Neumann algebra of $\rho(\mathfrak{B})'$ by~\cite[Thm. IV.8.32]{TakesakiI}, since $\mathfrak{B}$ is separable.
But as shown in Remark~\ref{rem:equivalent commutants}, we have that
\begin{equation}
  \rho(\mathfrak{B})'  = \rho^{\mathfrak{K}} (\mathfrak{K}^{\Lambda_a})'  
\end{equation}
meaning that both commutants have the same maximally abelian subalgebras.  Since $\mathfrak{K}^{\Lambda_a}$ is separable as well, we can therefore decompose $\rho^{\mathfrak{K}}$ into irreducibles over the same decomposition of $\mathcal{H}$, giving
$$\rho^{\mathfrak{K}} \cong \int_X^\oplus \rho^{\mathfrak{K}}_x d \mu (x).$$
We would like to be able to decompose $\rho^{\Lambda_a}$ over this decomposition of $\mathcal{H}$ as well, however we cannot simply apply the same theorem since $\mathfrak{B}^{\Lambda_a}$ is not separable.  However, since $\rho^{\cone_a}$ is a locally normal representation, $\rho^{\mathfrak{K}}$ is non-degenerate on each $K(\Lambda_1, \Lambda_2)$ for $\Lambda_1 \Subset \Lambda_2$ and $\Lambda_1, \Lambda_2 \in \mathcal{C}^{\Lambda_a}_\mathbb{Q}$ by Lemma \ref{lem: non-degenerate reps}.  This means that $\rho^{\mathfrak{K}}_x$ must be non-degenerate on $K(\Lambda_1, \Lambda_2)$ for all $x$ outside of a null set.  To see why, suppose there were a positive measure set $Y \subset X$ for which $\rho_x^\mathfrak{K}$ was degenerate.  Then we could find some non-zero $\psi \in \mathcal{H}$ that is supported only on $Y$ when expressed with respect to the decomposition $\mathcal{H} = \int_X^\oplus \mathcal{H}_x d\mu(x)$, namely $\psi = \int_X^\oplus \psi_x d\mu(x)$.  Then for every $A \in K(\Lambda_1,\Lambda_2)$, we would have that
$$\rho^\mathfrak{K}(A) \psi = \int_X^\oplus \rho^\mathfrak{K}_x (A) \psi_x d \mu(x) = 0,$$
which violates non-degeneracy of this representation.  Moreover, since this holds for every pair $\Lambda_1 \Subset \Lambda_2$ in $\mathcal{C}_\mathbb{Q}^{\Lambda_a}$, which is countable, we have that $\rho_x^\mathfrak{K}$ is non-degenerate on every such $K(\Lambda_1, \Lambda_2)$ for almost every $x$.  Hence, by Lemma~\ref{lem: non-degenerate reps}, $\rho_x^\mathfrak{K}$ can be extended to a unique locally normal representation $\rho_x^{\Lambda_a}$ of $\mathfrak{B}^{\Lambda_a}$ for almost every $x$.

We claim that the direct integral $\int_X^\oplus \rho_x^{\Lambda_a} d\mu (x)$ is a well defined locally normal representation of $\mathfrak{B}^{\Lambda_a}$ that extends $\rho^\mathfrak{K}$.  Before proving this claim, we explain its consequences.  First, by the uniqueness of the locally normal extension of $\rho^\mathfrak{K}$ to $\mathfrak{B}^{\Lambda_a}$, we have
$$\rho^{\Lambda_a} = \int_X^\oplus \rho_x^{\Lambda_a} d \mu(x).$$
Moreover, we have
$$\int_X^\oplus \rho_x d\mu(x) = \rho = \rho^{\Lambda_a} |_\mathfrak{B} = \int_X^\oplus \rho_x^{\Lambda_a}|_\mathfrak{B} d \mu(x),$$
so $\rho_x^{\Lambda_a} |_\mathfrak{B} = \rho_x$ for all $x$ outside a null set.

Since almost every $\rho_x^{\Lambda_a}$ is locally normal, $\rho_x|_{\mathfrak{B}(\Lambda)}$ must be unitarily implemented for almost every $x$ and every $\Lambda \in \mathcal{C}^{\Lambda_a}$.  In other words, we have shown
$$Z(\Lambda_a) = \{ x \in X \mid \rho_x |_{\mathfrak{B}(\Lambda)} \text{ is unitarily implemented for every } \Lambda \in \mathcal{C}^{\Lambda_a} \}$$
is a full measure set.  Since our choice of $\Lambda_a$ was arbitrary and $\mathcal{C}_\mathbb{Q}$ is countable, we see that $\bigcap_{\Lambda_a \in \mathcal{C}_\mathbb{Q}} Z(\Lambda_a)$ is a full measure set as well.  Since every cone $\Lambda$ is in $\mathcal{C}^{\Lambda_a}$ for some $\Lambda_a \in \mathcal{C}_\mathbb{Q}$, we have that $\rho_x |_{\mathfrak{B}(\Lambda)}$ is unitarily implemented for almost every $x$ and every $\Lambda \in \mathcal{C}$, and therefore $\rho_x$ satisfies SSC($\varphi$) for a full measure set of $x$ in $X$.

All that is left is to prove the claim that $\int_X^\oplus \rho_x^{\Lambda_a} d \mu(x)$ is well defined and locally normal.  We begin by showing that for any $\Lambda \in \mathcal{C}^{\Lambda_a}$, $A \in \mathfrak{B}(\Lambda)''$, and measurable fields $\{\xi_x\}_{x \in X}$ and $\{ \eta_x \}_{x \in X}$ on $\{ \mathcal{H}_x \}_{x \in X}$, the mapping
$$x \mapsto \langle \eta_x, \rho_x^{\Lambda_a} (A) \xi_x \rangle_{\mathcal{H}_x}$$
is a measurable function.  To do so we must find a sequence $A_n$ in $\mathfrak{K}^{\Lambda_a}$ that converges strongly to $A$.  By scaling, it is enough to assume $\| A \| = 1$. As in the proof of Lemma~\ref{lem: non-degenerate reps}, it suffices to consider $\Lambda \in \mathcal{C}^{\Lambda_a}_\mathbb{Q}$.

For any set $S \subset B(\mathcal{H})$, let $\overline{S}^s$ denote the strong closure of $S$, and let $(S)_1$ denote the set of all elements of $S$ with norm at most $1$.  Since for $\Lambda_1 \in \mathcal{C}^{\Lambda_a}_\mathbb{Q}$ with $\Lambda \Subset \Lambda_1$, $\mathfrak{B}(\Lambda)'' \subset \mathcal{N}(\Lambda,\Lambda_1) = K(\Lambda,\Lambda_1)''$, we have
$$(\mathfrak{B}(\Lambda)'')_1 \subset ( K(\Lambda,\Lambda_1)'')_1 = \overline{(K(\Lambda,\Lambda_1))_1}^s,$$
where the equality follows from the Kaplansky density theorem.  Since $\mathcal{H}$ is separable, the strong topology is metrizable on bounded sets by~\cite[Prop. II.2.7]{TakesakiI}, and we can use the above inclusion to obtain a sequence $A_n \in (K(\Lambda, \Lambda_1))_1$, rather than just a general net, converging strongly to $A$.  Moreover, since $K(\Lambda, \Lambda_1)'' \subset \mathfrak{B}(\Lambda_1)''$, and $\rho_x^{\Lambda_a}$ is a normal extension of $\rho_x^\mathfrak{K}$ on $\mathfrak{B}(\Lambda_1)''$, we must also have that $\rho_x^\mathfrak{K}(A_n)$ converges strongly to $\rho_x^{\Lambda_a}(A)$.  Thus, for every $x$, we have
$$\lim_{n \to \infty} \langle \eta_x , \rho_x^\mathfrak{K}(A_n) \xi_x \rangle_{\mathcal{H}_x} = \langle \eta_x , \rho_x^{\Lambda_a}(A) \xi_x \rangle_{\mathcal{H}_x}.$$
Furthermore, since $\{\rho_x^\mathfrak{K}\}_{x \in X}$ is a measurable family, the mapping
$$x \mapsto \langle \eta_x , \rho_x^{\mathfrak{K}}(A_n) \xi_x \rangle_{\mathcal{H}_x}$$
is measurable for every $n$, and hence
$$x \mapsto \langle \eta_x, \rho_x^{\Lambda_a} (A) \xi_x \rangle_{\mathcal{H}_x}$$
is a pointwise limit of measurable functions, and is therefore measurable.  As a result, the integral
$$\int_X^\oplus \rho_x^{\Lambda_a}|_{\mathfrak{B}(\Lambda)''} d \mu (x)$$
is a well defined representation of $\mathfrak{B}(\Lambda)''$, for each $\Lambda \in \mathcal{C}^{\Lambda_a}$, and is normal by~\cite[Thm. V.5.1]{TakesakiI}.  We therefore obtain a locally normal family of representations with respect to $\mathcal{C}^{\Lambda_a}_\mathbb{Q}$ with extends to a family with respect to $\mathcal{C}^{\Lambda_a}$, and so $\int_X^\oplus \rho_x^{\Lambda_a} d \mu (x)$ is a locally normal representation of $\mathfrak{B}^{\Lambda_a}$.
\end{proof}

Recall that we are assuming throughout that $\pi_1$ has properly infinite cone algebras, which follows for example if $\omega_1$ is a gapped ground state, and therefore satisfies the conditions of Theorem~\ref{thm:direct integral}.
With this fact and Theorem~\ref{thm:direct integral} in hand we are ready to prove the following.

\begin{thm}[Theorem~\ref{Thm: Type I}]\label{Thm:E}
    Let $\pi_1$ satisfy the approximate split property, Definition~\ref{Approximate Split Property}, and let $\sigma$ be a representation satisfying $\ssc(\pi_1)$. If in addition $\pi_1$ has at most countably many irreducible superselection sectors then $\sigma$ can be decomposed as a countable direct sum of irreducible representations satisfying $\ssc(\pi_1)$.  Moreover, if $\sigma$ is a factor representation, then $\sigma(\A_1)''$ is Type~I.
\end{thm}

\begin{proof}
    Let $\sigma : \mathfrak{A}_1 \rightarrow B(\mathcal{H})$ be a representation that satisfies SSC($\pi_1$).
    We decompose $\sigma$ again into irreducible representations
    $$\mathcal{H} \cong \int_X^\oplus \mathcal{H}_x d \mu (x) \ \text{and} \ \sigma \cong \int_X^\oplus \sigma_x d \mu (x)$$
    for a measure space $(X, \Sigma, \mu)$.  By Theorem \ref{thm:direct integral}, $\sigma_x$ satisfies SSC($\pi_1$) for all $x$ in a full measure subset of $X$.  Thus, by possibly removing a set of zero measure, we may without loss of generality take $\sigma_x$ to satisfy SSC($\pi_1$) for every $x \in X$.  Moreover, $\mathfrak{A}_1$ is separable and UHF, so each $\sigma_x$ is a faithful representation on a separable, infinite dimensional Hilbert space, and thus we have $\mathcal{H}_x \cong \mathcal{K}$ for all $x \in X$ and some fixed Hilbert space $\mathcal{K}$.  Since there are only countably many unitary equivalence classes of irreducibles satisfying SSC($\pi_1$), we index representatives from each of these classes by a countable set $J$, so that for each $x \in X$ there exists $j \in J$ for which $\sigma_x \cong \sigma_j$, where $\sigma_j : \mathfrak{A}_1 \to B (\mathcal{K})$ is irreducible.  Note this means we can partition $X$ as
    $$X = \bigcup_{j \in J} \{ x \in X \mid \sigma_x \cong \sigma_j \}.$$
    Let us denote $S_j = \{ x \in X  \mid \sigma_x \cong \sigma_j \}$.  Then each $S_j$ is measurable by Lemma~60 of \cite{KLM01}.  Letting $I = \{ j \in J \mid \mu (S_j) > 0\}$, we can therefore write
    $$\sigma \cong \int_X^\oplus \sigma_x d \mu (x) \cong \bigoplus_{i \in I} \int_{S_i}^\oplus \sigma_x d \mu (x).$$

    Let us now study the representation $\int_{S_i}^\oplus \sigma_i d \mu(x)$ on the Hilbert space $L^2(S_i, \mu|_{S_i} ; \mathcal{K})$ that acts on the fibre $\mathcal{K}$ by $\sigma_i$.  Since $\sigma_x \cong \sigma_i$ for every $x \in S_i$, \cite[Thm. IV.8.28]{TakesakiI} gives a measurable field of unitaries $U_x : \mathcal{H}_x \to \mathcal{K}$ for which $U_x \sigma_x(\cdot) U_x^* = \sigma_i(\cdot)$, and so $\int_{S_i}^\oplus \sigma_x d \mu(x) \cong \int_{S_i}^\oplus \sigma_i d \mu(x)$.  On the other hand, $\int_{S_i}^\oplus \sigma_i d \mu(x)$ only acts non-trivially on the fibre $\mathcal{K}$, so it is unitarily equivalent to
    $$A \mapsto \sigma_i (A) \otimes \mathbb{I} \in B(\mathcal{K}) \otimes \mathbb{I} \subset B(\mathcal{K} \otimes L^2(S_i, \mu|_{S_i}) ),$$
    and is therefore a Type I factor representation by Proposition~\ref{prop: description of Type I factors}.  Any Type I factor representation on a separable Hilbert space can be written as a direct sum of countably many copies of some irreducible representation, in this case $\sigma_i$, and so we can write
    $$\int_{S_i}^\oplus \sigma_x d \mu (x) \cong \bigoplus_{k = 1}^{d_i} \sigma_i,$$
    where $d_i$ is the (possibly infinite but countable) dimension of $L^2(S_i, \mu |_{S_i} )$.  Ergo,
    $$\sigma \cong \bigoplus_{i \in I} \bigoplus_{k=1}^{d_i} \sigma_i,$$
    which establishes our first claim that $\sigma$ is a countable direct sum.  Moreover, since each of the $\sigma_i$'s are inequivalent, we see that $\sigma$ will be a direct sum of inequivalent Type I factor representations whenever $I$ contains more than one element, in which case $\sigma$ could not be a factor representation.  Thus, if $\sigma$ is a factor representation, $I$ must be a singleton, meaning
    $$\sigma \cong \bigoplus_{k=1}^{d_i} \sigma_i,$$
    and is therefore Type I.
\end{proof}

\begin{rem}\label{rem: decomposition}
    In addition to justifying the Type I condition, Theorem~\ref{Thm: Type I} establishes sufficient conditions under which we can decompose an arbitrary reducible representation that satisfies the SSC into a direct sum of irreducible superselection sectors.  Indeed, any representation $\rho$ can be decomposed as a direct integral of irreducibles over some measure space,
    $$\rho \cong \int_X^\oplus \rho_x d \mu(x).$$
    However, even if $\rho$ satisfies the SSC, there is no a priori reason to believe that
    \begin{enumerate}
        \item $\rho_x$ satisfies the SSC for all $x$ in some full measure set in $X$, or
        \item the direct integral can be written as a direct sum.
    \end{enumerate}
    Firstly, (i) may not hold if the SSC fails for each $\rho_x$ individually and only becomes satisfied after the representations are averaged over an integral.  Furthermore, (ii) will hold if the Hilbert space is finite dimensional.
    It will still hold for infinite dimensional representations if the measure is atomic, however there is no reason why that would be true in general.
    
    We stress here that these issues arise not only in the context of stacking, but more generally in the context of superselection sector theory.
    Note that Theorem~\ref{Thm: Type I} implies that it is sufficient to consider only irreducible sectors.
    However, a general sector may still be a countable direct sum of irreducibles.
    In applications, it is often useful to restrict to sectors which are \emph{dualizable}; physically, this means there is a `conjugate sector'.
    By restricting to dualizable sectors, it follows that it is enough to consider only \emph{finite} direct sums~\cite{LongoRoberts97}, i.e., the category of dualizable superselection sectors becomes semi-simple.
    This is particularly useful in the context of \emph{fusion rules}.
    Briefly, one of the main results of Doplicher--Haag--Roberts theory is that we can endow the superselection sectors with a monoidal product, which we will write as $\widehat{\otimes}$ (this is \emph{not} the tensor product of representations!).
    The general construction can be found in~\cite{ogata_derivation_2021} for the models we are considering.
    A natural question is then how the tensor product $\pi_i \widehat{\otimes} \pi_j$ of two irreducible representations satisfying the SSC behaves.
    In particular, we get
    \[
            \pi_i \widehat{\otimes} \pi_j \cong \bigoplus_{k \in I} N^{k}_{ij} \pi_k.
    \]
    The set $\{ N^k_{ij} \}$ are called the \emph{fusion rules} of the theory.
    If we restrict to dualisable sectors only, this is a finite direct sum, and each $N_{ij}^k$ is a non-negative integer.
\end{rem}

\subsection{Topological entanglement entropy and the approximate split property}
\label{sec:teesplit}

Here we discuss the validity of the approximate split property, which is essential to ensure that almost all irreducible factors in the decomposition of a representation satisfying the SSC do satisfy the SSC themselves, see Theorem~\ref{thm:direct integral}.

We first recall some basic definitions. Let $\omega$ be a state on $\mathfrak{A}$.
If $\Lambda \subset \Gamma$ is a finite set, $\omega|_{\A (\Lambda)}$ is a state on some full matrix algebra and hence can be represented by a density operator $\rho_\Lambda \in \A (\Lambda)$ such that $\omega(A) = \Tr(\rho_\Lambda A)$ for all $A \in \A (\Lambda)$.
For disjoint, finite regions $A,C \subset \Gamma$, we denote $AC = A\cup C$ and define the \emph{quantum mutual information} as
\[
    I_\omega(A:C) := S(\rho_A) + S(\rho_C) - S(\rho_{AC})  = S(\rho_{AC}, \rho_A \otimes \rho_C).
    \]
Here $S(\rho) := -\operatorname{Tr} \rho \log \rho$ is the von Neumann entropy of the density operator $\rho$, and $S(\rho,\sigma) := \operatorname{Tr}\rho(\log \rho - \log \sigma)$ is the quantum relative entropy.

We will show that the vanishing of the mutual information
\begin{equation}
    \label{eq:zeromut}
    I_\omega(A:C) = 0
\end{equation}
whenever $A$ and $C$ are sufficiently far removed implies the approximate split property. Note that $I_\omega(A:C) = 0$ if and only if $\rho_{AC} = \rho_A \otimes \rho_C$, namely $\omega$ is a product state for observables separated far enough.  Here, with sufficiently far removed we mean that there is some $\xi > 0$ such that equation~\eqref{eq:zeromut} is satisfied for any pair of finite sets $A$ and $C$ with $d(A,C) > \xi$.

Before we get to the approximate split property, let us mention that there are several interesting models with topological order for which condition~\eqref{eq:zeromut} is satisfied. This is true in, for example, the toric code, if $\omega$ is its frustration free ground state. This can be seen from the explicit description of $\omega$ given in~\cite{AlickiFH07}. In fact, \eqref{eq:zeromut} follows from the following strict form of the area law which is expected to hold for the renormalization fixed points of topologically ordered phases~\cite{KitaevP06,LevinW06}: There are $\alpha$ and $\gamma$ such that for (large enough) compact and simply connected regions $A$,
\begin{equation}\label{eq:TEE}
    S(\rho_A) = \alpha \vert \partial A\vert - n_A \gamma.
\end{equation}
Here $\vert\partial A\vert$ is the length of the boundary $A$ and $n_A$ is the number of disconnected components of $\partial A$. While $\alpha$ depends on the microscopic details of the Hamiltonian, the topological entanglement entropy $\gamma$ is related to the quantum dimension of the anyons in the model.

\begin{figure}[ht]
  \begin{subfigure}[b]{.4\linewidth}
    \centering
    \includegraphics[height=3cm]{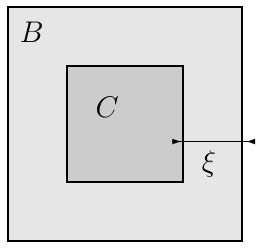}
    \caption{The setting of equation \ref{eq:axiom0}.}
    \label{fig:Axiom}
  \end{subfigure}\begin{subfigure}[b]{.6\linewidth}
    \centering
    \includegraphics[height=3cm]{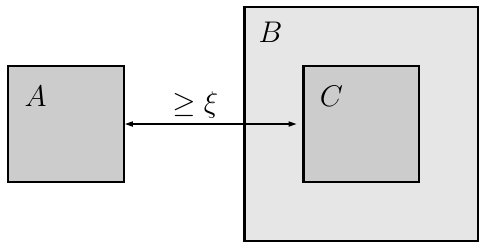}
    \subcaption{The vanishing of $I_\omega(A:C)$ whenever dist$(A,C)\geq \xi$.}
    \label{fig:QMI}
  \end{subfigure}
  \caption{The entanglement entropy and the mutual information.}
  \label{fig:Axiom and QMI}
\end{figure}

More generally, it follows from one of the axioms of~\cite{ShiKK20} that are postulated to hold for general gapped ground states.
One considers an annulus $B$ and the disk $C$ in its interior.
Their Axiom A0 says that
\begin{equation}
    \label{eq:axiom0}
        S(\rho_{BC}) + S(\rho_{C}) - S(\rho_B) = 0
\end{equation}
for regions $B,C$ as in Figure~\ref{fig:Axiom}.

To go from equation~\eqref{eq:axiom0} to~\eqref{eq:zeromut}, one can use strong subadditivity of the entropy~\cite{LiebR73} in the form
\[
        S(\rho_A) + S(\rho_B) \leq S(\rho_{AC}) + S(\rho_{BC}).
\]
We consider two finite sets $A,C$ separated by a distance of at least $\xi$. Around $C$, but disjoint from $A$, we can choose a region $B$ such that $BC$ is a region for which Axiom~\eqref{eq:axiom0} holds, see Figure~\ref{fig:QMI}. We then have
\[
    \begin{split}
        I_\omega(A:C) &= S(\rho_A) + S(\rho_C) - S(\rho_{AC}) \\
                &= S(\rho_A) + S(\rho_B) - S(\rho_{BC}) - S(\rho_{AC}) \leq 0.
    \end{split}
\]
We used~\eqref{eq:axiom0} in the second equality and the strong subadditivity to conclude the inequality.
Since $I_\omega(A:C) \geq 0$, this implies that $I_\omega(A:C) = 0$, and hence equation~\eqref{eq:zeromut}.

Just like the strict area law~\eqref{eq:TEE}, the strict equalities~\eqref{eq:zeromut} and~\eqref{eq:axiom0} hold for the renormalization fixed points and systems related to them by finite depth quantum circuits, but in general is not expected to hold for \emph{all} points in the same phase. 
However, its consequence, namely the approximate split property, is stable under the action of locally generated automorphisms and therefore holds throughout the phase, see~\cite{naaijkens2022split}.

\begin{rem}
    For infinite dimensional operator algebras, the von Neumann entropy is not a good quantity as it is often infinite.
    Following Araki~\cite{araki1975relative}, the \emph{relative} entropy can be defined in a general von Neumann algebraic setting, so that one could hope to make sense of $I_\omega(A:C)$ for infinite regions $A$ and $C$.
    Here however one has to be careful what exactly is a bipartite system.
    More precisely, if $\omega_{AB}$ is a state on some von Neumann algebra $\mathcal{M}$, there need not be a natural decomposition $\mathcal{M} = \mathcal{M}_A \overline{\otimes} \mathcal{M}_B$, so that there is no natural meaning of $\omega_A \otimes \omega_B$ as a state.
    In fact, the split property is equivalent to saying that there is a $*$-isomorphism between $\mathcal{M}_A \vee \mathcal{M}_B$ and $\mathcal{M}_A \overline{\otimes} \mathcal{M}_B$.
\end{rem}

We now prove that the approximate split property, Definition~\ref{Approximate Split Property}, is a consequence of the strict equality (\ref{eq:zeromut}). The following theorem is a generalisation of a similar proof for the toric code~\cite{naaijkens2011localized}.
To state the result, we need a slightly stronger assumption than approximate Haag duality, namely \emph{bounded spread Haag duality} \cite{MR4927814} which is a variant of \emph{weak algebraic duality}~\cite{Jones2024-nd}.

Our definition of bounded spread Haag duality is similar in spirit to that of~\cite{MR4927814}, in the sense that the commutant is \emph{strictly} localized in some larger cone.
Whereas in~\cite{MR4927814} the cone is enlarged by some fixed distance perpendicularly to the boundary, we require that the bigger cone is obtained by translating the cone over at least a minimum distance $R$, and slightly widening the opening angle.
This is the version of Haag duality that is shown to hold for a large class of models in~\cite{PerezGarcia} (see the proof of Corollary~4.1.5 there).
The arguments below can be adapted to the definition of~\cite{MR4927814} with ease.

\begin{defn}
\label{def:boundedspread}
Consider the notation of Definition~\ref{def:HaagD}.
Then $(\mathcal{H}, \pi_0)$ satisfies \emph{bounded spread Haag duality} if it satisfies approximate Haag duality with $f_\delta(t) = 0$, $U = \id$, and $R\geq 0$ that is independent of $\varepsilon$ and the opening angle of the cone.
\end{defn}

As remarked above, this implies that for any cone $\Lambda$, we can find some slightly wider, translated cone $\Lambda'$ with $\Lambda \subset \Lambda'$ and such that $\pi_0(\A(\conec))'\subset \pi_0(\A(\cone'))''$.

\begin{defn}
    Let $\xi > 0$ and let $\cone$ be a cone. Define $R_{\xi,\mathrm{min}}$ be the smallest number such that
    \[
        \text{dist}(\Lambda, (\Lambda-R_{\xi,\mathrm{min}}\textbf{e}_\Lambda)^c) \geq \xi.
    \]
    We define $\Lambda_{\xi,\mathrm{min}} := \Lambda - R_{\xi,\mathrm{min}}\textbf{e}_\Lambda$ and call it the \textit{minimal cone relative to }$\Lambda$.
\end{defn}
That is, $\Lambda_{\xi,\mathrm{min}}$ is the cone $\Lambda$ translated over the minimum distance along its axis that guarantees that the boundary of $\Lambda$ and $\Lambda_{\xi,\mathrm{min}}$ are at least distance $\xi$ apart. Of course, $R_{\xi,\mathrm{min}}$ depends only on the opening angle $\varphi$ of the cone, namely $R_{\xi,\mathrm{min}}=\xi/\sin\varphi$.

\begin{thm}
    \label{thm:split}
    Suppose that $\omega$ is a pure state satisfying condition~\eqref{eq:zeromut} for all regions $A$ and $C$ that are at least a distance $\xi > 0$ apart. 
    Assume furthermore that $\pi_\omega$ satisfies bounded spread Haag duality.
    For any cone $\cone$, there is a Type I factor $\mathcal{N}$ such that $\pi_\omega(\mathfrak{A}(\Lambda))'' \subset \mathcal{N} \subset \pi_\omega(\mathfrak{A}(\widehat\Lambda))''$ for all $\widehat\Lambda\supset (\Lambda_{\xi,\textrm{min}})_\varepsilon - R \textbf{e}_\Lambda$ for some $\varepsilon>0$, where $R\geq 0$ is the constant in Definition~\ref{def:boundedspread}.
\end{thm}

\begin{proof}
The condition that $I_\omega(A:C) = 0$ implies that $\rho_{AC} = \rho_A \otimes \rho_C$.
Hence $\omega(ac) = \omega(a) \omega(c)$ for all $a \in \A (A)$ and $c \in \A (C)$ for all finite sets $A \subset \Lambda$ and $C \subset \Lambda_{\xi,\textrm{min}}^c$.
Hence it follows that this is true for all $a \in \A (\Lambda)_{\textrm{loc}}$ and $c \in \A (\Lambda_{\xi,\textrm{min}}^c)_{\textrm{loc}}$.

Write $(\mathcal{H}_\omega, \pi_\omega, \Omega)$ for the GNS representation of $\omega$.
Let $\mathcal{R}_{\Lambda} := \pi_\omega(\A (\Lambda))''$ and $\mathcal{R}_{\Lambda_{\xi,\rm{min}}^c} := \pi_\omega(\A (\Lambda_{\xi,\rm{min}}^c))''$.
The map $a \mapsto \langle \Omega, a \Omega \rangle$ is a $\sigma$-weakly continuous extension of $\omega$ from $\pi_\omega(\A)$ to $\caB(\caH_\omega)$, which we also denote by $\omega$.
Consider $a \in \mathcal{R}_{\Lambda}$ and $c \in \mathcal{R}_{\Lambda_{\xi,\rm{min}}^c}$ with norm bounded by one.
From Kaplansky's density theorem (and Theorem II.2.6 of~\cite{TakesakiI}), there are nets $a_\lambda \in \A (\Lambda)_{\textrm{loc}}$ and $c_\mu \in \A (\Lambda_{\xi,\rm{min}}^c)_{\textrm{loc}}$ such that $a_\lambda \to a$ and $c_\mu \to c$ in the $\sigma$-weak topology.
We then have
\[
        \omega(a c_\mu) = \omega(\lim_\lambda a_\lambda c_\mu) = \lim_\lambda \omega(a_\lambda c_\mu) = \lim_\lambda \omega(a_\lambda) \omega(c_\mu) = \omega(a) \omega(c_\mu).
\]
Here we used that for fixed $b$, $a \mapsto ab$ is continuous in the $\sigma$-weakly topology, and that $\omega$ is a normal state, and hence $\sigma$-weak continuous.
Taking the limit over $\mu$ in a similar way shows that $\omega(ac) = \omega(a) \omega(c)$ for all $a \in \mathcal{R}_{\Lambda}$ and $c \in \mathcal{R}_{\Lambda_{\xi,\rm{min}}^c}$.
But this implies that the product state $\omega \otimes \omega$ on the \emph{algebraic} tensor product $\mathcal{R}_{\Lambda} \odot \mathcal{R}_{\Lambda_{\xi,\rm{min}}^c}$ can be extended to a $\sigma$-weakly continuous state on $\mathcal{R} := \mathcal{R}_{\Lambda} \vee \mathcal{R}_{\Lambda_{\xi,\rm{min}}^c}$.
Note that by locality, we have $\caR_{\Lambda} \subset \caR_{\Lambda_{\xi,\rm{min}}^c}'$ and that from a similar argument as in Lemma~\ref{lem:factors} it follows that $\caR$ is a factor.
It then follows by a result of Takesaki~\cite{Takesaki58} that $\mathcal{R}$ is $*$-isomorphic to the \emph{von Neumann} tensor product $\mathcal{R}_{\Lambda} \overline{\otimes} \mathcal{R}_{\Lambda_{\xi,\rm{min}}^c}$.

Since the $\caR_{\Lambda}$ are factors (cf.\ the proof of Lemma~\ref{lem:factors}), it follows from Theorem 1 and Corollary 1(iv) of~\cite{DAntoniL83} that there exists a Type I factor $\mathcal{N}$ such that $\mathcal{R}_{\Lambda} \subset \mathcal{N} \subset \mathcal{R}_{\Lambda_{\xi,\textrm{min}}^c}'$.
But then we have for all $\varepsilon > 0$
\[
\mathcal{R}_\Lambda 
\subset 
\mathcal{N} 
\subset 
\mathcal{R}_{\Lambda_{\xi,\textrm{min}}^c}' 
\subset
\mathcal{R}_{(\Lambda_{\xi,\textrm{min}})_\varepsilon - R \textbf{e}_\Lambda},
\]
where the last inclusion follows from bounded spread Haag duality.
By choosing $\varepsilon$ small enough, we can ensure that the algebra on the right hand side is contained in $\mathcal{R}_{\widehat{\Lambda}}$, which completes the proof.
\end{proof}

\begin{cor}\label{cor: ASP with xi and R}
Under the hypotheses of Theorem~\ref{thm:split}, the representation $\pi_\omega$ satisfies the approximate split property with parameter $d$ for any $d >\xi + R$, where $\xi$ is the entropy parameter of~\eqref{eq:zeromut} and $R$ is the spread of Definition~\ref{def:boundedspread}.
\end{cor}

\begin{proof}
Let $\Lambda_1 \Subset \Lambda_2$, so that $\Lambda_1 \subset \Lambda_2$, no
boundary ray of $\Lambda_2$ has the same direction as a boundary ray of
$\Lambda_1$, and $\operatorname{dist}(\Lambda_1,\Lambda_2^c) > d$. Write
$\Lambda_i = \Lambda_{z_i,\theta_i,\varphi_i}$ (for $i \in \{1,2\}$) and abbreviate $\mathbf{e}_2 :=
\mathbf{e}_{\theta_2}$; we assume that $0<\varphi_2 \leq \pi/2$ (the argument for the other case follows the same geometric intuition). Because $d > \xi + R \geq \xi + R\sin\varphi_2$, we may pick
\begin{equation}\label{eq: choice of tau}
  \tau \in \left( \frac{\xi}{\sin\varphi_2} + R,\ \frac{d}{\sin\varphi_2} \right)
\end{equation}
for which $\Lambda_1 \ \subset\ \Lambda_2 + \tau\mathbf{e}_2$. We then set 
\[
  \Lambda' =\ \Lambda_{z_2+\tau\mathbf{e}_2,\theta_2,\varphi_2 - \eta} \ \subset\
  \Lambda_2 + \tau\mathbf{e}_2 \subset\Lambda_2,
\]
for $\eta$ small enough such that $\Lambda_1 \subset \Lambda'$. Possibly reducing $\eta$ further, we may assume that
\[
  \tau \ >\ \frac{\xi}{\sin(\varphi_2 - \eta)} + R
  \ =\ R_{\xi,\min}(\Lambda') + R .
\]
Consequently the cone $\widehat{\Lambda'} := (\Lambda'_{\xi,\min})_{\varepsilon}
- R\,\mathbf{e}_2$ has apex
\[
  z_2 + \Big( \tau - \frac{\xi}{\sin(\varphi_2-\eta)} - R \Big)\mathbf{e}_2
  \ \in\ \Lambda_2 ,
\]
lying on the axis of $\Lambda_2$ strictly in front of $z_2$, and opening
half-angle $\varphi_2 - \eta + \varepsilon < \varphi_2$ provided $\varepsilon< 
\eta$. In particular, $\widehat{\Lambda'} \subset \Lambda_2$. We now apply Theorem~\ref{thm:split} to the cone $\Lambda'$ and to
this $\widehat{\Lambda'}$, obtaining a Type I factor $\mathcal{N}$ for which
\[
  \pi_\omega(\mathfrak{A}(\Lambda_1))''
  \ \subset\ \pi_\omega(\mathfrak{A}(\Lambda'))''
  \ \subset\ \mathcal{N}
  \ \subset\ \pi_\omega(\mathfrak{A}(\widehat{\Lambda'}))''
  \ \subset\ \pi_\omega(\mathfrak{A}(\Lambda_2))'' ,
\]
the first and last inclusions holding by isotony.
\end{proof}

\begin{figure}
    \centering

\begin{tikzpicture}[x=1.25cm,y=1.25cm,every node/.style={inner sep=1pt,fill=white}]
\clip (-0.25,-0.25) rectangle (6.2,5.8);

\pgfmathsetmacro{\phiTwo}{53}                 \pgfmathsetmacro{\etaAng}{20}                 \pgfmathsetmacro{\epsAng}{8}                  \pgfmathsetmacro{\phiPrime}{\phiTwo-\etaAng}  \pgfmathsetmacro{\phiHat}{\phiPrime+\epsAng}  \pgfmathsetmacro{\R}{9}                       

\foreach \x in {0,0.5,1,1.5,2,2.5,3,3.5,4,4.5,5,5.5,6}{
        \foreach \y in {0,0.5,1,1.5,2,2.5,3,3.5,4,4.5,5,5.5,6}{
        \filldraw (\x, \y) circle (0.01cm);
        }}

\draw[thick] (3,0.5) -- ++({90-\phiTwo}:\R) (3,0.5) -- ++({90+\phiTwo}:\R);
\filldraw (3,0.5) circle (0.06cm);
\node[text=black] at (2.75,0.25) {$z_2$};
\node[text=black] at (0.25,3) {$\Lambda_2$};

\draw[thick] (1.9,6) -- (2.4,4.1) -- (3.6,8);
\filldraw (2.4,4.1) circle (0.06cm);
\node[text=black] at (2.5,5.5) {$\Lambda_1$};

\draw[dashed, thick, black] (3,0) -- (3,2);
\draw[->, very thick, black] (3,2) -- (3,3.5);
\node[text=black] at (3,3.75) {$\theta_2$};

\draw[thick, gray] (3,2) -- ++({90-\phiTwo}:\R) (3,2) -- ++({90+\phiTwo}:\R);
\node[text=gray,anchor=west] at (4.5,4) {$\Lambda_2+\tau\mathbf{e}_2$};
\draw[thick,gray] (3,2) ++({90-\phiTwo}:1) arc[start angle={90-\phiTwo}, end angle=90, radius=1];
\node[text=gray] at (3.3,3.2) {$\varphi_2$};

\draw[thick,blue] (3,2) -- ++({90-\phiPrime}:\R) (3,2) -- ++({90+\phiPrime}:\R);
\draw[thick,blue] (3,2) ++({90-\phiTwo}:1.9) arc[start angle={90-\phiTwo}, end angle={90-\phiPrime}, radius=1.9];
\node[text=blue] at (4.5,3.61) {$\eta$};
\filldraw[blue] (3,2) circle (0.06cm);
\node[text=blue] at (4.75,5.2) {$\Lambda'$};
\node[text=blue,anchor=east] at (2.8,2) {$z_2+\tau\mathbf{e}_2$};

\draw[thick,red] (3,1) -- ++({90-\phiHat}:\R) (3,1) -- ++({90+\phiHat}:\R);
\filldraw[red] (3,1) circle (0.06cm);
\node[text=red] at (6,4.85) {$\widehat{\Lambda'}$};
\draw[thick,red] (3,1) ++({90-\phiHat}:0.8) arc[start angle={90-\phiHat}, end angle=90, radius=0.8];
\node[text=red,anchor=west] at (3.8,1.75) {$\varphi_2-\eta+\varepsilon$};
\end{tikzpicture}
    \caption{A visualisation of the considered cones in the proof of Corollary~\ref{cor: ASP with xi and R}}
    \label{fig:ASP with xi and R}
\end{figure}

\begin{rem}\label{rem:Validity}
The assumptions necessary for Theorem~\ref{thm:split} hold for a large class of models.
Levin--Wen string-net models for an input fusion category $\mathcal{C}$ which has multiplicity-free fusion rules (i.e. each simple anyon can appear at most once when decomposing the fusion product of two simple anyons) satisfy an area law of the form~\eqref{eq:TEE}, see~\cite{LevinW06}, and hence the condition~\eqref{eq:zeromut}.\footnote{It is conjectured that this holds without the multiplicity-free assumption, but as far as we are aware, this has not been written down in the literature.
}
Alternatively, condition~\eqref{eq:zeromut} follows from the LTO1 axiom by~\cite[Lemma 2.27]{NPW}, which is known to hold for all Levin--Wen string-net models~\cite[\S 4]{JonesNaaijkensPenneysWallick}.
They also satisfy bounded spread Haag duality (see the proof of~\cite[Cor. 4.1.5]{PerezGarcia}).
From Theorem~\ref{thm:split} it then follows that the approximate split property holds.

The stability result of~\cite{naaijkens2022split} then implies that the approximate split property holds for all states in the same phase as a Levin--Wen model.
These are conjectured to exhaust all commuting projector topologically ordered models with in 2D where the local algebras are full matrix algebras, which are all believed to be non-chiral.
(The latter condition is necessary, see for example~\cite{HaahSubalgebra}).
Indeed, Kim and Ranard~\cite{KimRanard} show that every model satisfying the `bootstrap axioms' (which as we have noted imply our entropy condition) and a `gappable boundary condition' (see~\cite[Def. 6]{KimRanard}) can be mapped to a Levin--Wen string-net model using a finite depth quantum circuit.
Since they are connected to the model we started with by a FDQC, they will also have the approximate split property.
\end{rem}

\subsection{Invertible states}

The classification of representations satisfying the superselection criterion of a stacked system yields the triviality of invertible states, in the sense of Definition~\ref{Invertible state definition}, which was the original motivation of introducing this class: If a state is invertible, its GNS representation has trivial superselection sectors.
We first recall the following definition.

\begin{defn}
    We say two states on the same $\mathrm{C}^*$-algebra are equivalent if their corresponding GNS representations are unitarily equivalent.
\end{defn}

Unitary equivalence of pure states of quantum spin systems has a particularly simple physical interpretation, namely that the states are ``indistinguishable at infinity'', see e.g.~\cite[Corollary 2.6.11]{Bratteli_Robs}.

The claim we want to show is that if $\omega \stack \overline{\omega}$ has trivial superselection structure, then so must $\omega$.
This follows from the following elementary result on representations of $\rm{C}^*$-algebras.

\begin{prop}\label{prop: equivalent states (direct implication)}
Let $\rho_i, \sigma_i$ be irreducible representations of $\mathfrak{A}_i$ respectively. Then $\rho_1 \stack \rho_2 \cong \sigma_1 \stack \sigma_2$ if and only if $\rho_1 \cong \sigma_1$ and $\rho_2 \cong \sigma_2$. 
\end{prop}

\begin{proof}
    ($\Leftarrow$) Let $U_1, U_2$ denote the unitaries implementing the equivalence $\rho_1 \cong \sigma_1$ and $\rho_2 \cong \sigma_2$. Then $U_1 \stack U_2$ implements the equivalence $\rho_1 \stack \sigma_1 \cong \rho_2 \stack \sigma_2$.
    
    ($\Rightarrow$) We have a unitary $W$ such that $W (\rho_1 \stack \rho_2)(A)W^* = (\sigma_1 \stack \sigma_2)(A)$ for all $A \in \mathfrak{A}_1 \stack \mathfrak{A}_2$. If $A = A_1 \stack \mathbb{I}_{\mathcal{H}_2}$, then $W (\rho_1(A_1)\stack \mathbb{I})W^* = \sigma(A_1) \stack \mathbb{I}$ for any $A_1 \in \mathfrak{A}_1$. Hence, the amplifications of $\rho_1$ and $\sigma_1$ are unitarily equivalent by $W$.
    Since an amplification of a representation is quasi-equivalent to the original representation, it follows that $\rho_1$ and $\sigma_1$ are quasi-equivalent. Since both are in fact irreducible, quasi-equivalence is promoted to unitary equivalence, giving $\rho_1 \cong \sigma_1$. The same argument with $1\leftrightarrow 2$ yields the second equivalence. 
\end{proof}

Recall (Definition~\ref{Invertible state definition}) that an invertible state is by definition pure with a pure inverse. 

\begin{cor}[Corollary~\ref{cor:invstate}]\label{cor: D} 
    Let $\omega$ be an invertible state of a quantum spin system $\A$ and let $\pi$ be its GNS representation. For all irreducible $\rho$ satisfying $\ssc(\pi)$, we have $\rho\cong\pi$.
\end{cor}
\begin{proof}
    Let $\overline{\mathfrak{A}}$ be a quasi-local algebra on $\Gamma$ and $\overline{\omega}$ be a state on $\overline{\mathfrak{A}}$ for which $\omega \stack \overline{\omega}$ is related to a product state under the action of an LGA.  We write $\overline{\pi}$ for the GNS representation of $\overline{\omega}$.
    
    By~\cite{naaijkens2022split}, the only irreducible superselection sector for $\pi \stack \overline{\pi}$ has $\pi \stack \overline{\pi}$ as a representative.
    Suppose that $\rho$ satisfies $\ssc(\pi)$, is irreducible and non-trivial, in the sense that it is not equivalent to $\pi$.
    Then $\rho \stack \overline{\pi}$ is irreducible and satisfies $\ssc(\pi \stack \overline{\pi})$ by Theorem~\ref{thm:SSC}, hence it is equivalent to $\pi \stack \overline{\pi}$. But this is in contradiction with Proposition~\ref{prop: equivalent states (direct implication)}.
\end{proof}

\section{Stacking as a categorical product}\label{Deligne}
So far, we have mostly considered the superselection sectors only.
The main result of the DHR approach is that these sectors have a rich structure, namely they are braided $\rm{C}^*$ (or in our case, in fact $\rm{W}^*$-) categories.
A $\rm{W}^*$-category is a category where (after adding finite direct sums if necessary) the Hom-spaces are $W^*$-algebras.\footnote{The categories we are interested in have finite direct sums. One can also define $\mathrm{W^*}$-categories intrinsically by requiring that $\Hom(X,Y)$ is the predual of a Banach space for each pair of objects $X,Y$. See~\cite[Cor. 1.2]{Yamagami} for the equivalence of these definitions.}
We refer the interested reader to~\cite{GLR,Yamagami,2411.01678} for complete details, but we aim to provide a concrete definition of the categories we are interested in below. In this section, we prove that whole structure is compatible with stacking in the sense of Theorem~\ref{thm:deligne}.

Throughout this section we consider quantum systems $\A_1$ and $\A_2$ with irreducible reference representations $\pi_1$ and $\pi_2$ satisfying Assumption~\ref{Assum:GGS} (properly infinite cone algebras), Assumption~\ref{Assum:approxhd} (approximate Haag duality), and the assumptions of Theorem~\ref{Thm: Type I} (approximate split property and countably many sectors).
The derivation of the monoidal structure, including the braiding, in this setting is worked out in~\cite{ogata_derivation_2021}.
We recall the main points here.

Any representation $\pi$ satisfying SSC($\pi_1$) has a unitarily equivalent representation represented on the Hilbert space of the reference representation $\pi_1$.
In fact, fix a cone $\Lambda$, and let $V_\Lambda : \mathcal{H}_\pi \to \mathcal{H}_{\pi_1}$ be the unitary setting up the equivalence in~\eqref{SSC}, i.e. $V_{\Lambda} \pi(A) V_{\Lambda}^* = \pi_1(A)$ for all $A \in \A_1(\Lambda^c)$.
Define a representation $\pi_\Lambda(A) := V_\Lambda \pi(A) V_\Lambda^*$.
Then $\pi_\Lambda$ is a representation on the same Hilbert space as the reference representation $\pi_1$, and $\pi_\Lambda(A) = \pi_1(A)$ for all $A \in \A_1(\Lambda^c)$.
We also say that $\pi_{\Lambda}$ is \emph{localized} in $\Lambda$.
This motivates the following definition.
\begin{defn}\label{def:sscat}
The category $\Delta_{\A_1}$ has as objects
\[
    \left\{ \pi : \A_1 \to B(\mathcal{H}_{\pi_1}) : \pi \textrm{ satisfies SSC(}\pi_1\textrm{) and } \pi(A) = \pi_1(A) \textrm{ for all } A \in \A_1(\Lambda^c) \right \},
\]
and as morphisms the intertwiners between such representations.
\end{defn}
\begin{rem}
This category is the same as the category $\mathcal{O}_{\Lambda}$ of~\cite[Eq. (5.2)]{ogata_derivation_2021}, see also Theorem 5.2 there.
It turns out to be independent of the choice of cone $\Lambda$ (up to unitary equivalence), and is also equivalent to the category of representations satisfying SSC$(\pi_1)$, but is easier to work with.
This has certainly been well-known for a long time, but is often used only implicitly.
For a discussion of this point, see e.g.~\cite{2505.07960} or~\cite{MR4927814}.
Since the choice of $\Lambda$ is not important, we will not use it in our notation $\Delta_{\A_1}$.
\end{rem}

The essential idea to define the monoidal structure (and braiding) is to pass from representations to endomorphisms.
Fix a cone $\Lambda_a$ such that the $\Lambda$ cone is admissible with respect to $\Lambda_a$, see equation~\eqref{eq:admissable}.
We write $\A_i^{\Lambda_a}$ for the auxiliary algebras~\eqref{eq:auxalgebra} for $\A_i$, and $\A_{12}^{\Lambda_a}$ for the auxiliary algebra\footnote{There are approaches to sector theory that do not require the choice of an auxiliary algebra, see e.g.~\cite{MR4927814,2505.07960}, but so far details have only been worked out under the assumption of \emph{bounded spread} Haag duality.} obtained from the stacked system.
Recall that we can identify $\pi_i(A)$ with $A$ for all $A \in \A_i$.
Then it can be shown (using approximate Haag duality in an essential way) that any $\pi \in \Delta_{\A_i}$ extends uniquely to a $*$-endomorphism $\rho_\Lambda$ of $\mathfrak{A}^{\Lambda_a}_i$, that is, $\rho_\Lambda(\pi_i(A)) = \pi(A)$ for all $A \in \A_i$.
For every admissible cone $\Lambda'$, this $\rho_\Lambda$ is weakly continuous on $\pi_i(\A_i(\Lambda'))''$.
This allows us to identify objects in $\Delta_{\A_i}$ with certain endomorphisms of $\Delta_{\A_i}$, and we will do so below when convenient.
We refer the reader to~\cite[Lemma~2.13]{ogata_derivation_2021} for details.

The simple objects in $\Delta_{\A_i}$ correspond to irreducible representations that satisfy SSC($\pi_i$).
Theorem~\ref{thm:ssstacked} and Proposition~\ref{prop: equivalent states (direct implication)} give an injection of simple objects in $\Delta_{\A_1}$ to simple objects in the stacked theory $\Delta_{\A_1 \stack \A_2}$.
\emph{A fortiori}, there is an injection mapping pairs of simple objects in $\Delta_{\A_1}$ and $\Delta_{\A_2}$ to $\Delta_{\A_1 \stack \A_2 }$.
Theorem~\ref{thm:ssstackedirreducible} in turn shows that this map is a surjection.
Not surprisingly, this map can be lifted to non-irreducible sectors, and this can be done in a way that is compatible with the monoidal structure and braiding.
We now construct a new $\rm{W}^*$-braided tensor category $\boxt$ which is equivalent to the category of the decoupled stacked system sector theory $\stacc$ via a `stacking' functor $\iota$.
The category $\boxt$ is a variant of the Deligne tensor product of abelian categories (see e.g.~\cite{Tensor_cats}) suitable for categories with infinite direct sums.
A similar construction for $\rm{W}^*$-categories has been considered in~\cite{2411.01678} in an abstract setting.
Our definition of $\boxt$ is a special case of this.
For the benefit of readers unfamiliar with the categorical construction, we prefer to give a less elegant but concrete description here in terms of representations satisfying the superselection criterion, and remark below how it relates to the abstract construction of~\cite{2411.01678}.

We start by defining a category consisting of simple objects only.
Note that this category in general is not closed under fusion (unless we have only abelian sectors in each of the layers).
We then later add in direct sums of the simple objects to obtain the category we want.
\begin{defn}
We define the category $\Delta_{\A_1} \boxtimes_0 \Delta_{\A_2}$ as follows: the objects are pairs $(\rho_{\A_1},\rho_{\A_2})$, where the $\rho_{\A_i} \in \Delta_{\A_i}$, $i=1,2$, are irreducible.
The morphisms are defined as $\Hom( (\rho_{\A_1}, \rho_{\A_2}), (\sigma_{\A_1}, \sigma_{\A_2})) := \Hom_{\Delta_{\A_1}}(\rho_{\A_1}, \sigma_{\A_1}) \otimes \Hom_{\Delta_{\A_2}}(\rho_{\A_2}, \sigma_{\A_2})$, where the tensor product is that of vector spaces.
\end{defn}

\begin{lem}\label{lem: full and faithful functor}
There is a full and faithful $*$-functor $F_0 : \Delta_{\A_1} \boxtimes_0 \Delta_{\A_2} \to \Delta_{\A_1 \stack \A_2}$.
\end{lem}
\begin{proof}
Let $(\rho_{\A_1}, \rho_{\A_2}) \in \Delta_{\A_1} \boxtimes_0 \Delta_{\A_2}$. By Theorem~\ref{thm:SSC}, $\rho_{\A_1} \stack \rho_{\A_2}$ satisfies the superselection criterion and hence $\rho_{\A_1} \stack \rho_{\A_2} \in \Delta_{\A_1 \stack \A_2}$.
Hence we can define the functor $F_0$ on objects by 
\begin{equation*}
    F_0( (\rho_{\A_1}, \rho_{\A_2})) := \rho_{\A_1} \stack \rho_{\A_2}.
\end{equation*}
If $S \in \Hom(\rho_{\A_1}, \sigma_{\A_1})$ and $T \in \Hom(\rho_{\A_2}, \sigma_{\A_2})$, we set
\[
    F_0( (S,T) ) := S \stack T \in \Hom( \rho_{\A_1} \stack \rho_{\A_2},  \sigma_{\A_1} \stack \sigma_{\A_2}).
\]
This functor $F_0$ clearly is linear, preserves the $*$-operation, and is faithful.
It is also full, since $\rho_{\A_1} \stack \rho_{\A_2}$ and $\sigma_{\A_1} \stack \sigma_{\A_2}$ are irreducible, and hence $\Hom_{\Delta_{\A_1 \stack \A_2}}(F_0((\rho_{\A_1},\rho_{\A_2})), F_0((\sigma_{\A_1},\sigma_{\A_2})))$ is either one-dimensional or the zero vector space by Schur's lemma. 
By Proposition~\ref{prop: equivalent states (direct implication)}, this space is one-dimensional if and only if $\rho_{\A_i} \cong \sigma_{\A_i}$.
The claim then follows.
\end{proof}

This result means that we can regard $\Delta_{\A_1} \boxtimes_0 \Delta_{\A_2}$ as a full subcategory of $\Delta_{\A_1 \stack \A_2}$, and we will do so from now on.

\begin{rem}
\label{rem:unique}
Alternatively, the functor can be defined using the endomorphism picture.
Since it is unclear if $\mathfrak{A}^{\Lambda_a}_1 \otimes \mathfrak{A}^{\Lambda_a}_2 = \mathfrak{A}_{12}^{\Lambda_a}$ (where the tensor product is that of $\rm{C}^*$-algebras represented on Hilbert spaces), this is less straightforward.
Note that if $\rho_{\A_1}$ is an endomorphism obtained from some representation satisfying SSC($\pi_1)$, it follows that $A \mapsto \rho_{\A_1}(\pi_1(A))$ satisfies the superselection criterion, and $\rho_{\A_1}$ is uniquely determined by $\rho_{\A_1}(\pi_1(A))$ for all $A \in \A_1$.
The same is of course true for the other stacked layer, and we can take the stacked representation.
As shown in the proof above, this is in $\Delta_{\A_1 \otimes \A_2}$, and we can \emph{uniquely} assign a $*$-endomorphism $\rho_{\A_1 \stack \A_2}$ to it.
It follows that we can define the functor on endomorphisms instead.
\end{rem}

From Theorem~\ref{Thm: Type I}, we know that a representation satisfying SSC($\pi_i$) decomposes as a countable (in particular, possibly infinite) direct sum of irreducible representations satisfying SSC($\pi_i$). It remains to be seen that infinite direct sums are well-defined in the category $\Delta_{\A_i}$. 
Let us first recall the definition of a subobject and a direct sum in $\rm{W}^*$-categories.
\begin{defn}[\cite{GLR}]\label{def:directsum}
    A category $\mathcal{C}$ has \emph{subobjects} if for every $\rho \in \mathcal{C}$ and $P = P^2 = P^* \in \Hom(\rho,\rho)$, there is an object $\sigma \in \mathcal{C}$ and $V \in \Hom(\sigma, \rho)$ such that $V V^* = P$ and $V^*V = \mathbb{I}$.

    If $I$ is a set, and $\rho_i \in \mathcal{C}$ for $i \in I$ a collection of objects.
    Then $\rho$ is the \emph{direct sum} of $\{\rho_i\}_{i \in I}$ if there are $V_i \in \Hom(\rho_i, \rho)$ such that
    \[
        V_i^* V_j = \delta_{i,j} \mathbb{I} \quad\quad\quad \sum_{i \in I} V_i V_i^* = \mathbb{I},
    \]
    where convergence of the sum is in the strong operator topology.
    We write $\rho = \bigoplus_{i \in I} \rho_i$ for the direct sum.
\end{defn}
Direct sums, if they exist, are unique up to unitary isomorphism.
Alternatively, one can define direct sums in $\rm{W}^*$-categories in terms of a universal property, see e.g.~\cite{Fritz_2019}.

\begin{prop}
The category $\Delta_{\A_1}$ has subobjects and countable direct sums.
\end{prop}

\begin{proof}
    It has already been shown that the category has subobjects and finite direct sums (see \cite[Lemma 5.7]{ogata_derivation_2021}). 
    It remains to show that we also have countable direct sums, which is proven in a similar way.
    Since $\pi_1(\A_1(\cone))''$ is properly infinite by assumption, there exist countably many isometries $V_i \in \pi_1(\A_1(\cone))''$ with orthogonal ranges such that 
    \[V_i\ad V_j = \delta_{i,j} \id \; \text{ and } \; \sum_iV_i V_i \ad \xrightarrow{\textrm{SOT}}
     \id,\]
     see the Halving Lemma and the proof of Theorem 6.3.4 of~\cite{kadison1997fundamentals2}.
    Then we can define the infinite direct sum $\sigma(A) := \sum_i V_i \sigma_i(A)V_i\ad$, with the sum converging in the strong operator topology.
    This is again a representation of $\A_1$. Moreover, $\sigma_i(A) = \pi_1(A)$ for all $i$ and $A \in \A_1(\Lambda^c)$ and so $\sigma(A) = \pi_1(A)$ for all $A \in \A_1(\Lambda^c)$ since $V_i\in\pi_1(\A_1(\cone))''\subset \pi_1(\A_1(\cone^c))'$ by locality. Since the $\sigma_i$ satisfy the superselection criterion, a short computation gives that the same is true for $\sigma$, and hence $\sigma \in \Delta_{\A_1}$.
    The verification that this is indeed a direct sum of the $\sigma_i$ in the category is analogous to the finite case.
\end{proof}
A category which has subobjects and direct sums is called \emph{Cauchy complete} in~\cite{2411.01678}.
Hence $\Delta_{\A_1}$ is Cauchy complete in that sense, with the caveat that we only require \emph{countable} (and not arbitrary) direct sums.

Since $\Delta_{\A_1 \stack \A_2}$ has countable direct sums, we can add them to our category $\Delta_{\A_1} \boxtimes_0 \Delta_{\A_2}$, which we recall we identify with a full subcategory of $\Delta_{\A_1 \stack \A_2}$.

\begin{defn}
We write $\Delta_{\A_1} \boxtimes \Delta_{\A_2}$ for the full subcategory of $\Delta_{\A_1 \stack \A_2}$ obtained by adding all (finite or countable) direct sums of objects to $\Delta_{\A_1} \boxtimes_0 \Delta_{\A_2}$.
\end{defn}

Since $\Delta_{\A_1} \boxtimes_0 \Delta_{\A_2}$ has subobjects (the only subobjects are trivial), by construction the category $\Delta_{\A_1} \boxtimes \Delta_{\A_2}$ also has subobjects.

\begin{rem}
This is a special case of the general construction in~\cite{2411.01678}. 
Briefly, we can define a tensor product of two Cauchy complete $\rm{W}^*$-categories $\mathcal{C}, \mathcal{D}$.
This category will in general not include all direct sums, but one can take a `completion' to add them.
Formally, if $\mathcal{C}$ is a $\rm{W}^*$-category, we can define a new category with as objects formal direct sums $\bigoplus_{i \in I} \rho_i$ with morphisms $\bigoplus_{i \in I} T_i$, such that the set $\{T_i \}_{i \in I}$ is bounded.
Similarly, if $\mathcal{C}$ does not have subobjects, we can add them as well.
The resulting category is denoted $\mathcal{C}^{\overline{\oplus}}$ and is called the \emph{Cauchy completion} of $\mathcal{C}$ (it does not matter if we first add direct sums, or first add subobjects).
Applying this to the tensor product of two Cauchy-complete $\rm{W}^*$-categories as mentioned earlier gives \cite[Definition~3.33]{2411.01678}. 
When applied to $\Delta_{\A_1}$ and $\Delta_{\A_2}$, this precisely gives a category that is equivalent (as $\rm{W}^*$-categories) to $\Delta_{\A_1} \boxtimes \Delta_{\A_2}$, when we restrict to adding at most countable direct sums only. In the notation of~\cite{Yamagami}, this means we are working in the category $\mathscr{SR}ep$ of normal representations on \emph{separable} Hilbert spaces only.
\end{rem}

We now make some observations that will be useful later.
Suppose that $\rho \in \Delta_{\A_1}$ and $\sigma \in \Delta_{\A_2}$.
Then it follows from Theorem~\ref{Thm: Type I} that $\rho \cong \bigoplus_{i \in I} \rho_i$ and $\sigma \cong \bigoplus_{j \in J} \sigma_j$, with the $\rho_i$ and $\sigma_j$ irreducible.
Let $\{V_i \}_{i\in I}$ and $\{W_j\}_{j\in J}$ be the corresponding isometries as in Definition~\ref{def:directsum}.
Then $Z_{ij} := V_i \otimes W_j \in \A_{12}^{\Lambda_a}$ forms a (at most) countable set of isometries such that $Z_{ij}^* Z_{k \ell} = \delta_{i,k} \delta_{j,\ell} \,\mathbb{I} \otimes \mathbb{I}$  and $\sum_{i,j} Z_{i,j} Z_{i,j}^*$ converges to $\mathbb{I} \otimes \mathbb{I}$ in the strong operator topology.

It follows that we can identify the pair $(\rho,\sigma)$ with an object in $\Delta_{\mathfrak{A}_1} \boxtimes \Delta_{\mathfrak{A}_2}$, and we will denote it by $\rho \boxtimes \sigma$.
From the construction it follows that $\boxtimes$ distributes over direct sums in both variables.
That is, the construction of $\boxt$ is compatible with the direct sums in the original categories in the expected way.
More concretely, we can define a functor $\overline{F} : \Delta_{\A_1} \times \Delta_{\A_2} \to \boxt$ by setting $\overline{F}( (\rho,\sigma) ) = \rho \boxtimes \sigma$ (and similarly for the morphisms in the category).
This functor is bilinear, and satisfies
\[
        \overline{F}\left( \bigoplus_{i \in I} \rho_i, \bigoplus_{j \in J} \sigma_j \right) = \left( \bigoplus_{i \in I} \rho_i \right) \boxtimes \left( \bigoplus_{j \in J} \sigma_j \right) \cong \bigoplus_{i \in I, j \in J} \rho_i \boxtimes \sigma_j = \bigoplus_{i \in I, j \in J} \overline{F}( (\rho_i, \sigma_j) )
\]
for all (at most countable) direct sums.

By definition $\boxt$ is a full subcategory of $\Delta_{\A_1 \stack \A_2}$.
Hence a natural question is whether they are equivalent, that is, if the inclusion functor is essentially surjective.
Since $\pi_1,\pi_2$ both satisfy the approximate split property, so does $\pi_1\otimes\pi_2$, by Lemma~\ref{lem:double commutants}. Theorem~\ref{thm:ssstackedirreducible} and Proposition~\ref{prop: equivalent states (direct implication)} imply that $\pi_1\otimes\pi_2$ has countably many sectors. Hence by Theorem~\ref{Thm: Type I}, it follows that every object in $\Delta_{\A_1 \stack \A_2}$ can be written as a countable direct sum of irreducibles.
By Theorem~\ref{thm:ssstackedirreducible}, each of this irreducibles is equivalent to $\rho \stack \sigma$, with $\rho$ and $\sigma$ irreducible.
Since this is an object of $\Delta_{\A_1} \boxtimes_0 \Delta_{\A_2}$, we have the following result.
\begin{prop}\label{prop:embedding}
The inclusion functor $\iota : \boxt \hookrightarrow \Delta_{\A_1 \stack \A_2}$ is an equivalence of $\rm{W}^*$-categories. 
\end{prop}

\begin{rem}
Again, this can be proven more abstractly, and it is enough to check that $\Delta_{\A_1} \boxtimes_0 \Delta_{\A_2}$ is a subcategory of so-called generators for $\Delta_{\A_1 \otimes \A_2}$, see~\cite[Lemma 3.16]{2411.01678}.
\end{rem}

Note that the categories $\Delta_{\A_i}$ come with a distinguished object, namely the reference representation $\pi_i$.
Hence we can define full and faithful functors $\Delta_{\A_i} \to \Delta_{\A_1}\times \Delta_{\A_2}$ by setting $\rho \mapsto (\rho, \pi_2)$ (resp. $\sigma \mapsto (\pi_1, \sigma)$).
Similarly, we can define inclusions $F_i : \Delta_{\A_i} \to \boxt$ by setting $F_1(\rho) := \rho \boxtimes \pi_2$ (resp. $F_2(\sigma) := \pi_1 \boxtimes \sigma$).
Finally, stacking with the reference representation of the other layer leads to full and faithful functors $\iota_i : \Delta_{\A_i} \hookrightarrow \Delta_{\A_1 \stack \A_2}$.
This gives the following commutative diagram:
\begin{equation}
\label{eq:cd}
\begin{tikzcd}
	& {\Delta_{\mathfrak{A}_1} \times \Delta_{\mathfrak{A}_2} }\\
	{\Delta_{\mathfrak{A}_1}} & \boxt & {\Delta_{\mathfrak{A}_2}} \\
	& {\Delta_{\mathfrak{A}_1 \stack \mathfrak{A}_2}}
	\arrow["\cong", "\iota"', dashed, from=2-2, to=3-2]
	\arrow["{\bar{F}}"{description}, from=1-2, to=2-2]
	\arrow[hook, from=2-1, to=1-2]
	\arrow["{F_1}"{description}, from=2-1, to=2-2]
	\arrow[hook', from=2-3, to=1-2]
	\arrow["{F_2}"{description}, from=2-3, to=2-2]
    \arrow["\iota_1"', from=2-1, to=3-2]
    \arrow["\iota_2", from=2-3, to=3-2]
\end{tikzcd}
\end{equation}
Here, the functors in the bottom half of the diagram (including the horizontal ones) are all $\rm{W}^*$-functors.

To complete the picture, it remains to consider the monoidal structure (`fusion') and the braiding on $\boxt$.
To define this, if $\pi \in \Delta_{\A_1}$, write $\pi^a$ for its extension to an endomorphism of $\A_1^{\Lambda_a}$.
Then for $\rho,\sigma \in \Delta_{\A_1}$, we can define $\rho \widehat{\otimes} \sigma := \rho^a \circ \sigma$.
Here we use $\widehat{\otimes}$ for the tensor product in the category, to distinguish it from the (many) other tensor products we have considered so far.
It can be shown that $\rho \widehat{\otimes} \sigma$ an object in $\Delta_{\A_1}$. To define the monoidal product on the morphisms, consider $S \in \Hom_{\Delta_{\A_1}}(\rho, \rho')$ and $T \in \Hom_{\Delta_{\A_1}}(\sigma, \sigma')$.
Using approximate Haag duality, one can show that $S,T \in \A_{1}^{\Lambda_a}$.
This is in fact true for any intertwiner between representations $\rho$ and $\sigma$ (strictly) localized in two cones $\Lambda_1$ and $\Lambda_2$ in the sense of Definition~\ref{def:sscat} as long are both cones are admissible (this is important when defining the braiding).
Using this observation, we can define $S \widehat{\otimes} T = S \rho^a(T)$.
This gives a strict monoidal category $(\Delta_{\A_1}, \widehat{\otimes}, \pi_1)$.
Finally, we can define a braiding $\epsilon_{\rho,\sigma} \in \Hom_{\Delta_{\A_1}} (\rho \widehat{\otimes}  \sigma, \sigma \widehat{\otimes} \rho)$.
The idea behind the definition of the braiding is that we move one of the representations (say, the one in the second variable) far away, show that we can then interchange both in the monoidal product, and move the representation back.
Again we refer the reader to \cite[Section~IV]{ogata_derivation_2021} for the full details.

There is a natural monoidal product and braiding on $\boxt$, coming from the corresponding operations of $\Delta_{\A_1}$ (resp. $\Delta_{\A_2}$).
For pairs $\rho_{\A_1}\boxtimes \rho_{\A_2}$ and $\sigma_{\A_1} \boxtimes \sigma_{\A_2}$, we define
\[
    \left(\rho_{\A_1} \boxtimes \rho_{\A_2}\right) \widehat{\boxtimes} \left(\sigma_{\A_1} \boxtimes \sigma_{\A_2}\right) := 
    \left(\rho_{\A_1} \widehat{\otimes} \sigma_{\A_1}\right) \boxtimes \left(\rho_{\A_2} \widehat{\otimes} \sigma_{\A_2}\right).
\]
The monoidal product on morphisms is also given component-wise.
These constructions can be extended to the full category $\boxtimes$ by defining $\widehat{\boxtimes}$ on countable direct sums.
For example,
\[
\left(\bigoplus_{i \in I} \rho_i \boxtimes \sigma_i \right) \widehat{\boxtimes} \left(\bigoplus_{j \in J} \tau_j \boxtimes \zeta_j \right) := 
\bigoplus_{i \in I, j \in J} (\rho_i \widehat{\otimes }\tau_j) \boxtimes (\sigma_i \widehat{\otimes }\zeta_j).
\]
The monoidal unit is given by $\pi_1 \boxtimes \pi_2$.
Using the construction above, similarly the braiding $\varepsilon^i_{\rho_{\A_i}, \sigma_{\A_i}}$ of $\Delta_{\A_i}$ extends to $\boxtimes$, by setting
\[
    \varepsilon^\boxtimes_{\rho_{\A_1} \boxtimes \rho_{\A_2}} := \varepsilon^1_{\rho_{\A_1}, \sigma_{\A_1}} \boxtimes \varepsilon^2_{\rho_{\A_2}, \sigma_{\A_2}},
\]
and extended as earlier to all of $\boxt$.
This makes $\boxt$ into a braided $\rm{W}^*$-category $(\boxt, \widehat{\boxtimes}, \pi_1 \boxtimes \pi_2)$.\footnote{Unlike $\Delta_{\A_1}$, the category $\boxt$ is no longer a \emph{strict} tensor category because direct sums are only unique up to unitary isomorphisms.}

\begin{rem}
    We emphasize that the category $\boxt$ represents a pairing of \textit{decoupled} quantum systems. That is, there are no `string excitations' which connect the two layers $\Delta_{\A_1}$ and $\Delta_{\A_2}$. With this in mind, the constructed monoidal product and braiding morphism respect this property and are given as formal tensor products of the product or braiding in individual layers.
\end{rem}

In view of Remark~\ref{rem:unique}, note that $\A_1^{\Lambda_a} \stack \A_2^{\Lambda_a} \subset \A_{12}^{\Lambda_a}$.
We then have the following lemma.
\begin{lem}\label{lem:extension}
    Let $\rho_i \in \Delta_{\A_i}$.
    Then the following equation holds for all $A \in \A_1^{\Lambda_a} \stack \A_2^{\Lambda_a}$:
    \[
        (\rho_1^a \stack \rho_2^a)(A) = (\rho_1 \stack \rho_2)^a(A).
    \]
    Here on the left-hand side, we first extend to the auxiliary algebra and then stack, while on the right-hand side we do it the other way round (note that the extension there is to $\A_{12}^{\Lambda_a})$.
\end{lem}
\begin{proof}
Recall from the definition of $\rho_i^a$ (see~\cite[Lem. 2.13]{ogata_derivation_2021}) that it is enough to define $\rho_i^a$ on cone algebras $\pi_i(\mathfrak{A}(\Lambda'))''$ for all admissible cones $\Lambda'$.
Given such a cone, it follows from the selection criterion that there is a unitary $V_i \in B(\mathcal{H}_{\pi_i})$, where $V_i$ is a charge transporter that transports $\rho_i$ to a representation that is contained in the cone $\Lambda_a$ (and sufficiently far away from $\Lambda'$).
We then have $\rho_i^a(A) := V_i A_i V_i^*$ for $A_i \in \A_i(\Lambda')''$.
But the unitary $V_1 \stack V_2$ is such a charge transporter for $\rho_1 \stack \rho_2$.
Hence for $A_i \in \A_i(\Lambda')''$, we have
\[
    (\rho_1 \stack \rho_2)^a(A_1 \stack A_2) = (V_1 \stack V_2) (A_1 \stack A_2) (V_1 \stack V_2)^* = V_1 A_1 V_1^* \stack V_2 A_2 V_2^* = (\rho_1^a \stack \rho_2^a)(A_1 \stack A_2).
\]
The result then follows for general $A \in \A_1^{\Lambda_a} \stack \A_2^{\Lambda_a}$ by using that all maps are $*$-endo\-morphisms. 
\end{proof}

By Proposition~\ref{prop:embedding}, $\boxt$ is a full subcategory of $\Delta_{\A_1 \stack \A_2}$, which has its own monoidal product and braiding defined through the sector theory.
It turns out that these agree with those on $\boxt$.

\begin{prop}
The embedding functor $\iota: \boxt \to \stacc$ is a braided monoidal functor.
\end{prop}
\begin{proof}

To see that $\iota$ is a monoidal functor with respect to the tensor products defined, it is sufficient to show this on the simple objects, as $\iota$ is a normal $*$-functor (in the language of $\rm{W}^*$-categories), which implies it sends direct sums to direct sums.
Because of Remark~\ref{rem:unique}, it is enough to show that
\[
        (\rho_1 \stack \rho_2)^a \circ (\sigma_1 \stack \sigma_2)(A) = \left( \rho_1^a \circ \sigma_1 \stack \rho_2^a \circ \sigma_2 \right)(A)
\]
for all $A \in \A_1 \stack \A_2$.
But this follows from Lemma~\ref{lem:extension}, since $\sigma_i(A_i) \in \A_1^{\Lambda_a}$.
It is then readily checked that the tensor products on the morphisms are the same as well.

Finally, we would like to see that this functor is braided with respect to the braiding in $\boxt$.
Again it is enough to verify it for `simple tensors' of the form $\rho \boxtimes \sigma$, using naturality of the braiding and that $\iota$ preserves direct sums.
Hence the problem reduces to verifying that
\[
    \varepsilon^1_{\rho_{\A_1}, \sigma_{\A_1}} \boxtimes \varepsilon^2_{\rho_{\A_2}, \sigma_{\A_2}}
    =
    \varepsilon_{\rho_{\A_1} \stack \sigma_{\A_1}, \rho_{\A_2} \stack \sigma_{\A_2}}.
\]
To see this, recall that the braiding is defined by choosing a sequence of unitaries (i.e., a sequence of charge transporters) that moves one of the excitations.
But by the same observation as in the proof of Lemma~\ref{lem:extension}, choosing these sequences for each of the stacked layers and then stacking the unitaries gives an equivalent sequence for the stacked representations.
These can then be used to define the braiding $\varepsilon$ of $\Delta_{\A_1 \stack \A_2}$, and the equality above follows.
\end{proof}

Recall that just as for ordinary functors, if we have a braided monoidal full and faithful functors between braided categories which is also essentially surjective, this implies that the two categories are \emph{braided} equivalent.
Hence the results of this section can be summarised as the following theorem.

\begin{thm}[Theorem~\ref{thm:deligne}]\label{thm:C}
The categories $\boxt$ and $\Delta_{\A_1 \stack \A_2}$ are equivalent as braided $\rm{W}^*$-categories.
\end{thm}

And so we have constructed a braided tensor category which is equivalent to the stacked system sector theory. The following remarks demonstrate the advantages of having done so.

\begin{rem}
\label{rem:ogata_mixed}
    The work in \cite{ogata2025mixedstatetopologicalorder} establishes a faithful braided functor between the individual layer category $\Delta_{\A_1}$ and the stacked system category $\stacc$ in the case where the reference representation on $\A_2$ is the GNS representation of a product state (and hence has trivial superselection sectors) via a *-isomorphism $\Theta:\pi_\varphi(\A_1)'' \to \pi(\A_1)''$ with $\varphi$ the restriction of a state on ${\A_1}\otimes_s \A_2$ to the algebra $\A_1$. This construction can be extended to the stacking functor $\iota$ via the inclusion $F_1: \Delta_{\A_1} \hookrightarrow \Delta_{\A_1} \boxtimes \Delta_{\A_2}$ given by pairing with the trivial (reference) representation, as defined above.

We need to see that $\iota \circ F_1 :\Delta_{\A_1} \to \Delta_{\A_1 \stack \A_2}$ preserves the monoidal structure of the categories. Since a composition of braided monoidal functors gives a braided monoidal functor, it only remains to see that the inclusion $F_1$ is indeed a functor of this nature. 
To see it is monoidal, consider $\rho_1 \widehat{\otimes}\rho_2 \in \Delta_{\A_1}$. Then we have \[F_1(\rho_1\widehat{\otimes}\rho_2) = (\rho_1\widehat{\otimes}\rho_2) \boxtimes \pi_2 = (\rho_1 \boxtimes \pi_2) \widehat{\boxtimes} (\rho_2 \boxtimes \pi_2)\] by using the monoidal product in this category. Hence, $F_1(\rho_1 \widehat{\otimes}\rho_2) = F_1(\rho_1)  \widehat{\boxtimes}  \;F_1 (\rho_2).$

The proof that $F_1$ preserves the braiding is similar.
Therefore, also the composition  $\iota \circ F_1$ carries over the monoidal structure of the categories. A similar argument can be made for $F_2: \Delta_{\A_2} \to \Delta_{\A_1} \boxtimes \Delta_{\A_2}$.
That is to say, the functors in the lower part of diagram~\ref{eq:cd} are all braided monoidal functors.
\end{rem}

\begin{rem}
The above results show that the sectors of the stacked system are related to the sectors of the two layers.
In applications, it is often desirable to  restrict to the full subcategory $\Delta^f_{\A_1}$ of sectors which are dualizable, namely, they have conjugates.
Having dual (conjugate) objects is usually assumed in algebraic descriptions of the theory of anyons~\cite[Appendix E]{KitaevAppendices}, and indeed is part of the requirements of a UMTC.
Physically, this means that we only consider anyons for which there is a conjugate charge, namely an anti-particle.
It follows that all Hom-spaces in the category are finite dimensional, which implies that every sector in the category can be written as a finite direct sum of irreducible sectors~\cite{LongoRoberts97}.
In this case, we have that $\Delta^f_{\mathfrak{A}_1} \boxtimes \Delta^f_{\mathfrak{A}_2} \cong \Delta^f_{\mathfrak{A}_1 \stack \mathfrak{A}_2}$, where in the construction of $\boxtimes$ we now only add \emph{finite} direct sums.
In this case, $\boxtimes$ is essentially the Deligne product of monoidal categories (apart from the technical point that our categories strictly speaking are not abelian as we exclude the zero representation for convenience).
\end{rem}
 
\section{Examples}\label{Examples}
\subsection{Stacking quantum double models}
\label{sec:qdexample}
Quantum double models~\cite{kitaev2003fault} form an important class of examples of anyonic systems, and therefore it is useful to analyze the stacking operation specifically in this context.
Indeed, the quantum double model of any finite group has only finitely many superselection sectors~\cite{bols2025classification}, satisfies bounded spread Haag duality~\cite{PerezGarcia}, and satisfies the approximate split property by Corollary~\ref{cor: ASP with xi and R}, where the existence of a finite $\xi$ follows for example from the area law for these models~\cite{Bachmann17,Cui20} (cf.\  Remark~\ref{rem:Validity}), or can be verified directly by showing that the ground state is a product state for sufficiently far removed regions~\cite[\S 5]{naaijkens2015haag}. 
Hence, our main results apply directly to this situation.
As an example, we show that the quantum double model for direct product groups $G \times H$ can be naturally interpreted in terms of the stacking operation.
Namely, the quantum double model for $G \times H$ is equivalent to the stacking of the model for $D(G)$ and that for $D(H)$, where $D(G)$ is the quantum double of the group algebra $\mathbb{C}[G]$.
If we write $\Delta_G^f$ for the braided category of (dualizable) sectors of the quantum double model for $G$, this analysis implies that $\Delta_{G \times H}^f \cong \Delta^f_G \boxtimes \Delta^f_H$ as unitary braided monoidal categories.

This result can be understood purely algebraically.
One can check that $D(G \times H) \cong D(G) \otimes D(H)$ as quasi-triangular Hopf algebras.
This implies the result that $\textbf{Rep}_f D(G \times H) \cong \textbf{Rep}_f D(G) \boxtimes \textbf{Rep}_f D(H)$, where $\textbf{Rep}_f$ is the (unitary braided) category of finite dimensional unitary $D(G)$-modules (cf.~\cite[Prop. 1.11.2]{Tensor_cats} or~\cite[Sect. 4.2]{Mueger03}).

We begin by reviewing some basic concepts, as well describing the setting and conventions we will use for the remainder of this section.  Fix two finite groups, $G$ and $H$, and denote their direct product by $K = G \times H : = \{ g \times h \mid g \in G , h \in H \}$.  For $J = G,H,K$, let $\mathcal{H}_J$ denote the Hilbert space formed by formal linear combinations of group elements,
$$\sum_{j \in J} \lambda_j |j \rangle, \ \lambda_j \in \mathbb{C},$$
where the group elements of $J$ are taken to be an orthonormal basis.  Hence, we may write
$$\mathcal{H}_J = \text{span} \{ | j \rangle \mid j \in J \}.$$
Since $K= G \times H$, it is clear that we may make the identification $\mathcal{H}_K \equiv \mathcal{H}_G \otimes \mathcal{H}_H$ by identifying $|g \times h \rangle \equiv | g \rangle \otimes |h \rangle$ for any $g \in G$ and $h \in H$.

In addition to forming a basis for $\mathcal{H}_J$, we also identify each group element $j \in J$ with the unitary defined by $j | \tilde{j} \rangle := | j \tilde{j} \rangle$ for any $\tilde{j} \in J$, and then extended to all of $\mathcal{H}_J$ by linearity.  This is indeed a unitary since each group element simply permutes the defining basis of $\mathcal{H}_J$ by acting on the underlying group on the left.  The group $\mathrm{C}^*$-algebra of $J$ is defined as the subalgebra $C^* (J) \subset B(\mathcal{H}_J)$  that is generated by the unitaries corresponding to the elements of $J$.  By our identification $|g \times h \rangle \equiv | g \rangle \otimes |h \rangle$, it is clear we can make the identification $g \times h \equiv g \otimes h$ as unitaries acting on $\mathcal{H}_K$.  Thus, we have that $C^*(K) = C^*(G) \otimes C^*(H)$.

Let $\mathfrak{A}_J$ denote the quasi-local algebra on the lattice $\Gamma$ whose on-site algebras are given by copies of $B(\ell^2(J)) \cong M_{|J|}(\mathbb{C})$.  Here we will take our sites to be the edges of $\Gamma$ rather than the vertices or plaquettes, as is customary for the quantum double model.
What is more, the edges will be taken to be directed.
Since $|K| = |G| \cdot |H|$, we can identify $M_{|K|}(\mathbb{C}) \cong M_{|G|}(\mathbb{C}) \otimes M_{|H|}(\mathbb{C})$.
We do this identification in the obvious way, namely such that the unitary $(g,h)$ corresponds to $g \otimes h$ under the isomorphism.
Using this identification, with a slight abuse of notation, we have that $\mathfrak{A}_K = \mathfrak{A}_G \stack \mathfrak{A}_H$.

For each vertex $v$ and plaquette $p$, denote by $S_v$ and $S_p$ the set of all edges incident with $v$ and the set of edges forming the boundary of $p$ respectively.  Let $A_J^v (j)$ denote the vertex operator on the local algebra $\mathfrak{A}_J(S_v)$ associated to $j \in J$, and $B_J^p$ the plaquette operator on the local algebra $\mathfrak{A}_J(S_p)$. 
Schematically, these operators can be defined on a basis as follows:
\begin{equation}
A_J^v(j)\,
\tikzmath[x=1.25cm, y=1.25cm]{
    \draw[mid] (0,0) -- (1,0) node[midway,below] {$g_1$};
    \draw[mid] (0,0) -- (0,1) node[midway,right]  {$g_2$};
    \draw[mid] (-1,0) -- (0,0) node[midway,below] {$g_3$};
    \draw[mid] (0,-1) -- (0,0) node[midway,right]  {$g_4$};
    \draw[fill] (0,0) circle (3pt);
    \node[below right] at (0,0) {$v$};
}
=
\tikzmath[x=1.25cm, y=1.25cm]{
    \draw[mid] (0,0) -- (1,0) node[midway,below] {$j g_1$};
    \draw[mid] (0,0) -- (0,1) node[midway,right]  {$j g_2$};
    \draw[mid] (-1,0) -- (0,0) node[midway,below] {$g_3 j^{-1}$};
    \draw[mid] (0,-1) -- (0,0) node[midway,right]  {$g_4 j^{-1}$};
    \draw[fill] (0,0) circle (3pt);
    \node[below right] at (0,0) {$v$};
}
\quad\quad
B_J^p\,\,
\tikzmath[x=1.25cm, y=1.25cm]{
    \draw[mid] (0,0) -- (1,0) node[midway,below] {$g_1$};
    \draw[mid] (1,0) -- (1,1) node[midway,right]  {$g_2$};
    \draw[mid] (0,1) -- (1,1) node[midway,above] {$g_3$};
    \draw[mid] (0,0) -- (0,1) node[midway,left]  {$g_4$};
    \draw[fill] (0,0) circle (3pt);
    \draw[fill] (1,0) circle (3pt);
    \draw[fill] (1,1) circle (3pt);
    \draw[fill] (0,1) circle (3pt);
    \node at (0.5,0.5) {$p$};
}
\,
=
\,
\delta_{e,g_1 g_2 g_3^{-1} g_4^{-1}}
\,
\tikzmath[x=1.25cm, y=1.25cm]{
    \draw[mid] (0,0) -- (1,0) node[midway,below] {$g_1$};
    \draw[mid] (1,0) -- (1,1) node[midway,right]  {$g_2$};
    \draw[mid] (0,1) -- (1,1) node[midway,above] {$g_3$};
    \draw[mid] (0,0) -- (0,1) node[midway,left]  {$g_4$};
    \draw[fill] (0,0) circle (3pt);
    \draw[fill] (1,0) circle (3pt);
    \draw[fill] (1,1) circle (3pt);
    \draw[fill] (0,1) circle (3pt);
    \node at (0.5,0.5) {$p$};
}
\end{equation}
Here, $g_1, \dots, g_4 \in J$, and $e$ is the unit element of $J$.
We refer the reader to a standard reference, such as \cite{kitaev2003fault} for further details.
Also denote
$$A_J^v = \frac{1}{|J|} \sum_{j \in J} A_J^v (j).$$
The Hamiltonian for the quantum double model of $J$ is given by
$$H_J = \sum_v (\mathbb{I} - A_J^v) + \sum_p (\mathbb{I} - B^p_J).$$
Of course, in the infinite volume case this sum does not converge, but $i [H_J, O]$ is well-defined for local operators $O$ and defines a derivation that generates the dynamics of the system, and hence we can talk about ground states.
We say a state $\omega_J$ on $\mathfrak{A}_J$ is \textit{frustration-free} if $\omega_J(A_J^v) = \omega_J(B^p_J) = 1$ for every vertex $v$ and plaquette $p$.  It can be shown, though we will not do so here, that there is only one frustration-free ground state $\omega_j$ on $\mathfrak{A}_J$~\cite{NaaijkensPhD,naaijkens2015haag,Cui20}.

\begin{prop}
    \label{prop:qdstack}
    Let $\omega_G$ and $\omega_H$ be the unique frustration-free ground states on $\mathfrak{A}_G$ and $\mathfrak{A}_H$ respectively.  Then the frustration-free ground state on $\mathfrak{A}_K$ is given by $\omega_K = \omega_G \stack \omega_H$.
\end{prop}

\begin{proof}
    Let $v$ be a vertex and $p$ a plaquette.  For every $\phi \in \bigotimes_{e \in S_v} \mathcal{H}_G$ and $\psi \in \bigotimes_{e \in S_v} \mathcal{H}_H$, we have
    $$A_K^v (| \phi \rangle \otimes | \psi \rangle ) = \frac{1}{|K|} \sum_{g \times h \in K } A_K^v (g \times h) (| \phi \rangle \otimes | \psi \rangle)$$
    $$= \frac{1}{|G|} \frac{1}{|H|} \sum_{g \in G} \sum_{h \in H} A_G^v (g) | \phi \rangle \otimes A_H^v (h) | \psi \rangle = A_G^v | \phi \rangle \otimes A_H^v | \psi \rangle.$$
    Moreover, for $\phi \in \bigotimes_{e \in S_p} \mathcal{H}_G$ and $\psi \in \bigotimes_{e \in S_p} \mathcal{H}_H$, a similar computation yields
    $$B_K^p (| \phi \rangle \otimes | \phi \rangle) = B_G^p | \phi \rangle \otimes B^p_H | \psi \rangle.$$
    Thus, as quasi-local operators on the whole lattice we have that $A_K^v = A_G^v \stack A_H^v$ and $B_K^p = B^p_G \stack B^p_H$, and so
    $$\omega_G \stack \omega_H (A^v_K) = \omega_G(A^v_G) \omega_H(A^v_H) = 1 \ \text{and} \ \omega_G \stack \omega_H (B^p_K) = \omega_G(B^p_G) \omega_H(B^p_H) = 1.$$
    By uniqueness of the frustration-free ground state, we therefore have $\omega_K = \omega_G \stack \omega_H$.
\end{proof}

The category of superselection sectors for the quantum double model has been worked out explicitly~\cite{bols2025classification,bols2025category}.
It can be shown that every irreducible sector is dualizable, and hence it is natural to only consider finite direct sums of irreducible sectors.
The corresponding category, which we denote $\Delta_G^f$, is then unitary braided equivalent to $\textbf{Rep}_f D(G)$.
Hence if we apply Theorem~\ref{thm:deligne} to the result of Proposition~\ref{prop:qdstack}, it follows that $\textbf{Rep}_f(D(K)) \cong \textbf{Rep}_f(D(G))\boxtimes\textbf{Rep}_f(D(H))$, where each of the categories is equivalent to the corresponding DHR category.
This reproduces the result from Hopf algebra theory mentioned earlier.

If we are just interested in the superselection sectors (and not the fusion rules and braiding), this result can be understood more directly.
We will now sketch the idea.
By the above and Proposition \ref{prop: GNS of product}, we see that if $(\pi_J, \mathcal{H}_J, \Omega_J)$ is the GNS triple corresponding to the frustration-free ground state on the quantum double model $\mathfrak{A}_J$ for $J=G,H,K$, then
$$(\pi_G \stack  \pi_H, \mathcal{H}_G \otimes \mathcal{H}_H, \Omega_G \otimes \Omega_H) = (\pi_K , \mathcal{H}_K , \Omega_K).$$
Our goal will therefore be to understand the representations satisfying SSC($\pi_K$) in terms of those satisfying SSC($\pi_G$) and SSC($\pi_H$).

A central role in the analysis of the sector structure is played by so-called ribbon operators $F^{g_1, g_2}(\xi)$, where $\xi$ is a `ribbon' (we refer to e.g.~\cite{kitaev2003fault} or~\cite{Yan_2022} for the precise definition).
These operators create a pair of (conjugate) excitations at the endpoints of the ribbon.
Representatives of superselection sectors are then obtained by a limiting procedure, where one end of the ribbon is sent to infinity.
It is known that irreducible representations of $D(G)$ are in one-one correspondence with pairs $(C, R)$, where $C$ is a conjugacy class of $G$, and $R$ an irreducible representation of the centralizer of an element in $C$~\cite{Gould}.
And indeed, the ribbon operators can be grouped into multiplets corresponding to such pairs $(C,\rho)$, and these lead precisely to representatives of each irreducible sector~\cite{bols2025classification}.
We will now write $\underline{\Delta}_J$ for the set of equivalence classes of irreducible sectors.

Our ultimate goal will be to show that $\underline{\Delta}_K = \underline{\Delta}_G \stack \underline{\Delta}_H : = \{ [\rho_G \stack \rho_H ] \mid [\rho_G] \in \underline{\Delta}_G , [ \rho_H ] \in \underline{\Delta}_H \}$.  Indeed one can accomplish this by unpacking the definition of the ribbon operators and showing that for any $g_1,g_2 \in G$ and $h_1,h_2 \in H$,
$$F^{(g_1 \times h_1, g_2 \times h_2)} (\xi)= F^{(g_1,g_2)} (\xi) \stack F^{(h_1,h_2)} (\xi).$$
However, since the $\underline{\Delta}_J$'s are finite sets, the result can be established by simply counting the possible representations.

\begin{prop}
   Let $G,H$ be finite groups and $K = G\times H$. Then $|\underline{\Delta}_K| = |\underline{\Delta}_G||\underline{\Delta}_H|.$
\end{prop}

\begin{proof}
    For each group $J$, let $C_J$ denote the set of conjugacy classes of $J$ and for each $c \in C_J$ let $R_c$ denote the set of unitary equivalence classes of irreducible representations of $N_c$.  For any $g \times h \in K$,
    $$\{ (\tilde{g} \times \tilde{h})^{-1} (g \times h ) (\tilde{g} \times \tilde{h}) \mid \tilde{g} \times \tilde{h} \in K \} = \{ \tilde{g}^{-1} g \tilde{g} \times \tilde{h}^{-1} h \tilde{h} \mid \tilde{g} \in G, \tilde{h} \in H \}$$
    $$= \{ \tilde{g}^{-1} g \tilde{g} \mid \tilde{g} \in G \} \times \{ \tilde{h}^{-1} h \tilde{h} \mid \tilde{h} \in H \},$$
    and so $C_K = \{ c \times c' \mid c \in C_G, c' \in C_H \}$.  This implies immediately that $|C_K| = |C_G||C_H|$.  Moreover, for any $g \in c \in C_G$ and $h \in c' \in C_H$,
    $$\{g' \times h' \in K \mid (g' \times h') (g \times h )= (g \times h) (g' \times h') \}$$
    $$= \{g' \in G \mid gg'=g'g \} \times \{ h' \in H \mid hh' = h'h \},$$
    so $N_{c \times c'} = N_c \times N_{c'}$.  It is a standard fact from the representation theory of finite groups that the irreducible representations of a direct product of groups are just the tensor product of the representations of the individual groups.  Thus, $|R_{c \times c'}| = |R_c| |R_{c'}|$.  Therefore the computation
    $$|\underline{\Delta}_K| = \sum_{\tilde c \in C_K} |R_{\tilde c}| = \sum_{c \in C_G , c' \in C_H} |R_c||R_{c'}| = \Big( \sum_{c \in C_G} |R_c| \Big)  \Big( \sum_{c' \in C_H} |R_{c'}|  \Big) = |\underline{\Delta}_G||\underline{\Delta}_H|$$
    concludes the proof.
\end{proof}

\begin{cor}
   Let $G,H$ be finite groups and $K = G\times H$. Then  $\underline{\Delta}_K = \underline{\Delta}_G \stack \underline{\Delta}_H$.
\end{cor}

\begin{proof}
By Theorem~\ref{thm:SSC}, if $\rho_G,\rho_H$ are irreducible representations satisfying SSC($\pi_G$), respectively SSC($\pi_H$), then $\rho_G\otimes \rho_H$ satisfies SSC($\pi_K$). Since $U_G\in\caU(\caH_G),U_H\in\caU(\caH_H)$ give a unitary $U_G\otimes U_H \in \caU(\caH_G\otimes \caH_H)$ and we have that $\caH_G\otimes \caH_H = \caH_K$, this 
gives a well-defined map $\iota: \underline{\Delta}_G \times \underline{\Delta}_H \to \underline{\Delta}_K$.
By Proposition~\ref{prop: equivalent states (direct implication)}, this map is injective.
It follows that we get an inclusion $\underline{\Delta}_G \stack \underline{\Delta}_H \subset \underline{\Delta}_K$.
Finally,
    $$| \underline{\Delta}_G \stack \underline{\Delta}_H | =| \{[\rho_G \otimes \rho_H] \mid [\rho_G] \in \underline{\Delta}_G \ \text{and} \ [\rho_H] \in \underline{\Delta}_H \} | = |\underline{\Delta}_G | |\underline{\Delta}_H | = | \underline{\Delta}_K |,$$
    and since these are finite sets, we have $\underline{\Delta}_K = \underline{\Delta}_G \stack \underline{\Delta}_H$.
\end{proof}

\subsection{Morita-equivalent Levin--Wen models}
We now turn to Levin--Wen string-net models~\cite{LevinW06,GenStringNets}, which are commuting-projector Hamiltonians associated with a unitary fusion category $\mathcal{D}$.
Let $\omega$ be the unique frustration-free ground state of the model (the existence of which follows from~\cite[Thm. 4.8 and Rem. 2.25]{JonesNaaijkensPenneysWallick} or~\cite{QiuWang}).
Then its GNS representation satisfies approximate Haag duality by~\cite{PerezGarcia}.
Hence we can define the braided $\rm{C}^*$-category of superselection sectors $\Delta_\omega$.
This model is recently shown to have finitely many irreducible superselection sectors~\cite{BolsKjaerI} and satisfies the assumptions in Sect.~\ref{sec:teesplit} (see Remark~\ref{rem:Validity} for details).
Hence this model has the Type I property.\footnote{Since we will only need stacking with trivial product states, alternatively one could use the results in~\cite{ogata2025mixedstatetopologicalorder} instead of the Type I property.}

An important question is how $\Delta_\omega$ depends on the input category $\mathcal{D}$. While $\mathcal{D}$ determines the microscopic description of the model, in particular the dimension of the local Hilbert spaces at each site and the local interactions, different local descriptions may lead to equivalent emerging `global' properties, such as the category $\Delta_\omega$ of superselection sectors. Because the local dimensions for different $\caD$'s are in general different, there cannot always be an LGA connecting the ground states of such different models. However, a connecting automorphism may be constructed after the addition of degrees of freedom at each site, namely by stacking with a trivial system: One then speaks of \emph{stable equivalence}, see again~\cite{KitaevSRE}. Recalling~\cite{naaijkens2022split} that product states have trivial superselection structure, our results confirm that such a stabilization leaves the sector theory invariant indeed. The Levin--Wen examples illustrate that it is needed; see also~\cite[Appendix~B]{FKM} for the simplest example of non-interacting fermions.

To state the result, we first recall the notion of Morita equivalence for tensor categories~\cite{MR1966524}.
\begin{defn}
    Two unitary fusion categories $\mathcal{D}_1, \mathcal{D}_2$ are called \emph{Morita equivalent} if they have equivalent centers: $\mathcal{Z}(\mathcal{D}_1) \cong \mathcal{Z}(\mathcal{D}_2)$.
    Here the equivalence is as unitary braided tensor categories.
\end{defn}

Here $\mathcal{Z}(\mathcal{D})$ is the quantum double (or Drinfeld center) of $\mathcal{D}$, a generalisation of the quantum double construction for Hopf algebras to tensor categories.
See e.g.~\cite{MR1966525} for the definition in the case of the $\rm{C}^*$-categories we are considering.
The definition of Morita equivalence given above is equivalent to the one given in~\cite{MR1966524} by~\cite[Thm. 8.12.3]{Tensor_cats}.

\begin{prop}
    The ground states of Levin--Wen string-net models with Morita equivalent input categories have the same superselection structure, namely $\Delta_{\omega_1} \cong \Delta_{\omega_2}$.
\end{prop}

\begin{proof}
Let $\omega_1$, $\omega_2$ be the ground states of Levin--Wen models from Morita equivalent categories $\mathcal{D}_1$ and $\mathcal{D}_2$. By~\cite{Lootens_2022}, there is a finite-depth quantum circuit mapping between ground states of two Morita equivalent string-net models $\caD_1, \caD_2$.
Since the total dimension of Hilbert spaces for the two models need not match, one first adds ancillary qudits at each site.
In the present setting, this amounts to stacking the systems with a product state $\varphi_j,j=1,2$ to obtain algebras $\A_{\om_1} \stack \A_{\varphi_1}$ and $\A_{\om_2} \stack \A_{\varphi_2}$ of the same local dimension.
On the stacked system one can define an FDQC $\alpha = \alpha_1 \circ \dots \circ \alpha_n$ such that $(\omega_1 \stack \varphi_1) \circ \alpha = \omega_2 \stack \varphi_2$
Here each `layer' $\alpha_i$ is an automorphism given by conjugating with unitaries with mutually disjoint support (and the size of the support is uniformly bounded). Such an automorphism $\alpha$ can be seen as a particularly simple LGA. We then have the following commuting diagram, where each equivalence is an equivalence of braided $\rm{C}^*$-categories:

\[\begin{tikzcd}
	{\Delta_{\omega_1}} && {\Delta_{\omega_2}} \\
	\\
	{\Delta_{\omega_1 \otimes_s\varphi_1}} && {\Delta_{\omega_2 \otimes_s \varphi_2}}
	\arrow["\cong", dashed, from=1-1, to=1-3]
	\arrow["{\cong}"', from=1-1, to=3-1]
	\arrow["\alpha"', from=3-1, to=3-3]
	\arrow["\cong"', from=3-3, to=1-3]
\end{tikzcd}\]
The vertical equivalences follow from Theorem~\ref{thm:C} and the comment at the beginning of the proof.
The horizontal equivalence at the bottom is a consequence of the stability result of~\cite{ogata_derivation_2021} applied to the automorphism $\alpha$ of~\cite{Lootens_2022}. This completes the proof.
\end{proof}

\begin{rem}
    As pointed out in the introduction, this shows that $\caZ(\caD)$ completely characterizes the sector theory, but it does not require a priori knowledge of it. A proof that the superselection sectors for the Levin--Wen model with as input some unitary fusion category $\mathcal{D}$ is given by $\mathcal{Z}(\mathcal{D})$ has recently been announced~\cite{BolsKjaerI,BolsKjaerII}.
\end{rem}

\begin{rem}
A natural question is to determine under which additional condition two states $\omega_1,\omega_2$ (potentially defined on different systems) having equivalent superselection category are stably equivalent, namely there is an LGA connecting the two states, potentially after stacking with trivial systems.
For a special class of models (which roughly correspond to the `RG flow fix points') this follows from the results in~\cite{KimRanard}, which shows that the ground states of these models can be connected to a Levin--Wen string-net state. Although its very definition in a general setting remains unclear, see~\cite{sopenko2024} for recent progress, a non-vanishing chiral central charge is believed to be an obstruction. As far as we are aware, the general case remains conjectural.
\end{rem}

\subsection{Symmetry enriched toric code}
As a final example, we consider the \emph{symmetry enriched} toric code model presented in~\cite{Kawagoe25} (see also~\cite{OgataGarreRubio}), where there is an additional global symmetry.\footnote{We do not consider the most general case of symmetry enrichment, which also allows for permutations of the anyons (the ``$e$-$m$ swap'' for the toric code)~\cite{BarkeshliBondersonChengWang}.}
Consider the square lattice $\mathbb{Z}^2$.
As in the usual treatment of the toric code, we put a qubit $\mathbb{C}^2$ on each edge.
We give the edges an orientation by always having them point upwards or to the right.
If $e$ is an edge, we write $\partial_0 e$ for the vertex it points away from, and $\partial_1 e$ for the vertex it points to.
Furthermore, we write $e \ni v$ if the edge $e$ either points away from or towards $v$.
In the symmetry enriched version, put an additional qubit $\mathbb{C}^2$ at each \emph{vertex} in $\mathbb{Z}^2$.
To distinguish the degrees of freedom on the edges and vertices, we write $\sigma_e^k$ for the Pauli matrices on the edges, and $\tau_v^k$ for those on the vertices.

Recall that the interactions in the toric code are defined in terms of \emph{star} and \emph{plaquette} operators.
For a vertex $v$, a star is the set of all edges either starting or ending in $v$ (that is, $e \ni v$).
Similarly, a plaquette is formed by the edges along a face of the lattice.
For a face $f$, we write $e \in f$ for these edges.
The star and plaquette operators are then defined as\footnote{This is the standard convention for the toric code. These operators are the same as obtained for the quantum double model in Sect.~\ref{sec:qdexample} for $G = \mathbb{Z}_2$ up to a shift by the identity.}
\[
    A_v = \bigotimes_{e \ni v} \sigma_e^x, \quad\quad\quad B_f = \bigotimes_{e \in f} \sigma_e^z.
\]
With this notation, we define the following interaction terms:
\[
\widetilde{B}_f := i^{-\sum_{e \in f} \sigma_e^x (\tau_{\partial_1 e}^z -\tau_{\partial_0 e}^z)/2} B_f,
\quad\quad\quad
\widetilde{Q}_v := \frac{\mathbf{I}+A_v}{2} \tau_v^x i^{-\tau_v^z \sum_{e \ni v} f(e,v) \sigma_e^x/2}.
\]
Here, $f(e,v)$ is equal to $+1$ if $e$ points away from $v$, and $-1$ if $e$ points toward $v$.
Note that these dynamics couple the vertex with the edge degrees of freedom.
It can be shown that this Hamiltonian has a unique (and hence pure) frustration free ground sate, which we denote by $\omega_{\textrm{SET}}$.

We have the following results, which gives an alternative proof of~\cite[Prop. 7.25]{Kawagoe25}, which does not require mimicking the toric code analysis in the symmetry enriched case.
\begin{prop}
    The category $\Delta_{\operatorname{SET}}$ is unitarily braided equivalent to $\Delta_{\operatorname{TC}}$, the category of superselection sectors of the toric code.
\end{prop}
\begin{proof}
We can consider the original toric code model and the sites at the vertices as two layers of a stacked system (cf. Remark~\ref{rem:differentlattices}).
The first layer is the usual toric code model (see~\cite{kitaev2003fault,naaijkens2011localized}) and we write $\omega_{\mathrm{TC}}$ for its unique frustration free ground state.
The toric code satisfies Haag duality and the distal split property, and there are only finitely many irreducible sectors~\cite{naaijkens2013kosaki,naaijkens2015haag}, and hence the Type I property is satisfied and our analysis applies.
For the stacked layer of vertices, define the on-site Hamiltonian $H_v := (\mathbb{I}-\tau^x_v)/2$.
This has a unique frustration free ground state, which we call $\omega_{\textrm{sym}}$.
Note that this is a product state, and hence has trivial superselection structure by~\cite{naaijkens2022split}.
From~\cite[Sect. 7.2.2]{Kawagoe25} there is a finite-depth quantum circuit $\alpha$ such that $\omega_{\textrm{SET}} \circ \alpha = \omega_{\textrm{TC}} \stack \omega_{\textrm{sym}}$.
The result then follows from Theorem~\ref{thm:C} and~\cite{ogata_derivation_2021} (note in particular that~\cite{ogata_derivation_2021} implies that $\omega_{\operatorname{SET}}$ satisfies at least approximate Haag duality).
\end{proof}

The model has an on-site $\mathbb{Z}_2$-symmetry.
Define the unitary $U^g_v := \tau^x_v$ for all vertices $v$, and $U^g_e := \mathbb{I}$ for all edges, where $g$ is the non-trivial element of $\mathbb{Z}_2$.
This defines an action $g \mapsto \beta_g$ of $\mathbb{Z}_2$ on the quasi-local algebra by conjugating on each site.
The Hamiltonian above can be shown to be invariant under this action.
The on-site symmetry allows us to talk about \emph{symmetry defect sectors}.
Briefly, this gives a category of sectors which is graded by the group $G$.
The component with the trivial grading is the category of sectors is the category $\Delta_{\operatorname{TC}}$, but the symmetry enriched toric code has a non-trivial graded component that distinguishes it from the toric code model.
We refer to~\cite{Kawagoe25} for the details. 
\vspace{\baselineskip}

\noindent\textbf{Copyright statement.}
For the purpose of open access, the authors have applied a CC BY public copyright licence to any Author Accepted Manuscript version arising.

\printbibliography

\end{document}